\documentclass[twoside]{article}

\usepackage[preprint]{aistats2027}
\usepackage[round]{natbib}

\usepackage{enumitem}
\setlist[itemize]{leftmargin=*}
\setlist[enumerate]{leftmargin=*}

\newcommand{\R}{\mathbb{R}}
\usepackage{dsfont, amsmath, amssymb, amsthm}
\usepackage{xcolor}
\usepackage{subcaption}
\usepackage{algorithm}
\usepackage{algpseudocode}

\newcommand{\ds}[1]{\textcolor{black}{#1}}

\newtheorem{definition}{Definition}
\DeclareMathOperator*{\argmin}{arg\,min}

\usepackage{hyperref, cleveref, graphicx}

\begin{document}

% If your paper is accepted and the title of your paper is very long,
% the style will print as headings an error message. Use the following
% command to supply a shorter title of your paper so that it can be
% used as headings.
%
%\runningtitle{I use this title instead because the last one was very long}

% If your paper is accepted and the number of authors is large, the
% style will print as headings an error message. Use the following
% command to supply a shorter version of the author names so that
% they can be used as headings (for example, use only the surnames)
%
%\runningauthor{Surname 1, Surname 2, Surname 3, ...., Surname n}

\twocolumn[

\aistatstitle{Fair Variable Selection}

\aistatsauthor{ D\'eborah Sulem \And Jack Jewson }

\aistatsaddress{ Faculty of Informatics \\ Universit\`a della Svizzera italiana \And Department of Econometrics and Business Statistics \\ Monash University} ]

\begin{abstract}
    %Algorithms are increasingly being used to help automate and improve data-driven decisions, but 
    Care must be taken to prevent algorithms used to help automate and improve data-driven decisions from learning discriminatory patterns from historical data and perpetuating their biases. 
    Statistical notions of fairness aim to mitigate either a model's \emph{disparate impact} on disadvantaged groups (ensuring \emph{group fairness}) or the resulting \emph{disparate treatment} of individuals with similar features  (ensuring \emph{individual fairness}). %Disparate impact and disparate treatment are two notions of (un)fairness that are generally incompatible except for trivial models.
    Simultaneously mitigating disparate impact and disparate treatment is generally impossible, necessitating a compromise.
    %\jack{Achieving both is challenging because ...}
    %There is increasing interest in using algorithms to aid decision making, and also in preventing potential discriminatory patterns that could be learnt from historical data. 
    %Conditional statistical parity (CDP) is a paradigm for statistical fairness that aims to mitigate both a model's \emph{disparate impact} on disadvantaged groups and the resulting \emph{disparate treatment} of individuals with similar features.
    %Conditional statistical parity (CDP) is one fairness constraint that can be enforced in the learning process to mitigate both \emph{disparate impact} on disadvantaged groups and \emph{disparate treatment} of individuals with similar features. % but from different demographic groups.
    %CDP relies on the choice of \emph{legitimate covariates}, a choice that no previous statistical methodology addresses. 
    %A key aspect of CDP is its requirement to select a set of \emph{legitimate covariates}, \ds{i.e., predictors that permit a decision maker to legitimately discriminate individuals, according to a legal framework or some ad-hoc notion of equity.}   \jack{need to define these, say why this is beneficial and then say why this is problematic}
    In this paper, we introduce the Fair Lasso %-type
    %\jack{Don't need bayesian in the title then} 
    and Fair $\ell_1$-ball prior as methods for selecting fair covariates in generalised linear models. %\emph{legitimate covariates}, 
    By targeting variables that are simultaneously strong predictors of the response and weakly dependent on the sensitive group memberships, we aim to achieve favourable trade-offs between
    %on the basis that they should be both strong predictors and weakly dependent on the group memberships. Within the framework of 
    %Our method is applicable to 
    %generalised linear models, 
    %our methods 
    %mitigating 
    disparate treatment and disparate impact.
    %and achieve favourable trade-offs between these two incompatible criteria. 
    %Additionally, 
    Our selected features can additionally be used as %the conditioning set of 
    \emph{legitimate features} in the paradigm of Conditional Demographic parity (CDP) when no prescriptive legal framework exists.
    %and can incorporate any prior distribution.  
\end{abstract}

\section{Motivation}

Statistical modeling and learning algorithms are increasingly used for making predictions and decisions that directly affect individuals. For instance, judges can use risk assessment tools to estimate the re-offending risk %of a convicted defendant and hence determine their sentence 
\citep{berk2021fairness}, banks use credit scoring algorithms to minimise the risk of loan defaults  \citep{hurlin2026fairness} %\citep{kozodoi2022fairness}
and hiring managers  use %CV 
screening algorithms to select candidates for interviews \citep{raghavan2020mitigating}.
%in an interview \citep{barocas-hardt-narayanan}. %\jack{get AI to give us some more references here - I know Nicolas Schreuder has done some stats privacy works and had a sentencing example somewhere }

However, it is widely recognised that biases and discrimination can emerge from such data-informed decisions, when the model's training data incorporate sampling bias (e.g., under-representing a demographic group) or reflect historical and societal inequalities %(in particular towards minority groups) %and discriminatory patterns that underpin our societies 
\citep{Castelnovo_2022}. %Without any proper methodology for dealing with those biases, 
%Such biases (or \emph{unfair patterns}) are simply reproduced or even amplified by learning algorithms and statistical models that are trained to seek the dominant data pattern.
%\jack{Can we say that this is reasonbale, most rraining algoruthms aim to predict past data, need to change hopw the model is trained to ensure fairness}
%\jack{Pre, in and post processing here}
\emph{Fair} learning, %from biased data implies 
%therefore, 
requires correcting for these biases either by pre-processing the data, changing the learning  objective or algorithm (in-processing) or post-processing the predictions. % obtained via a standard method. 
%This paper's focus is on in-processing methods.
%modifying the data, the statistical model \jack{we modify the learning algorithm/prior right? Does this count as the model?} or the objective function. 
%Choosing the approach to address a particular unfair data problem is complex since the unfairness mechanism %underlying those biases 
%is often unknown or vaguely defined. %Alternatively, mitigating the effect of those biases within the statistical methodology can be done via in-processing or post-processing a model's predictions. 
% The myriad of methods and arguments supporting their underlying philosophy illustrates the problem's complexity (see for instance the recent reviews of \cite{mitchell2021algorithmic, berk2023fair, he2025fairness}). \jack{What exactly are you trying to sy here? That a one size fits all approach is very hard?}

Often, the unfair data pattern comes from a dependence between a target outcome $Y$ and a group of so-called \emph{sensitive attributes} $S$ (e.g., gender, race, skin color, sexual orientation, or religion). Most commonly, $S$ is a categorical variable which divides a population of individuals into demographic groups. 
%, but in our framework it can be discrete or continuous. %, although this attribute is deemed irrelevant for predicting this outcome (e.g., loan repayment). 
Although, these attributes may be predictive of the outcome in the training data, this dependence is deemed non-causal and one that the trained model should not capture at prediction time. 
%
%Biased or unfair patterns are often defined in terms of the model's outcome distribution within multiple demographic groups, defined by a characteristic such as gender, race, skin color, sexual orientation, or religion. Such characteristics are often called \emph{sensitive} or \emph{protected} attributes, and, in many contexts such as hiring, university admissions, or sentencing, considered irrevelant for the prediction task at hand. 
Let $f(X, S)$ be a prediction model for $Y$ based on a $d$-dimensional vector of non-sensitive covariates $X$ and that could also depend on $S$.
%either directly or indirectly (see, e.g., \cite{scutari2022achieving} which subtracts from $X$ its orthogonal projection onto the span of $S$).
%A prediction model whose outcome %distribution 
If $f(X, S)$ is independent of $S$, %the sensitive attribute, 
then it satisfies
%implies satisfying a fairness constraint called %verifies a \emph{fairness} criterion known as 
\emph{demographic parity} (DP), also known as \emph{group fairness}.
%\jack{Are these two names for the same thing?}.

\begin{definition}[Demographic parity / group fairness]
%Given a sensitive attribute $S$ and covariates $X$, 
A prediction model $f(X,S)$ %\jack{properly defined? Would you ever use $S$ in the model?} 
satisfies demographic parity (DP) (or, group fairness) if
\begin{align*}
    f(X,S) \perp S.
\end{align*}
\end{definition}
Intuitively, DP requires equal distribution $f(X,S) \mid S=s$ for all possible values $s$ of $S$, and is thus said to remove \emph{disparate impact} on demographic groups. % defined via $S$.
Note that %the definition of 
DP 
%does not involve the outcome $Y$, 
%therefore in practice, it 
can be satisfied without knowing $Y$. 
%even when %no reliable 
%ground-truth labels are not available, for instance, at prediction time on new data.
%for a dataset without observed outcomes.
 %\jack{What do you mean here, that the data is bad, or the model is bad?}. 
%Besides, we note that the presence of $S$ as an input of $f$ can in some scenarios help to achieve DP, e.g., by removing the dependence on $S$ due to $X$ (see, e.g., \cite{scutari2022achieving} which subtracts from $X$ its orthogonal projection onto the span of $S$).
%}

The utility of DP has been largely debated. Intuitively, it ensures that model-informed decisions lead to equally distributed resources across demographic groups 
%(thus preventing \emph{disparate impact}) 
and is often assimilated to \emph{affirmative action} \citep{dwork2012fairness}. More specifically, DP-fair models preserve the ranking of predicted outcomes within each demographic group but alter the global ranking by aligning the groups 
%\jack{However, `aligning' the groups} generally alters the global ranking of individuals' predicted outcomes
\citep{chzhen2020fairb}. %DP is often criticised for 
This can create situations where two similar individuals receive different predictions because they belong to different demographic groups.
%\jack{situations where two similar individuals in different demographic groups receive different predictions, as the group-specific ranking prevails over the global ranking. 
This results in \emph{disparate treatment} of individuals based on their demographic group and violates \emph{individual fairness} (IF). %\citep{dwork2012fairness}). %  \jack{can we give an example}, .
\begin{definition}[Individual Fairness] Assuming a distance metric $\mathbf{d}$%between two individuals with observed variables $\Tilde X_1 = (X_1, Y_1)$ and $\Tilde X_2 = (X_2, Y_2)$
, a model $f$ is individually fair if it is Lipschitz wrt $\mathbf{d}$, i.e., there exists $L>0$ such that for any $\{X_1, Y_1, S_1\}$, $\{X_2, Y_2, S_2\}$, 
\begin{align*}
    %\mathbf{d}(\Tilde X_1, \Tilde X_2) \approx 0 \implies f(\Tilde X_1) \approx f(\Tilde X_2).
        | f(X_1, S_1)- f(X_2, S_2) | \leq L \mathbf{d}(\{X_1, Y_1\}, \{X_2, Y_2\}).
\end{align*}
\end{definition}
Note that in this definition $\mathbf{d}$ %the distance between individuals intentionally 
may depend on the covariates $X$ and/or the %observed 
outcome $Y$, but importantly not on the sensitive attribute $S$. % which is in most cases considered irrelevant in fairness-sensitive prediction tasks.
%\jack{Should we note that the sensitive attribute definitely doesn't enter the distance here} 
%In \cite{dwork2012fairness} (see also \cite{gillen2018online}), individual fairness is defined as a Lipschitz constraint on $f$ wrt the metric $\mathbf d$, but 
%Intuitively, individual fairness requires that two similar individuals regardless of their sensitive attribute %(for which $\mathbf{d}(\Tilde X_1, \Tilde X_2)$ is small) 
%have similar predictions 
%$f(X_1)$ 
%\jack{Should be $f(X_1)$ right} 
%and $ f(X_2)$ 
%and therefore avoids \emph{disparate treatment}.
However, this definition depends on the  choice of $\mathbf d$ and $L$, which has proven an obstacle to its practical use. % of this fairness criterion in a learning objective.  %\cite{berk2017convexframeworkfairregression} proposed distances that only depend on the observed outcome, e.g., $\mathbf{d}(Y_1, Y_2) = e^{-|Y_1 - Y_2|}$ for continuous outcomes. \jack{}

Group and individual fairness are generally incompatible \citep{dwork2012fairness} and finding an adequate and acceptable compromise between the two is still an open question. The incompatibility results from the fact that often not only the targeted outcome depends on the sensitive attribute, but the non-sensitive %standard
(or, relevant) features in $X$ (e.g., education, place of residence, family size, etc) do too. Removing this %indirect 
dependence without creating disparate treatment and breaking the model's accuracy %predictive power 
is particularly challenging.

A compromise lies in  the notion of %related fairness criterion (neither stronger nor weaker), known as 
\emph{conditional demographic parity} (CDP), which allows some covariates to legitimately account for differences in predicted outcome distributions amongst %the demographic 
groups, while equalizing the conditional distributions given these legitimate %(or `fair') 
covariates. %Specifically, CDP imposes conditional independence between the outcome and the sensitive attribute, after conditioning on some \emph{legitimate variables}. 
%The latter are covariates  that are, for example, known to be strongly associated with the outcome and for which altering their influence by requesting (unconditional) independence would be too prejudicial. 
For instance, a legitimate variable in a hiring assessment problem  could be the qualification level: %In this context, 
a CDP-fair predictive model then guarantees predictive parity for individuals sharing the % decision maker may want a model that has the same predictive properties for individuals with the 
same %level of 
qualifications. % and therefore limits individual unfairness.

\begin{definition}[Conditional Demographic Parity]
    Given a %sensitive attribute $S$, covariates $X$, and a 
    subset $X_{D}$ of \emph{legitimate covariates}% (where $\bar D \subset \{1, 2, \dots, d\}$)
    , a prediction model $f(X,S)$ satisfies conditional demographic parity (CDP) if
    \begin{align*}
    f(X,S) \perp S \mid X_{D}.
\end{align*}
\end{definition}
%
%
%Legitimate covariates are variables that can \emph{legitimately} explain differences between predicted outcome distribution between groups. For instance, if the predicted outcome is job hiring and $S$ corresponds to gender or race, a legitimate covariate could be the level of education. }

%\jack{[This paragraph is great, but no longer aligns with the idea that CDP is a secondary thing not the main goal]}
Previous work \citep{ritov2017conditional, ghassemi2025auditing} consider that legitimate covariates %$X_{\bar D}$ 
are imposed by a policy maker or legal framework.
%\ds{While the set of legitimate covariates can be imposed by a legal framework \citep{ghassemi2025auditing}, 
However, many prediction contexts lack such normative guidance.
%and the methods we propose in this paper can fill this gap.
%a principled selection method is currently lacking. This paper aims at filling this gap. 

%Our core idea is to define fair (or, legitimate) covariates as being 
Our core idea is to use variable selection as a means to learn fair prediction models that balance DP and IF, while providing a means to select legitimate covariates for %the 
CDP. %framework. 
In particular, we encourage the selection of covariates that are simultaneously
strong predictors of the outcome $Y$, encouraging IF,  %of the outcome and if they are 
and weakly dependent on the sensitive attributes $S$, encouraging DP. Such a definition is heuristically related to procedural fairness \citep{grgic2018beyond}.% and %results in CDP-fair models that
%strikes a trade-off between DP and individual fairness, and provides a methods to select legitimate covariates for the CDP framework. %(see Section \ref{sec:methodo} and experimental results in Section \ref{sec:experiments}). 

To operationalise this, we extend 
%standard approaches for variable selection such as
Lasso penalised likelihood estimation \citep{tibshirani1996regression} and the sparse $\ell_1$-ball prior Bayesian analog \citep{xu2020bayesian}. %In particular, 
Bayesian methods 
%are relevant here for their ability to 
quantify uncertainty on the set of selected covariates and as a result better control false positives \citep{rossell2018tractable}.
Within generalised linear models (GLMs), our Fair Lasso and Fair $\ell_1$-ball prior are directly interpretable. The selection of each covariate is penalised proportionally to how dependent it is on $S$ %the sensitive attributes, 
and the size of its corresponding GLM coefficient.
%regression coefficient within the GLM. 
The selected set of fair variables and %the fair 
model then depend on a target unfairness budget. 
%or relative improvement with respect to an unrestricted model.
%relies for the context of gene and a sIn addition to ensuring fairness, our proposed method is transparent and de-facto interpretable \jack{is this just coming from the interpretability of VS for linear models}: our variable selection method is based on a modified LASSO objective where each variable is given a different penalisation parameter set by a measure of marginal dependence between this variable and the sensitive attribute. \jack{I wonder if this high level notion should be defined before the bullet points}%Besides, the framework of generalised linear model allows a direct interpretation of the model's predictions. Overall, 
%This makes our approach applicable to a variety of learning contexts were fairness and interpretability are both needed.
More specifically, our contributions are:
\begin{itemize}[noitemsep]
    \item %In the context of generalised linear models
    %linear regression (and more generally, generalised linear models), 
    We introduce the Fair Lasso, an adaptation of the Lasso for fair
    %legitimate \jack{make sure legitimate is ver clearly defined before} 
    variable selection in GLMs, and %. We also intuitively 
    show that the implied fair model strikes a trade-off between DP and IF;   % that %selects the set of legitimate covariates (as previously defined) and 
    %also produces a CDP-fair %(generalised) linear  model. \jack{That also produce CDP, is it not designed to?}
    \item We design a Bayesian counterpart %of the previous variable selection methods, 
    that produces a Fair Posterior distribution over the inclusion and exclusion of covariates and parameters; %estimation 
    %: a \emph{fair} posterior distribution supported on a set of selected fair %legitimate  covariates; %and that we call CDP-fair posterior. % using the \emph{projection-posterior} (or \emph{immersion posterior}) framework \citep{pal2025bayesianhighdimensionallinearregression}. 
    %With the Bayesian posterior, we thus enhance our CDP-fair estimation method with uncertainty quantification.
    \item We show, via Pareto fronts and risk-fairness performance comparisons, %our approaches 
    %on synthetic data and two real-world benchmark datasets 
    that our methods  %and show that our method
    achieve advantageous trade-offs between predictive performance, DP and IF compared to the state-of-the-art. 
    %existing methods that only target one of these objectives.
    %and different (incompatible) notions of fairness compared to existing methods. \ds{to improve}
    %As both methods are able to incorporate some level of relaxation of the CDP criterion, allowing the model user to tune the trade-off between prediction accuracy and fairness. Additionally, the Bayesian approach allows uncertainty quantification in the context of fairness-constrained models.
\end{itemize}

\textbf{Outline:} In Section \ref{sec:litt}, we review related work; highlighting the scarcity of work related to fair feature selection, CDP, and uncertainty quantification. We then present our methodology in Section \ref{sec:methodo} and report our numerical results in Section \ref{sec:experiments} before concluding.

\section{Related work}\label{sec:litt}

%\jack{We then need to do a sligtly better job of setting up this compromise between DP and IU with CDP in the middle - for example its not obvious that unconstrained and IU align}

%Group and individual fairness are two prominent notions of statistical (or, algorithmic) fairness. Group fairness (also known as \emph{demographic parity} (DP)) aims to prevent \emph{disparate impact} of the predicted outcome on demographic groups, i.e., the distribution of this predicted outcome should be the same for all groups. In constrast, individual fairness aims to remove \emph{disparate treatment} of individuals, i.e., for two individuals which are \emph{similar} for the prediction task, their predicted outcome should be close. 
%%Existing work in the fairness literature has mostly focuses on designing algorithms that satisfy \emph{demographic parity} (DP),  \emph{equalised odds} (EO) or relaxed versions of these criteria (see, e.g., the recent review by \cite{he2025fairness}). %\jack{maybe cite the reviews form before if they corroborate this}.
%%\ds{These two notions of fairness reflect different world views and assumptions about the data generating process:  DP assumes that the observed outcome $Y$ is unfair to certain demographic groups, while EO prevents an algorithm to make  different predictions (and hence have different performance) for individuals which are similar for the prediction tasks (i.e., share the same outcome). 
%However,  are generally incompatible
%, therefore existing work generally targets only one of the two.
Existing work generally targets either DP or IF, and not the two simultaneously. 
%
%
%
%Recall that DP is a \emph{group parity} criterion, while EO \jack{does this have a citation} ensures a parity between similar individuals (similarity defined on the basis of the true outcome). Few works have investigated trade-offs between group and individual fairness, e.g., via CDP, fair feature selection, and uncertainty quantification (UQ) methods. \jack{I would say it s better to review the literature and conclude with this, that have this before even mentioning the literature.} We first review standard fairness approaches, then methods that relate to CDP and fair feature selection and finally methods combining fairness and UQ.
%Many works in the realm of statistical fairness focus on the demographic parity criterion, for its simplicity and relevance in specific applications. In addition, in regression contexts, 
%
Such methods %targeting group or individual fairness 
often rely on a penalised objective or constrained model spaces.
In the context of linear regression, \cite{calders2013controlling} use the mean difference between the predicted outcomes of two demographic groups as the penalisation function and %as a 
proxy for DP violation. For regression and classification, \cite{berk2017convexframeworkfairregression} define a measure of violation of  group (resp., individual) unfairness that results in a convex penalised objective. For nonparametric regression, \citep{gouic2020projection, chzhen2020fairb} showed that the optimal DP-fair predictor under the  $L_2$-risk %is analytically tractable and can be expressed as 
is a Wasserstein barycenter and %an estimate 
can be obtained by post-processing a base predictor with no fairness constraint. %This analysis is extended in  \cite{chzhen2022minimax} for satisfying an approximate notion of DP. %and achieving trade-off between accuracy and fairness is sought
\cite{oneto2020general} design a generic constrained empirical risk minimisation framework for approximate fair learning based on discretising the outcome and sensitive attributes. Targeting DP for (G)LMs, \cite{komiyama2018nonconvex} and \cite{scutari2022achieving} employ as penalisation the proportion of outcome variance explained by the sensitive attribute. % and measures a violation of DP. %then the latter quantity is used as a penalisation function.
%In \cite{calders2013controlling}, the penalisation function is the distance between residuals across groups - a penalisation which is in fact null for the OLS estimate in linear regression. \jack{reverse the ordering so they are presented in chronological order}
%The \emph{separation} fairness criterion (imposing conditional independence between the predicted outcome distribution and the sensitive attribute, given the ground truth outcome) can also be (approximately) enforced within the generalised fair empirical risk minimisation of \citep{oneto2020general}. A related criterion to \emph{separation} consists of enforcing equalised residuals across groups - a criterion which is achieved \emph{for free} by the OLS estimate in the standard linear regression setting \citep{calders2013controlling}.
%Most of these methods rely on defining a regularisation function measuring the violation of the chosen type of fairness criterion. In that way, the penalised estimate achieves a trade-off between accuracy and a specific notion of fairness. Nonetheless, DP-related criteria and individual fairness criterion have their limitations and are generally incompatible. 

%\ds{paragraph on individual fairness}

Finding a trade-off or compromise between DP and IF is challenging. One possible approach is \emph{counterfactual fairness} (CF) criterion \citep{kusner2017counterfactual}, which relies on %the 
causal inference. %framework. 
Intuitively, CF %achieves a compromise between group and individual fairness by 
equalise counterfactual distributions when the sensitive attribute is intervened upon. We call it a compromise between DP and IF since, although the model correction implied by CF is based on $S$, %the group membership,
it is individualised and explainable via the causal graph. However, this methodology requires knowledge or inference of the causal graph, %and strong assumptions,
which can make this approach impractical. %\jack{here or in related work section, so far you have built a stroy, this is obvious relevant but not sure it advances the story}

%Another approach lies in the Conditional Demographic Parity (CDP) criterion, which requires predicted outcome distributions to be equal across groups, conditionally on some predetermined \emph{legitimate} covariates (or \emph{discriminatory attributes}). 
All existing CDP methods require the set of \emph{legitimate} covariates %(or \emph{discriminatory attributes})
to be predetermined. %Methods achieving CDP are much rarer, perhaps due to the lack of principle way of choosing the set of legitimate features. 
\cite{ritov2017conditional} introduce a pre-processing step of the features so that any prediction model %built on the latter 
satisfies CDP. Assuming that the  legitimate covariates take discrete values, \cite{ghassemi2025auditing} define a measure of violation of CDP based on nested Wasserstein distances  and solve a penalised objective.
%where the penalisation is \emph{conditional demographic disparity} measure based on nested Wasserstein distances. \jack{complexity is nto on its own bad, is it computationally a problem, is it non-interpretable?} 
%\jack{Again, reorder chronologically} %and uses as a regularisation function in a computationally costly optimisation problem. However, in this method the set of legitimate features is assumed to be given.
However, in most %of their 
applications, there is no legal framework guiding the choice of legitimate covariates and the latter is made in an \emph{ad-hoc} way.

% Penalisation methods with fairness-related regulariser %critera specific to classication contexts (e.g., \emph{equality of odds}, \emph{equality of opportunity}, etc)
% allow to construct models achieving particular fairness-accuracy trade-offs. Most existing work targets the DP criterion \citep{scutari2022achieving, chzhen2020fairb} or classification-specific fairness criteria such as \emph{equality of odds} or \emph{equality of opportunity}. For CDP, \cite{ghassemi2025auditing} proposed a \emph{conditional demographic disparity} measure based on nested Wasserstein distance as a regulariser. Such criterion being non-regular, tractable approximations are needed. Nonetheless, in \cite{ghassemi2025auditing}, the set of legitimate features is assumed to be given.

%Besides, there are only few a methods for selecting \emph{fair} features.
%(not learning new fair features).  
%In this work we design CDP-fair methods that are approximately DP-fair. 
%Instead, in this work, we are interested in selecting features that strike an optimal trade-off between CDP and model accuracy. 
%Our goal is also loosely connected with 
\emph{Suppression methods} \citep{kamiran2012data, kamiran2013quantifying} are perhaps the first attempts for selecting \emph{fair} features: they remove variables that have high correlation with $S$%the sensitive attribute%(and the sensitive attribute itself)
%\jack{?? is one outcome}
, without considerations of their quality as predictors. 
%Other fairness-aware feature selection methods include \citep{grgic2018beyond, khodadadian2021information}. 
\cite{grgic2018beyond} finds an optimal set of features balancing accuracy and fairness by exploring all possible subsets, % of features, 
a strategy which does not scale beyond a small number of covariates. \cite{belitz2021automating} improve this approach in classification by applying a \emph{greedy algorithm to fair feature selection.} In \cite{khodadadian2021information},  the fairness of a set of features is assessed via information measures, independently of the model and downstream task. %, i.e., independently of \emph{how} the features are used for predictions, and 
In contrast, our method %aims to 
selects fair 
%(or, legitimate) 
variables by balancing predictive accuracy for a specific task. %\jack{something about balancing predictive accuracy} 

We further highlight an important distinction between our work and  approaches to build fair features via transformations of the original covariates \cite{zemel2013learning, louizos2016variational}. By selecting the features rather than transforming them, 
%we build a set of approximately fair features without 
we do not alter their interpretability and ensure  transparency in what is determined ``legitimate'' or ``fair''.

%However, in previous work the \emph{suppressed} variables are assumed to be externally given. Our approach is to select the set of legitimate covariates that achieves an advantageous trade-off between predictive power and DP-fairness. As explained in the next section, with such set, one achieves CDP \emph{by design} by conditioning on those legitimate covariates.
%while keeping as much information about the outcome as possible. However here we select the variables using both their prediction power and their correlation with sensitive attribute.

In the absence of fairness considerations, variable selection is a standard task in statistical inference. Penalised  likelihood methods such as Lasso \citep{tibshirani1996regression} and their variants are largely employed in high-dimensional GLMs. %generalised linear models. 
In the Bayesian framework, sparse priors such as spike-and-slab \citep{mitchell1988bayesian, george1993variable} are used to produce posterior distributions over the inclusion and exclusion of variables. They are advantageous in producing sparse posteriors compared to shrinkage priors such as Bayesian Lasso \citep{park2008bayesian} for which only the mode %(maximum a-posteriori) 
is sparse. Recently, the $\ell_1$-ball prior \citep{xu2020bayesian} was introduced as a computationally convenient alternative to the spike-and-slab.  However, these methods 
%target improved out-of-sample performance and interpretability, and 
do not currently incorporate fairness constraints.

\section{Fairness in GLMs}\label{sec:methodo}

%\subsection{\ds{Fairness in Generalised Linear Models}}\label{sec:set-up}

We consider a learning set-up where observations are $n$ realisations $\mathcal{D}_n :=\{(x_i,y_i,s_i)\}_{1 \leq i \leq n}$ from a random vector $(X,S,Y)$.
%, where $X$ %$X \in \R^d$ (covariates) 
%is a $d$-dimensional vector of covariates, $Y$ a real-valued target outcome and  $S$ is a real-valued %(continuous or discrete) 
%sensitive attribute. Most commonly, $S$ is a categorical variable but in our framework it can be discrete or continuous. 
%it can represent membership to a demographic group, e.g., genders male and female can be represented by a binary sensitive attribute $S \in \{0,1\}$. %For simplicity of presentation, we assume $S$ is binary, though all of methodology can handle more than two sensitive groups and even $S$ to be continuous. However, 
%\ds{In this work, we assume that there is only one sensitive attribute.}
%We denote by $X$ %$X \in \R^d$ (covariates) 
%a $d$-dimensional vector of covariates and $Y$ a real-valued outcome. 
%\ds{extension to discrete outcomes and generalised linear moels?}. 
%and $Y \in \R$ (outcome):
% \begin{align*}
%     &S \sim Ber(p) \\
%     &X |S=s\sim N(\mu^s, \Sigma^s) \\
%     &Y|X,S \sim N(\alpha S + \beta^TX , \sigma^2).
% \end{align*}
We assume the following generalised linear data-generative model for the outcome $Y$: %\jack{Aren't we doing GLMs? Could introduce generally and then have special cases for the link function and error distribution}
\begin{align}\label{eq:glm}
    Y = g\left(\mu^*\right) + \epsilon, \qquad \mu^* = \beta_0^* + \alpha^* S + (\beta^*)^TX , %\qquad     \mathbb E[\epsilon \mid \mu] = 0, \qquad \mathbb V[\epsilon \mid \mu] = v(\mu), %\qquad \epsilon \sim N(0,\sigma^2),
\end{align}
where $\alpha^* \in \mathbb R$, $\beta_0^* \in \mathbb R$, $\beta^* \in \R^d$, $g$ is an invertible function $ \mathbb E[\epsilon \mid \mu^*] = 0$, and $\mathbb V[\epsilon \mid \mu^*] = v(\mu^*)$. %, %$\epsilon \sim N(0,\sigma^2)$ is a Gaussian noise independent of $X$ and 
%$\sigma^2 > 0$, and 
%$\epsilon$ is the noise 
%and
%and verifies:
% \begin{align*}
%     \mathbb E[\epsilon \mid \mu] = 0, \qquad \mathbb V[\epsilon \mid \mu] = v(\mu)
% \end{align*}
%where 
%$\mu = \alpha^\star S + (\beta^\star)^TX $.
%$\epsilon \perp (X,S)$. 
The Gaussian linear regression model corresponds to \eqref{eq:glm} where $g$ is the identity,  $v(\mu^*) = \sigma^2 > 0$ a constant, and $\epsilon \sim N(0,\sigma^2)$  %Equivalently, the linear regression model can be written as
%\begin{align}\label{eq:true-model}
%    Y|X,S \sim N(\beta_0^* + \alpha^\star S + (\beta^\star)^TX , \sigma^2),
%\end{align}
while for the logistic regression model, %is also an instance of  \eqref{eq:glm} where 
$g$ is the logistic (i.e., sigmoid) function, $v(\mu^*) = g(\mu^*)(1-g(\mu^*) )$ and $\epsilon \in \{1-g(\mu^*), -g(\mu^*)\}$ with $\mathbb P(\epsilon = 1-g(\mu^*)) = g(\mu^*)$. 
%Besides, we denote by  $\mathcal{L}(\mathcal{D}_n; \alpha, \beta_0, \beta) = \sum_{i=1}^n \ell(y_i, x_i ; \alpha, \beta_0, \beta) $ the negative log-likelihood of the data $\mathcal{D}_n$ under model \eqref{eq:glm} for a set of parameter $(\alpha, \beta, \beta)$. 
%\ds{Let us also denote by $f^*(X) = g(\beta_0^\star + (\beta^\star)^TX)$ the expected outcome when the effect of $S$ is discarded from the model.}
We also assume that $g$ is either globally Lipschitz or is locally Lipschitz and $\mu$ belongs to a bounded space.

%\ds{add true model $(\alpha^*, \beta^*)$}

%\ds{separate DP and IU}

%We denote by $f^*(X,S) = g(\beta_0^* +\alpha^\star S + (\beta^\star)^TX) = \mathbb E[Y \mid X,S] $ the true model with parameter $(\alpha^*, \beta_0^*, \beta^*)$ and by $f(X,S)$ a generic prediction model. 

%We now recall the definition of DP for a generic prediction model $f(X,S)$.

%\begin{definition}[Demographic parity / group fairness]
%Given a sensitive attribute $S$ and covariates $X$, a prediction model $f(X,S)$ %%\jack{properly defined? Would you ever use $S$ in the model?} 
%satisfies demographic parity (DP) (or, group fairness) if
%\begin{align*}
%    f(X,S) \perp S.
%\end{align*}
%\end{definition}

%Intuitively, DP requires equal distribution $f(X,S) \mid S=s$ for all possible values $s$ of $S$, and is thus said to remove \emph{disparate impact} on demographic groups defined via $S$.
%\jack{Do we want to talk about a true model here, or any model with $\alpha = 0$ }
%Under model \eqref{eq:glm}, i.e., 
We estimate a GLM where the direct effect of $S$ is suppressed, i.e., models of the form $f_\beta(X) = g(\beta_0 + \beta^T X)$.
%\jack{Let  $\mathcal{L}(\mathcal{D}_n;\beta_0, \beta) = \sum_{i=1}^n \ell(y_i, x_i ; \beta_0, \beta)$ be the negative log-likelihood of the data $\mathcal{D}_n$ under model \eqref{eq:glm} for a set of parameter $(\beta_0, \beta)$ with $\alpha = 0$.} 
As a result, $f_\beta(X)$ is
%$f(X,S) = g(\beta_0 + \alpha S + \beta^T X)$, with
%$\alpha = 0$ (i.e., $S$ is not included in model $f$), 
 partially independent of $S$, %$f(X,S)$, 
 i.e.,
%which means that the dependence between $Y$ and $S$ only comes from the dependence between $S$ and $X$. Formally,  $\alpha^\star = 0$ implies %translates into a conditional independence statement: 
$f_\beta(X) \perp S \mid X$.
We note that if the covariates $X$ are independent of $S$, 
%conditionally on $Y$, %then $Y \perp S$ and %
%i.e., $X \perp S \mid Y$, 
then $f_\beta(X) \perp S$, i.e., any such model 
%(in particular $f^*(X)$) 
%and thus $f(X,S) = g(\beta_0 + \beta^T X)$ 
is DP-fair. Otherwise, excluding $S$ from $f_\beta(X)$ is not generally enough to satisfy DP. %Intuitively, this means that 
%Excluding $S$ from the model is not generally enough to satisfy DP. %, unless covariates are all conditionally independent of $S$.}
%Otherwise, if $X$ is not independent of $S$ conditionally on $Y$, then even if $\alpha = 0$,  $f(X,S)=g(\beta_0 + \beta^T X)$  %can be dependent on $S$ if $X$ is not independent of $S$, therefore such $f$ 
% is generally not DP-fair. 
%Intuitively, this means that excluding $S$ from the model is not enough to satisfy DP, unless all covariates are all conditionally independent of $S$. %\jack{I would give the definition first and then relate back to the model}
%
%
%
%
This leads us to introduce a notion of DP fairness for a set of variables. %each variable. %, which will allow to design models that are DP-fair. 
\begin{definition}[DP-fair variable set]\label{def:dp-var}
A set of variables $X_D$ is DP-fair wrt sensitive attributes $S$ if $X_D \perp S$.
\end{definition}
It is straightforward to see that if $X_D$ is DP-fair,
%$D \subset \{1, \dots, p\}$ is such that $X_j$ is DP-fair for each $j \in D$, and denoting by $X_D$ the subset of variables indexed by $D$, 
then a model $f_\beta(X) = f_\beta(X_D) = g(\beta_0 + \beta_D^T X_D)$, i.e., $\beta_j = 0$ for $j\not\in D$, is DP-fair.
%\ds{and individually fair, for any distance metric $\mathbf{d}$ based on $X_D$. }
%\jack{[But don't we use $Y$ in our experiments for this distance? Would be good to have a reference for this approach ]}
%\ds{For evaluating individual fairness, an ideal distance metric  $\mathbf{d}$ would be based on %data-generating prediction after removing the sensitive attribute, 
%[can we call this something like the legitimate prediction and define in when we first define $f^{\ast}$]} 
%$Z^* =f^*(X)$, e.g., $\mathbf{d}(z,z') = (z-z')^2$. 
%$f^*(X)$, i.e. $f^*(X)$ is individually fair. However,
%\ds{However, if $\mathbf{d}$ is based on $Y^* = g(\beta_0^\star + (\beta^\star)^TX)$ and
%any model $f_\beta(X_D)$ where $X_D$ excludes some variables truly predictive of $Z^*$ (i.e., for which $\beta^*_j \neq 0$), then $f(X_D)$ is generally not individually fair.} 
For evaluating its IF, \cite{berk2017convexframeworkfairregression} argue that the distance metric  $\mathbf{d}$ should only depend on the observed response $Y$. Such a choice requires covariates that truly create variation in $Y$ to be included in the model, and irrelevant ones discarded. As a result,  if $X_D$ excludes variables truly predictive of $Y$, $f_\beta(X_D)$ will generally not be individually fair.

%\jack{Berk says to use $Y$ in the distance of IF, this implies covariates that create variation in $Y$, should be in the model, and irrelevant ones discarded, but we've established that predictive variables are probably not DP-fair.}

%\jack{[Better than this is underwhat circumstances is the GLM model IU] However, we believe that $X_D$ may exclude variables truly predictive of $Y$ and therefore that $f(X_D)$ results in disparate treatment and is, therefore, not individually fair.}

However, a compromise between DP and IF can be reached if a subset $X_D$ of \emph{approximately} DP-fair covariates that includes the \emph{most} relevant predictors of $Y$ can be found. The resulting $f_\beta(X_D)  = g(\beta_0 + \beta_D^T X_D)$ would then be \emph{approximately} DP-fair and individually fair. %, thus allowing for the mitigation of both disparate impact and disparate treatment.
Further, $f_\beta(X_D) = g(\beta_0 + \beta_D^T X_D)$ is CDP-fair with respect to variables $X_D$, and we argue that in the absence of a legal regulatory framework, balancing DP and IF provides a logical way to decide upon ``legitimate'' features.

Our goal is thus to select a set $X_D$ of approximately DP-fair covariates that are also strong predictors of $Y$. %$Z^*$ \ds{(and thus of $g(\mu^*)$ `if the ground-truth model is such that $\alpha^* = 0$).} 
%Those variables can be %interpreted as reflecting similarity between individuals for the prediction task and thus 
%considered \emph{legitimate} for the prediction task when no policy or legal framework prescribes this choice in the CDP criterion. 
%Then, as previously explained, a model $f(X) = f(X_D)$ %using only these selected variables 
%is CDP-fair and mitigate both disparate impact and treatment. 
Achieving this in the GLM  framework is equivalent to estimating a sparse parameter $\beta$ that selects approximately fair covariates, and achieves a trade-off between accuracy and fairness.
%the \emph{legimitate} (or fair) covariates amongst $X$: the $X_j$'s which are both strong predictors of $Y$ and approximately DP-fair. That is, we want to estimate $\beta$ such that there is a subset $D \subset \{1, 2, \dots, d\}$ such that $\beta_{D^c} = 0$, and $X_D$ achieves a trade-off between DP-fairness and statistical performance. 
In the next section, we define how we quantify unfairness in such models. 
%exhibited of a parameter % group of features included 
%in the particular context of generalised linear models.
%our space of fair parameters of a generalised linear model under a certain budget.

%\ds{only for non-randomised models such as GLM}
% \begin{align*}
%    | f(X_1,S_1) - f(X_2,S_2) | \leq L d((X_1,Y_1), (X_2,Y_2))
% \end{align*}
% where $L > 0$ is a constant.

%In particular, our goal will be to select such \emph{approximately} $DP$-fair predictors given a prescribed budget of DP-unfairness for the model $f$. 

%\begin{table}[h]
%\caption{Statistical notions of fairness (DP: Demographic Parity, CDP: Conditional Demographic Parity, IU: Individual Fairness) and their ability to remove disparate impact (DI) and/or disparate treatment (DT).} \label{tab:fair-def}
%\begin{center}
%\begin{tabular}{l|l|l}
%Criterion & Removes DI & Removes DT \\ \hline
%        DP & Yes & No \\
%        CDP & Partially & Partially \\
%        IU & No & Yes \\
%\end{tabular}
%\end{center}
%\end{table}

%\jack{Can we add a table, either here or in the appendix where we list the 4 criteria and then have tick and crosses for what they achieve?}

%\jack{ITS CDP THAT MOTIVATES VARIABLE SELECTION OVER SHRINKAGE!}

\subsection{Quantifying unfairness 
%\jack{Violations of DP
}\label{sec:space}

%\jack{[Pairwise scalable version of DP, pariwise doens't guarantee group, but it is tractable]}

%\jack{DP hard to enforce as constraint/doensn't give variable selection, we follow \cite{grgic2018beyond} and reduce mutatiual independnce to pairwise addiitve dependence and use ...}

%\jack{[Have we argued why we can't capture DP and IF exactly, IF I think so, DP?]}

%We recall that our goal is to select a set of fair (or, legitimate) covariates $D$ and estimate a  CDP-fair model $f(X) = g(\beta_0 + \beta^TX) = g(\beta_0 + [\beta_D]^T X_D)$ balancing performance and fairness. 
For model, $f_{\beta}(X) = g(\beta_0 + \beta^TX)$ we quantify its unfairness using the following computational convenient proxy
%define  an unfairness constraint function as
%\ds{DP unfairness (i.e., deviation from DP)} using \ds{a computational proxy}
%\jack{[Can we sell this as its violation of DP] - then combining with the likleihood in Sections 3.3 and 3.4 is what tackles IU}
%of DP-unfairness as \jack{do we want to present this as a constraint, rather than a penalty? - In Bayes its a constraint and we can interpret the LASSO like this too - we use it as a constrain in $B(\epsilon)$}
\begin{align}\label{eq:unfairness}
    \mathcal{U}(f) = \mathcal{U}(\beta) = \| \rho \circ \beta \|_1 := \sum_{j=1}^d \rho_j |\beta_j|,
\end{align}
where  $\circ$ denotes the entry-wise vector product and $\rho = (\rho_j)_{1 \leq j \leq d} = (\rho(X_j,S))_{1 \leq j \leq d}$ is a positive measure of dependence between each non-sensitive covariate $X_j$ and the sensitive attribute $S$. For instance, $\rho_j$ could be the absolute value of the marginal correlation between $X_j$ and $S$. Alternatively, if $S$ is discrete, a nonlinear $\rho$ could the Wasserstein distance between conditional distributions $X_j \mid S=s$, i.e., for binary $S$,
\begin{align}\label{eq:dis-distance}
    \rho_j = 
       \frac{1}{2} W_2^2( \mu_{X_j|S=0} ,  \mu_{X_j|S=1}),
\end{align}
% \begin{align}\label{eq:dis-distance}
%    \rho_j = \begin{cases}
%        \frac{1}{2} W_2^2( \mu_{X_j|S=0} ,  \mu_{X_j|S=1}) &\textrm{ if } S \textrm{ binary}\\
%        \min_{\nu} \sum_{s} W_2^2 (  \mu_{X_j|S=s}, \nu) &\textrm{ otherwise}
%    \end{cases},
% \end{align}
%\jack{these distances have nice interpretable forms right?}
%if $S$ binary, and 
%\begin{align*}
%   \rho_j = \min_{\nu} \sum_{s} W_2^2 (  \mu_{X_j|S=s}, \nu),
%\end{align*}
%if $S$ takes a countable number of values. In the previous expressions, 
with $W_2$ the 2-Wasserstein distance, and $\mu_{X_j|S=s}$ %an estimate of 
the conditional distribution of $X_j$ given $S=s$, $s \in \{0,1\}$.

Firstly, note that in \eqref{eq:unfairness} a variable only counts towards the model's unfairness if it is actually included in the model $f$, i.e., variables $X_j$ for which $\beta_j = 0$ do not contribute. Further, a well known property of the $\ell_1$-norm is that it favours situations where $\beta$'s are shrunk exactly 0, encouraging in variable selection. Secondly, for a model $f$ supported on a subset $X_D$, $\mathcal{U}(f)$ quantifies its unfairness in an additive way: each variable $X_j$ adds a ``cost'' proportional to its dependence on $S$ and it influence on the outcome in the model $|\beta_j|$. Although pairwise independence between variables $X_j, \: j \in D$ and $S$, does not guarantee mutual independence $X_D \perp S$, the additive form of $\mathcal{U}(f)$ is computationally convenient  and matches our heuristic of selecting variables that are weakly dependent on $S$.

Another interpretation of the penalty \eqref{eq:unfairness} in linear regression 
%($g(\mu) = \mu$) 
is that it
%A particular advantage of adopting $\hat{\rho}$ as the \jack{...} is that we can then interpret the unfairness penalty in \eqref{eq:pb2} as an 
upper bounds the covariance between the predicted outcome and the sensitive attribute, when %. Let 
$\rho_j = |Cov(X_j, S)|$ is the absolute %(population) 
covariance between $X_j$ and $S$.
%the $j$-th covariate and the sensitive attribute. Recall that our family of linear models is $f(x) = \beta^T x $. Then, 
By the triangle inequality,
%we can interpret $\mathcal{U}(\beta)$ as an upper bound on $|Cov(f(X,S), S)|$, the absolute covariance between the predictions and $S$. To see this, it suffices to use the bilinearity of the covariance and the triangle inequality:
\begin{align*}
    &|Cov(f_\beta(X), S)| %= |Cov(\beta_0 + X^T \beta, S)| = |\beta^T Cov(X, S)| 
    = \left|\sum_{j=1}^d \beta_j  Cov(X_j, S)\right|  \\
    \leq& \sum_{j=1}^d  \left|\beta_j  Cov(X_j, S)\right| = \sum_{j=1}^d |\beta_j | \rho_j = \| \rho \circ \beta \|_1= \mathcal{U}(f).
\end{align*}
%where we denote by $\mathcal{U}^*(f)$ the population penalty (in contrast to the empirical penalty  $\mathcal{U}(f) = \| \widehat \rho \circ \beta \|_1$ where $\widehat \rho$ is estimated from a finite sample). 
In the common set-up where the covariates %and outcome 
are standardised, 
% \begin{align*}
%    Corr(X_j, S) = \frac{ Cov(X_j, S)}{\sqrt{Var(S)}},
% \end{align*}
%and similarly for $Corr(f(X), S)$, 
%thus 
% up to a factor $1/\sqrt{Var(S)}$, 
setting $\rho_j = |Corr(X_j, S)|$ implies that $\mathcal{U}(f)$ is an upper bound on the covariance $|Cov(f_\beta(X), S)|$ up to a factor $\sqrt{Var(S)}$.
%In practice,  correlation can be preferred to more general measures of dependence for its interpretability and the moderate computational cost of estimating it.
In practice,  correlation provides an interpretable, scale-independent measure of dependence, which we will use in our experiments.
%from a finite sample. %we can also interpret the penalty as an upper bound on the (absolute) correlation between the predicted outcome and $S$. 
%\jack{can you not just reason in terms of correlations, rho's here always?}

Given the definition of $\mathcal{U}(f)$, we define a space of $\epsilon$-fair models %parameter 
as the weighted $\ell_1$-ball
\begin{align*}
  B(\epsilon) :=  \{ \beta \in \mathbb R^d : \mathcal{U}(f) \leq \epsilon \}
\end{align*}
where $\epsilon > 0$ is a given unfairness budget. 
Generally speaking,  this budget $\epsilon$ 
%(or fraction $r$)  
should be set by an external authority or stake-holder.
In their absence, a possible way 
%to set the unfairness budget in practice 
is to set $\epsilon =  r \mathcal{U}( f^*)$, where $r \in (0, 1)$ is a propotion and $\mathcal{U}( f^*)$ is the unfairness of the %true %unconstrained  (or best possible) model $f^*$. 
best possible model $f^*$ (see also \cite{chzhen2022minimax}). 
%Let $\widehat f^*(x)= (\widehat \beta^*)^T x$ with $\widehat \beta^*$ be the least-square estimate
    % \begin{align*}
    %                 \widehat \beta^* = \argmin_{\beta \in \R^d}  \|\mathbf X \beta - \mathbf y \|_2^2.
    % \end{align*}
%Then one could set $\epsilon =  r \mathcal{U}( f^*)$, corresponding to %so that the fair LASSO provide 
%an $r$-relative improvement (see also \cite{chzhen2022minimax}). %compared to the best %linear  predictor.model.

%From now on, we consider that $B(\epsilon)$ is our constrained parameter space and we introduce two constrained estimation methods of $\beta$: Fair Lasso and a fair Bayesian posterior.
%We note that 
In practice, the dependence measure $\rho$ is unknown and needs to be estimated from data. Given an estimate $\widehat \rho$,  our %constrained 
space of $\epsilon$-fair models then becomes
\begin{align}
  \widehat{B}(\epsilon) :=  \{ \beta \in \mathbb R^d : \widehat{\mathcal{U}}(f) = \|\widehat \rho \circ \beta\|_1 \leq \epsilon \}. \label{equ:FairParams}
\end{align}
%i.e., a weighted $\ell_1$-ball %\jack{This hasn't been introduced yet} 
%with radius $\epsilon$, where with a slight abuse of notation, we keep our notation $ B(\epsilon)$ and $\mathcal{U}(f)$ unchanged. 
%\jack{Hmm but we could also be underestimatng some values, we could just use this to talk abou tintervals for $\rho$}
%However, since $\widehat \rho_j$ will often be positive even when $X_j$ and $S$ are independent (and thus the population quantity $\rho_j = 0$). %Thus, the unfairness penalty may be conservative since part of the  budget may be spent on actually fair covariates.
%Therefore,  it can be relevant to set the budget $\epsilon$ depending on the sample size $n$, e.g., no smaller than $\epsilon_n \sim \frac{1}{\sqrt{n}}$ if $\widehat \rho_j = \rho_j + O(\frac{1}{\sqrt{n}})$. 
%Therefore, for any estimate such that $\widehat \rho_j = \rho_j + O(\frac{1}{\sqrt{n}})$ we recommend setting the fairness budget $\epsilon$  no smaller than $\epsilon_n \sim \frac{1}{\sqrt{n}}$. 
However, if $\widehat \rho_j = \rho_j + O(n^{-1/2})$ we recommend setting the fairness budget $\epsilon$  no smaller than $n^{-1/2}$, %$\epsilon_n \sim \frac{1}{\sqrt{n}}$, 
since $\widehat \rho_j$ is positive almost surely
%will often be positive 
even when $X_j$ and $S$ are independent and $\rho_j = 0$. %and $\rho_j = 0$. 
Note that the estimate  $\widehat \rho$ can be computed using  unlabeled data (e.g., an auxiliary dataset $(\mathbf{X}',\mathbf{s}')$). % as in \citep{chzhen2020fairb}.
%\jack{discuss in Section 3.2}} 
%We now provide penalised likelihood estimates and Bayesian posterior distributions operating on the constrained %parameter 
%space $\widehat{B}(\epsilon)$. 

%\jack{subsection introdcing rho first? Can say that out stratergy is going to be to encourage model fits that well predict the response but don;t contain predictors that are too related to the sensitive attribute. First we define this notion of relationship between predeictors and the sensitive attribute. }

%\ds{Additional section: violation of DP and space of budgeted fair coefficients. then statisticla estimation LASSO and Bayesian}

\subsection{Fair variable selection with Fair Lasso}\label{sec:plugin}

Let  $\mathcal{L}(\mathcal{D}_n;\beta_0, \beta) %= \sum_{i=1}^n \ell(y_i, x_i ; \beta_0, \beta)
$ be the negative log-likelihood of the data $\mathcal{D}_n$ under model \eqref{eq:glm} for parameters $(\beta_0, \beta)$ with $\alpha = 0$. 
%In this section we introduce our Fair Lasso estimate.
%, and for simplicity, here we consider the linear regression setting.} \jack{Is this necessary, we should probably just introduce a loss function/log-likelihood to estimate the parameters in Section 2} 
%\jack{Necessary?}Let us first denote by $\mathbf X = (x_i)_{1 \leq i \leq n} \in \R^{n \times d}$ the design matrix whose $i$-th row is $x_i$, $\mathbf{y} = (y_i)_{1 \leq i \leq n} \in \R^n$ the vector of observed outcomes, and $\mathbf{s} = (s_i)_{1 \leq i \leq n} \in \R^n$ the vector of sensitive attributes. 
%\jack{Lets have this right at the start of the section}
%However, contrary to the Lasso, the choice of the hyperparameter $\lambda$ in \eqref{eq:pb2} is not chosen via a statistical criterion (otherwise, a minimally fair model would always be chosen if the true model $f^*$ incurs a high unfairness penalty), but is %should be
%determined by the unfairness budget $\epsilon$. Specifically, $\lambda = \lambda(\epsilon)$ is set so that $ \hat \beta_{FL}$ is also the solution of
%The Fair Lasso is defined as the solution of
Given unfairness budget $\epsilon$ and constrained parameter space \eqref{equ:FairParams} a natural point estimate 
%for GLM parameter $\beta_0$ and $\beta$ 
is
\begin{align}\label{eq:pb1}
    %\hat \beta (\epsilon) = \arg
   % (\widehat \beta_{FL0}, \widehat \beta_{FL}) = \argmin_{(\beta_0, \beta) \in \mathbb R \times \R^d : \mathcal{ U}(\beta) \leq \epsilon} \|\mathbf X \beta + \beta_0 \mathds{1}_n - \mathbf y \|_2^2,
   %(\widehat \beta_{FL0}, \widehat \beta_{FL}) = \argmin_{(\beta_0, \beta) \in \mathbb R \times \R^d : \mathcal{ U}(\beta) \leq \epsilon} \mathcal{L}( \mathcal{D}_n; \alpha = 0, \beta_0, \beta),
   (\widehat \beta_{FL0}, \widehat \beta_{FL}) := \argmin_{\beta_0 \in \mathbb R, \beta \in \widehat{B}(\epsilon)} \mathcal{L}( \mathcal{D}_n; \beta_0, \beta),
\end{align}
%where $\alpha = 0$ is fixed so the sensitive attribute is not directly incorporated in the model. 
\eqref{eq:pb1} is a convex constrained optimisation, and, therefore, by strong duality, there exists $\lambda = \lambda(\epsilon)$ such that $(\widehat \beta_{FL0}, \widehat \beta_{FL})$ is also the solution of
\begin{align}\label{eq:pb2}
        \min_{(\beta_0, \beta) \in \R^{d+1}}  \frac{1}{n} \mathcal{L}(\mathcal{D}_n;\beta_0, \beta) + \lambda \|\widehat \rho \circ \beta \|_1.
\end{align}
%Recall that $\hat \rho = \hat \rho(\mathbf{X}, \mathbf{s}) = (\hat \rho_j)_{1 \leq j \leq p}$ is an estimate of %measure of 
%the dependence between each column of $\mathbf{X}$ and the sensitive covariates $\mathbf{s}$, for instance, %An illustrative case is to choose $\hat \rho_j$ as 
%the marginal empirical correlation between $\mathbf{X}_{\cdot j}$ and $\mathbf{s}$. 
This alternative formulation makes explicit the connection between our Fair Lasso estimate $(\widehat \beta_{FL0}, \widehat \beta_{FL})$ and the standard Lasso estimate
%Recall that the standard Lasso estimate 
\citep[][see \eqref{eq:lasso}]{tibshirani1996regression}. 
In fact, \eqref{eq:pb2} can be seen as a special case of the adaptive Lasso and solved using coordinate descent algorithms \citep{friedman2010regularization}. 
%for (adaptive) Lasso. % (see more details on the implementation are provided in Appendix ??). Secondly, like the standard LASSO, the $\ell_1$-penalty in the objective implies a sparse solution, provided that $\lambda$ is sufficiently far from 0 (or, equivalently, the unfairness budget $\epsilon$ is not too large) and the non-zero entries of $\hat \beta_{FL}$ gives us the set of legitimate covariates \jack{probably should soend some more itme on this as this is an aditional contribution}. By penalising each covariate by their unfairness, the selected set should contain the strong predictors with lowest DP unfairness
% Note that the latter estimate can be computed using  unlabeled data (e.g., an auxiliary dataset $(\mathbf{X}',\mathbf{s}')$) as in \citep{chzhen2020fairb}. \jack{discuss in Section 3.2}

Further, similar to Lasso, depending on the size of $\lambda$, $\widehat{\beta}_{FL}$ contains coordinates
%elements 
shrunk exactly to 0, thus providing variable selection, i.e., a subset of %covariates 
$X_j$ such that $  [\widehat \beta_{FL}]_{j}  \neq 0$, alongside parameter estimation.
%We note that similarly to the Lasso, 
Larger $\lambda$ (or smaller $\epsilon$) corresponds to sparser %Fair Lasso 
estimates. 
%On an intuitive level, 
Since each coefficient $\beta_j$ is penalised proportionally to %the DP unfairness 
the dependence between $S$ and the corresponding variable $X_j$, %the set selected by 
the Fair Lasso 
encourages the selection of 
%estimate %earning should  contains
covariates that are both strong predictors of $Y$ and have low DP 
unfairness. We call fair or legitimate the variables in the selected set, denoted %the latter 
by $D$. 
%As explained in Section \ref{sec:set-up}, 
The corresponding model 
%$\widehat f(x) := \widehat \beta_{FL}^T x$ 
is %thus 
CDP-fair with %$D$ the set of 
legitimate covariates $X_D$.

\begin{figure*}[h!]
    \centering
    \begin{subfigure}[b]{0.55\textwidth}
    \includegraphics[width=0.5\textwidth, trim = 1cm 0cm 1cm 1.3cm, clip]{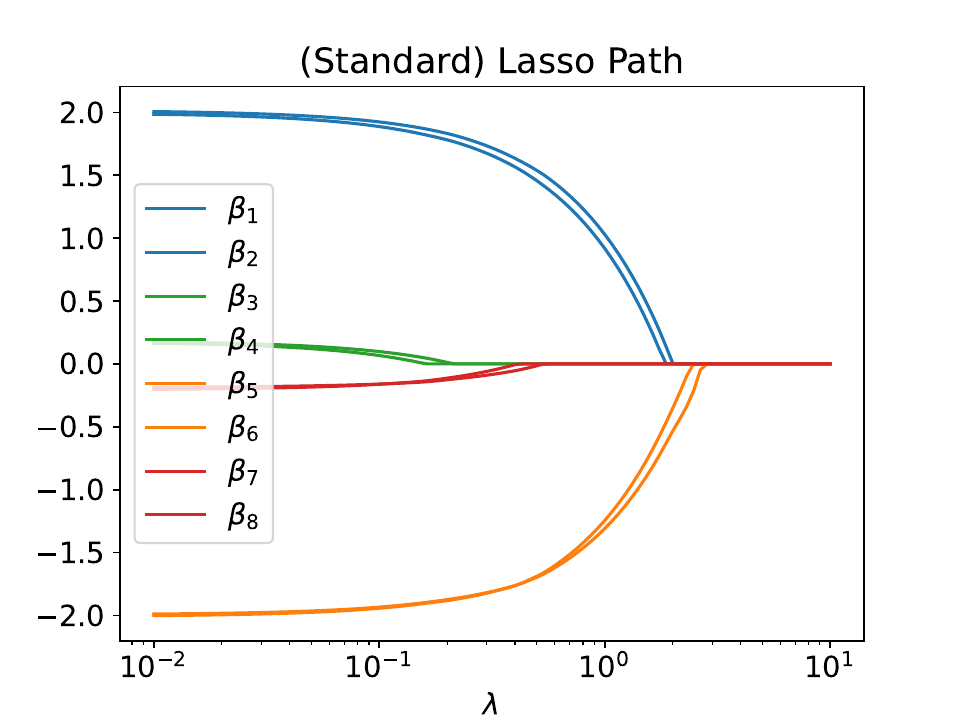}%
    \includegraphics[width=0.5\textwidth, trim = 1cm 0cm 1cm 1.3cm, clip]{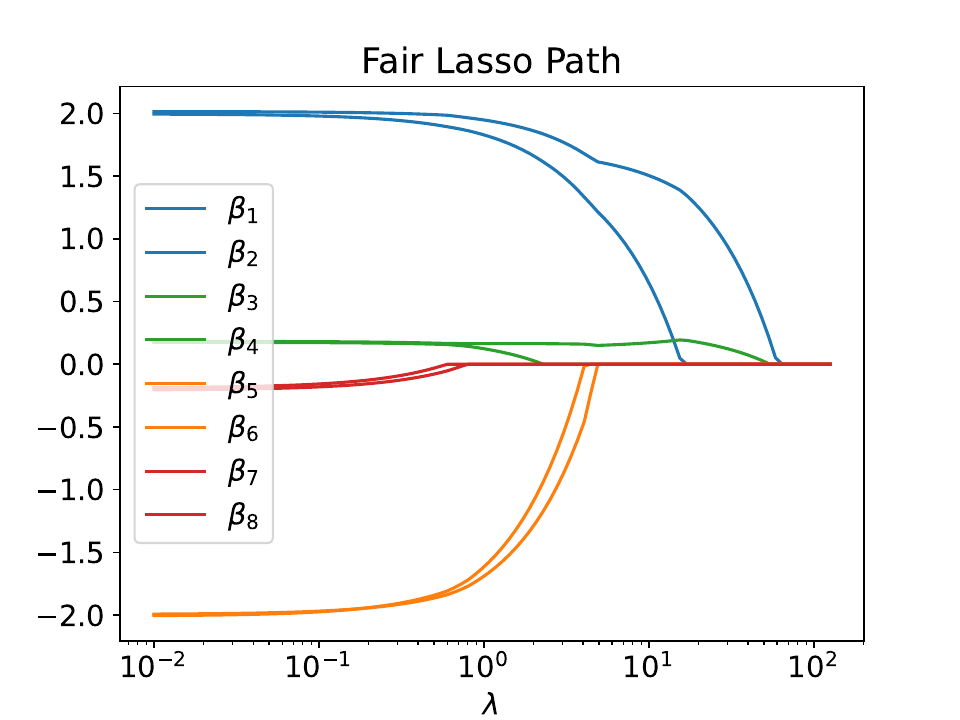}%
    \caption{Lasso (left) and Fair Lasso (right) paths}
    \label{fig:lassopath}
    \end{subfigure}%
    \begin{subfigure}[b]{0.45\textwidth}
    \includegraphics[width=0.5\linewidth]{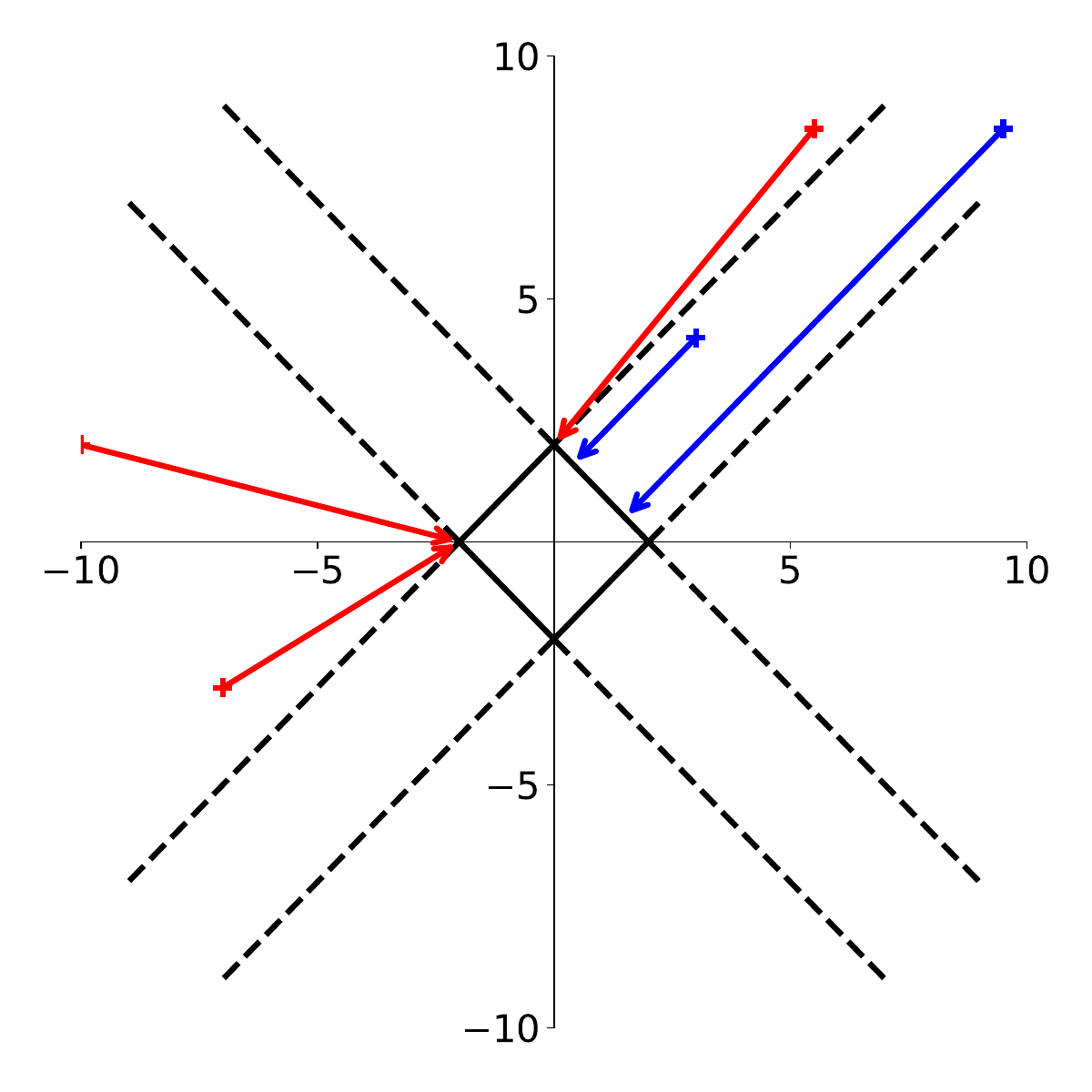}%
    \includegraphics[width=0.5\linewidth]{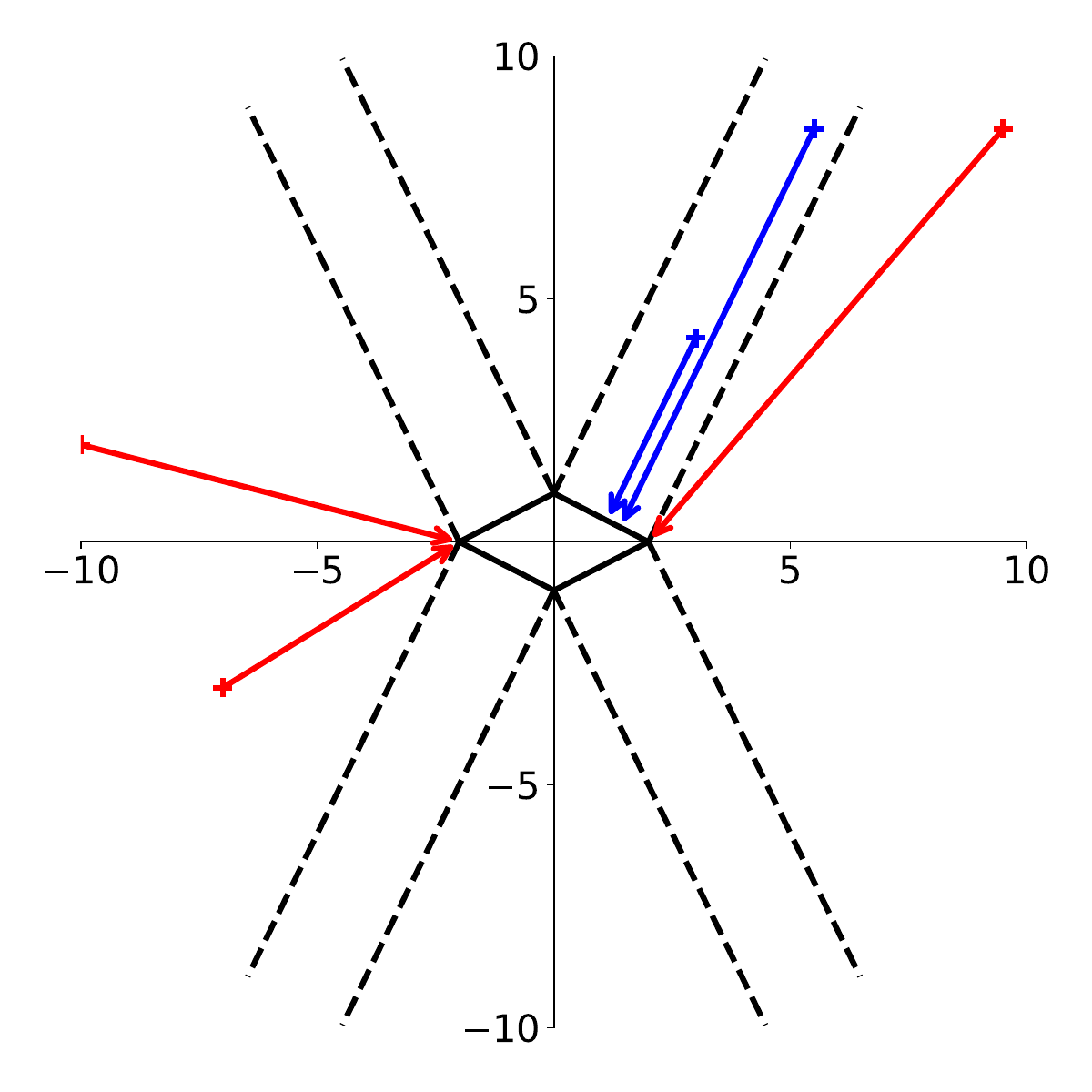}
    \caption{Standard (left) and weighted (right) $\ell_1$-ball in $\mathbb R^2$}
    \label{fig:proj}
    \end{subfigure}
    \caption{(a) Lasso  and Fair Lasso paths: %regression 
    coefficients $\beta_j$ %$\: j=1,\dots, 8$, 
    vs regularisation %penalisation parameter 
    $\lambda$. Coefficients are coloured by the type %properties 
    of the corresponding covariates: strong predictors  independent on $S$ (blue), weak predictors  independent on $S$ (green),  strong predictors  dependent on $S$ (orange) or weak predictors dependent on $S$ (red). %\jack{consider 4 coolours, strong and unfair etc.}
    (b) Illustration of standard (left) and weighted (right) $\ell_1$-ball projections in $\mathbb R^2$. Dashed lines separate the areas outside of the ball which are either projected on  edges or corners of the ball. 
    %and thus have an element of $\beta$ exactly 0. 
    Blue (resp. red) arrows are instances of projections onto edges (resp. corners). %$ of the ball. %and thus have an element of $\beta$ exactly 0. 
    In the weighted $\ell_1$-ball, $\rho_2 = 2 \rho_1$  and the area  projected onto $\beta_2 = 0$ is larger. % than in the standard ball.
    }
    \label{}
\end{figure*}

\Cref{fig:lassopath} plots the values of Fair Lasso with standard Lasso  coefficients  %$[\widehat \beta_{FL}]_j$ and $[\widehat \beta_{L}]_j$ 
along their regularisation paths, for simulated data described %in full 
in \Cref{sec:experiments}.
For Lasso, weak predictors %, $\beta_3, \beta_4, \beta_7, \beta_8$, 
are removed first as $\lambda$ increases and all strong predictors %, $\beta_1,\beta_2, \beta_5, \beta_6$, 
have similar paths. Alternatively, under Fair Lasso,  %the coefficients' paths are also separated by the \ds{dependence on $S$} of the corresponding predictors. Specifically, 
weak and  $S$-dependent predictors  %, $\beta_7$ and $\beta_8$,
are removed first and one of the two weak and $S$-independent predictors  %, $\beta_3$ and $\beta_4$,
is removed after the strong but $S$-dependent ones. %, $\beta_5$ and $\beta_6$. 
Thus, Fair Lasso is, as expected, prioritising $S$-independent and strong predictors, while achieving a trade-off between removing strong and $S$-dependent predictors and keeping weak but $S$-independent ones. Besides, %\Cref{fig:lassopath} shows that
coefficients corresponding to
%when an 
$S$-dependent covariates 
%is included, its corresponding coefficient is  
are relatively 
more shrunk than that of $S$-independent covariates with similar predictive strength. We note that the Fair Lasso can exclude not only variables that are dependent on $S$ (the \emph{unfair} ones) but also those who may be independent of $S$  (the \emph{fair} ones) since the empirical measure of dependence $\widehat \rho_j$ is non-zero almost surely. %See \Cref{sec:sim1} for full details of this simulation setting.

% Moreover, we compare the variables selected by Fair LASSO and LASSO via their selection paths (see Figure \ref{fig:lassopath}). For the LASSO, all weak predictors ($\beta_5- \beta_8$) are removed for similar values of the penalisation parameter $\lambda$ and the coefficients corresponding to all strong predictors ($\beta_1- \beta_4$) are similarly shrinked. In contrast, the Fair LASSO first removes the weak and unfair predictors $(\beta_7,\beta_8)$ then the weak and fair $(\beta_5,\beta_6)$ and the strong and unfair predictors $(\beta_3,\beta_4)$ at almost the same value of $\lambda$. Therefore, depending on the unfairness budget, the strong and unfair predictors can be either removed or selected but their corresponding regression coefficients are shrinked, and this shrinkage is larger than that of the strong and fair variables. 

The value of $\epsilon$, and thus $\lambda$ controls the trade-off between IF and DP and predictive performance. 
%On one hand, 
If $\lambda = 0$ ($\epsilon$ sufficiently large), %then  %f_{\widehat \beta_{FL}}(x) = \widehat \beta_{FL}^T x$ 
$\widehat \beta_{FL}$
reduces to the least-squares estimate, which is often IF %and should predict well, 
but not DP-fair (see Section \ref{sec:experiments}). As $\lambda$ increases ($\epsilon$ decreases) covariates %highly 
dependent on $S$ start to be dropped from the model, improving on DP  but harming IF and predictive performance if these covariates were truly predictive of the response.
%In the extreme case 
If $\lambda = \infty$ ($\epsilon = 0$), Fair Lasso only selects variables that are exactly DP-fair ($\widehat \rho_j = 0$) and the resulting model satisfies DP %will also be group fair 
but predictions will generally be poor. 

\subsection{Bayesian Fair Variable Selection with the Fair $\ell_1$-ball prior }\label{sec:fair-bayes}

A direct Bayesian extension of \eqref{eq:pb2}
%$\hat \beta_{FL}$ proposed in Section \ref{sec:plugin} 
places double-exponential priors on each $\beta_j$ %with scale parameter equal to $1/(\lambda\hat{\rho}_j)$
\citep{park2008bayesian}. While the mode of this posterior is %will agree with 
\eqref{eq:pb2}, samples from such a posterior will not necessarily belong to $\widehat B(\epsilon)$, %satisfy the fairness constraint
 or be sparse, meaning selection %estimation 
of the fair %or legitimate 
covariates is no longer possible. 
Instead, we construct a fair Bayesian posterior using a prior that is constrained to have 0 mass outside of $\widehat{B}(\epsilon)$. While directly proposing a prior $\pi_{\epsilon}(\beta)$ that satisfies this %joint 
constraint is difficult, we can instead achieve this by projecting a standard (unconstrained) prior  
onto the %space of parameters 
%$\epsilon$-fair 
weighted $\ell_1$-ball $\widehat{B}(\epsilon)$.
%(see Section \ref{sec:space}).
%satisfying $\mathcal{U}(f) < \epsilon$. 
% Here, we present our methodology based on projected prior, which overall achieves better performance than the projected posterior \citep{pal2025bayesianhighdimensionallinearregression} %\jack{citation}, 
% presented in Appendix \ref{app:projected-post} for completeness. \jack{Move to the end of the section, it is a little distracting here}

%An alternative is to square the $\ell_2$ norm so that the objective becomes analog to a Lasso objective:
    % \begin{align}\label{eq:pb4}
    %     \tilde \beta = \argmin_{\beta} \| \hat \beta - \beta \|_2^2 + \lambda_{\hat \beta} \|\hat \rho \circ \beta\|_1.
    % \end{align}

% For each sample $\hat \beta \sim p(\beta | y)$, choose $\lambda > 0$ such that
% \begin{align*}
%     \tilde \beta = \argmin_{\beta} \| \hat \beta - \beta \| + \lambda \|\hat \rho \circ \beta\|_1 
% \end{align*}
% verifies $\hat \gamma = \mathds{1}_{\hat \beta \neq 0}$, $\sum_{i \in \hat \gamma} \hat \rho_i \leq \epsilon$.

%\subsection{Fair posterior with $\ell_1$-ball prior}

%Yet an alternative to the projected posterior is to instead project the prior.

Specifically, let $\tilde{\beta}\in\mathbb{R}^d$ be unconstrained latent variables with prior density $\tilde{\pi}(\tilde{\beta})$, we construct the Fair $\ell_1$-ball 
%$\epsilon$-fair 
prior $\pi_{\epsilon}(\beta)$ by projecting samples from $\tilde{\beta} \sim \tilde{\pi}$ onto 
%the space of $\epsilon$-fair parameters
$\widehat B(\epsilon)$ defined in \eqref{equ:FairParams}.
%Specifically, given a prior distribution $\pi(\beta)$ (which can be for instance the conjugate prior or a prior informed by a domain expert), we construct a  $\epsilon$-fair prior $\Tilde \pi(\beta)$ by projecting $\pi$ onto the space of $\epsilon$-fair parameters $B(\epsilon)$ from \eqref{equ:FairParams}
%\begin{align*}
%    = \{\beta \in \mathbb R^d : \mathcal{U}(\beta)  = \|\widehat \rho \circ \beta\|_1 \leq \epsilon),
%\end{align*}
%i.e., vectors $\beta$ that respect the unfairness budget constraint.
Let $P_{\widehat B(\epsilon)}: \mathbb R^d \to \widehat B(\epsilon)$ the projection operator onto $\widehat B(\epsilon)$, 
\begin{align}\label{eq:projection}
    P_{\widehat B(\epsilon)}(\Tilde\beta) = \argmin_{ \beta \in \widehat B(\epsilon)} \| \Tilde \beta - \beta \|_2^2.
\end{align}
Then formally the prior distribution $\pi_{\epsilon}$ is defined as %the density of
\begin{align*}
    \pi_{\epsilon}(\beta\in A) = \int \mathds{1}_{P_{\widehat B(\epsilon)}(\Tilde \beta) \in A}\tilde{\pi}(\Tilde \beta)d\Tilde \beta, \qquad A \subseteq \mathbb R^d.
\end{align*}
This approach is an extension of the recently proposed $\ell_1$-ball prior \citep{xu2020bayesian} for Bayesian variable selection, which projects a base prior %our projection is onto a weighted $\ell_1$-ball, $\widehat B(\epsilon)$, with weights provided by $\widehat{\rho}$ rather than
onto the standard $\ell_1$-ball $\{ \beta : ||\beta||_1 \leq \epsilon\}$.
Here, for the same reasons as in \cite{xu2020bayesian}, \eqref{eq:projection} is convex, so has a unique solution (see Appendix \ref{app:weighted_proj}), is %continuous and 
differentiable almost everywhere, and $\pi_{\epsilon}(\cdot)$ has positive probability that any coordinate $\beta_j$ is 0, thus providing variable selection.
%The solution of \eqref{eq:projection} %projection of prior samples  onto $B(\epsilon)$ 
% can be easily found via a modification of the algorithm in 
% \cite{duchi2008efficient} (see Appendix \ref{app:weighted_proj}).
Figure \ref{fig:proj} illustrates the differences between the standard $\ell_1$-ball projection and the weighted one \eqref{eq:projection}. %the left hand plot illustrates that regions of $\tilde{\beta}$ that when projected to the $\ell_1$-ball (black) result in elements of $\beta$ being 0. The right hand plot shows that for the 
%coordinates with larger DP-unfairness $\widehat \rho_j$ (coordinate 2 in this case), 
%the space of parameters for which the coordinate with larger DP-unfairness $\widehat \rho_j$ (coordinate 2 in this case) is projected to 0 is enlarged compared to the standard $\ell_1$-ball
In particular, it shows %This implies 
that coordinates with small $\widehat \rho_j$ are more likely to be non-zero after projection. %, and therefore selected.

Given a further prior $\pi_0(\beta_0)$ for the intercept
(and possibly for other nuisance parameter such as the noise level in linear regression, which we omit here for simplicity),
% , e.g., a Gaussian prior $N(0, b^2)$ with hyperparameter $b^2 > 0$, and a prior % one also needs to set 
% $\pi_\theta(\theta)$ on any nuisance parameters, e.g., in linear regression an Inverse-Gamma distribution for noise level $\theta = \{\sigma^2\}$
% in linear regression, %and  we set a standard prior $\pi_\sigma(\sigma^2)$ 
our Fair Posterior distribution is 
\begin{align}\label{eq:fair-post}
    %\pi(\beta_0, \beta,  \sigma^2 \mid \mathbf X, \mathbf y) \propto \prod_{i=1}^n \ell (y_i ; \beta_0, \beta, \sigma^2, x_i) \tilde \pi(\beta) \pi_0(\beta_0)\pi_\sigma(\sigma^2),
    &d\pi_{\epsilon}(\beta_0, \beta  %\theta 
    \mid \mathbf X, \mathbf y) \propto \exp\left\{-\mathcal{L}(\mathcal{D}_n; \beta_0, \beta)\right\} d\pi_{\epsilon}(\beta) d\pi_0(\beta_0).%\pi_\theta(\theta),
\end{align}
%where the GLM's dependence on $\theta$ is hidden in $\mathcal{L}$, and in logistic regression $\theta = \emptyset$.
%with $\ell( y ; \beta_0, \beta, \sigma^2, x) = \frac{1}{\sqrt{2\pi \sigma^2}} e^{-(y-\beta_0 - \beta^T x)^2/(2 \sigma^2)}$ in the linear regression setting. In the logistic regression setting, 
%\begin{align*}
%    \pi(\beta_0, \beta \mid \mathbf X, \mathbf y) \propto \prod_{i=1}^n \ell(y_i ; \beta_0, \beta, x_i) \tilde \pi(\beta)\pi_0(\beta_0),
%\end{align*}
%with $\ell( y ; \beta_0, \beta, x) = (1 + e^{-(\beta_0 + \beta^Tx)})^{-y}  (1 + e^{\beta_0 + \beta^Tx})^{y-1}$. For other generalised linear model, only the likelihood part needs to be modified. 
%\jack{In \cite{xu2020bayesian}, the prior is projected on a standard $\ell_1$-ball which radius is given a prior distribution as well. }
%However, 
Contrary to \cite{xu2020bayesian}, where the sparsity level is learned by placing a prior on the $\ell_1$-ball radius, here %denoted by $\epsilon$, in \eqref{eq:fair-post}, 
this radius $\epsilon$ is the fixed unfairness budget.
Note that %, as with the $\epsilon$-fair prior, 
our Fair Posterior \eqref{eq:fair-post} is only supported on $\widehat{B}(\epsilon)$.
% , therefore every posterior sample, %is $\epsilon$-fair,  
% posterior parameter estimate, credible set, and the posterior predictive are also guaranteed to be $\epsilon$-fair.

The previously discussed continuity and differentiability of $P_{\widehat B(\epsilon)}$ is particularly convenient as it allows for the use of gradient-based MCMC algorithms to sample from $\pi_\epsilon(\beta_0, \beta,  %\theta 
\mid \mathbf X, \mathbf y)$ even though it is a mixture of continuous and discrete distributions.
For example, we use the NUTS \citep{hoffman2014no} extension of Hamiltonian Monte-Carlo \citep{duane1987hybrid} implemented in %the 
Stan %probabilistic programming languages 
\citep{carpenter2016stan}. Posterior sampling is done on the continuous unconstrained parameter $\tilde{\beta}$ and samples of these can then be mapped to fair parameter via $P_{\widehat{B}(\epsilon)}$. %\ds{stan is the language right?}

The Fair $\ell_1$-ball prior having positive mass on any element of $\beta$ being 0 allows the fair posterior \eqref{eq:fair-post} to be extended to a fair inclusion/exclusion posterior $\pi_\epsilon(\gamma \mid \mathbf X, \mathbf y)$, where $\gamma = (\gamma_j)_{j=1}^d = (\mathds{1}_{\beta_j \neq 0})_j$ parametrises whether covariate $X_j$ is used in the model or not.
%By defining $\gamma = (\gamma_i)_{i} = (\mathds{1}_{\beta_i \neq 0})$ the inclusion variables, the Fair  Posterior \eqref{eq:fair-post} also defines a distribution $\pi_\epsilon(\gamma \mid \mathbf X, \mathbf y)$, 
$\pi_\epsilon(\gamma \mid \mathbf X, \mathbf y)$ quantifies uncertainty over possible sets of fair %/legitimate 
covariates that comply with the unfairness budget. If a final selection is desired, e.g., for use in CDP, %framework, 
one could select the maximum-a-posteriori model or  threshold %based on 
marginal inclusion probabilities.
%i.e. $\pi_\epsilon(\gamma_i \mid \mathbf X, \mathbf y) \geq 0.90$. 
%the fair variables as those which have $\geq 0.90$ marginal inclusion probability (i.e., $\pi_\epsilon(\gamma_i \mid \mathbf X, \mathbf y) \geq 0.90$, 
%or the set of variables which is the union of variables in $\gamma$ for $\gamma$ in 90\% posterior credible set.

Finally, we remark that an alternative approach to designing a %fair 
prior supported on $\widehat B(\epsilon)$ is to project the samples from an unconstrained posterior as in \cite{pal2025bayesianhighdimensionallinearregression}. We detail this approach in Appendix \ref{app:projected-post} for completeness, and show in our numerical experiment that our %Fair Posterior %methodology based on  
Fair $\ell_1$-ball %projected 
prior %which 
overall achieves better performance than the projected posterior.
Here the unprojected parameters appear in the likelihood and thus after projected, fit the data less well.

\section{Numerical experiments}\label{sec:experiments}

We test our Fair Lasso (\textbf{FL}) and Fair Posterior (\textbf{FP}) methods %from Sections \ref{sec:plugin} and \ref{sec:fair-bayes} 
on simulated and real-world data and compare to existing fair methods, namely the Mean Difference  (\textbf{MD}) estimate \citep{calders2013controlling}, the Fair Ridge Regression Model \textbf{(FRRM)} \citep{scutari2022achieving}, %a Bayes posterior, 
the sparse posterior induced the standard $\ell_1$-ball prior \textbf{(SP)} \citep{xu2020bayesian} and the projected-posterior (\textbf{PP}) (descriptions provided in Appendix \ref{app:baselines}). There we also report a comparison to the greedy Fair Feature Selection method by \citep{belitz2021automating} in classification and to the standard posterior. %\jack{[why in App?]}
%, using accuracy and fairness metrics quantifying violation of individual fairness (i.e., disparate treatment)  and violation of group fairness (i.e., disparate impact). %\jack{This pairing is useful and way maybe not clear beforehand}. 
%For simplicity and 
To be able to compare to existing methodology, we only consider settings with binary sensitive attribute $S$, though our framework also allows for continuous $S$.

\subsection{Performance metrics}

On a held-out test set, we evaluate the accuracy of %the %model's resulting from 
our proposed and baseline methods via the mean squared-error (\textbf{MSE}) or binary cross-entropy loss (\textbf{BCE}) %area-under-the-ROC curve (\textbf{AUC}) 
in respectively linear and logistic regression.  We quantify their unfairness using the following metrics (%with a slight abuse of notation we also 
denoting by $n, n_1, n_0$ the sample and group sizes in the test set):
\begin{enumerate}[noitemsep]
    \item \textbf{Individual unfairness:} we use the definition of \cite{berk2017convexframeworkfairregression}: $\text{IU}(f) =$
    \begin{align*}
    \frac{1}{n_1 n_0} \sum_{i : s_i=1, j:s_j=0} \mathbf c(y_i, y_j) (f(x_i, s_i) - f(x_j, s_j))^2,
    \end{align*}
    with $\mathbf c$ a similarity function that only depends on the observed outcome and the %specific 
    prediction task:
    \begin{itemize}[noitemsep]
        \item Regression: we set $\mathbf c(y, y'; \xi) = \xi e^{-\xi (y-y')^2}$ with $\xi > 0$ a constant;
        \item Classification: for the simulated data where we have access to $p= \mathbb P(Y=1\mid X,S)$, we set $\mathbf c(p, p'; \xi) = \xi e^{-\xi (p-p')^2}$. % (and denote the corresponding unfairness measure by \textbf{IU2}). 
        For the real data where only $y$ is known, we set  $\mathbf c(y, y') = \mathds{1}_{y = y'}$. % (and denote the corresponding unfairness measure by \textbf{IU}).
    \end{itemize}
    %is chosen as $\mathbf d(y, y'; \xi) = \xi e^{-\xi (y-y')^2}$ with $\xi > 0$ a constant for the regression settings and  $\mathbf d(y, y') = \mathds{1}_{y \neq y'}$ for the classification contexts. 
    %This metric measures %how far the model is from removing
    %the amount of disparate treatment.
    \item \textbf{DP Unfairness:} %in regression settings, 
    we use the definition of \cite{chzhen2020fairb} %where %the
    %violation of DP is measured via 
    using the 1-Wasserstein distance:
    \begin{align*}
        \text{WDU}(f)  = W_1 (\hat \mu_{f(X,S)|S=1}, \hat \mu_{f(X,S)|S=0})
    \end{align*}
    where %$W_1$ is the 1-Wasserstein distance, 
    $\hat \mu_{f(X,S)|S=k}$, $k \in \{0,1\}$,  is the empirical conditional distribution of $f(X,S)$ given $S=k$. %$k \in \{0,1\}$, 
    % i.e.,
    % \begin{align*}
    %     \hat \mu_{f(X,S)|S=k} (\cdot) = \frac{1}{n_k} \sum_{i: s_i=k} \delta_{f(x_i,s_i)}(\cdot),
    % \end{align*}
    % with $\delta_y$ the Dirac measure at $y$. 
    % In classification settings, we measure the violation of DP via the Parity Gap defined as
    % \begin{align*}
    %     PG(f) = \frac{1}{n_1} \sum_{i:s_i=1} \mathds{1}_{f(x_i,s_i) > 0.5} - \frac{1}{n_0} \sum_{j:s_j=0} \mathds{1}_{f(x_j,s_j) > 0.5}
    % \end{align*}
    %This metric measures %how far the model is from removing 
    %the amount of disparate impact.
    %Group fairness via DP and correlation analysis: how from removing disparate impact
    %\item \ds{Conditional demographic disparity: how well the fairness criterion is achieved?}
\end{enumerate}

%For the Bayesian methods, we compare the performance of the posterior mean as well as the expected value of the previous metrics under the posterior distribution. 

\subsection{Simulated experiments}

%\jack{Figure 2 shows the effect of moving from the LASSO to the fairLASSO. Could we have this in Section 3.2}

%\jack{then Figure 4 shows that changing $\lambda/\epsilon$ allows for trading off DP Unfairness, and Individual Fairness)}

%\jack{Then we compare the performance with other methods using what is now Figure 3}

%\subsubsection{Simulation 1: linear regression}{\label{sec:sim1}}

We sample the sensitive attributes $s_i \overset{i.i.d.}{\sim} Ber(0.5)$ %from a Bernoulli distribution with mean $0.5$ 
and %we generate 
the design matrix 
 $\mathbf X \in \R^{n \times 8}$
%$\mathbf X \in \R^{n \times (d_1+d_2+d_3+d_4)}$ 
with 4 types of covariates: strong and independent of $S$ ($X_1$ and $X_2$), weak and  independent of $S$ ($X_3$ and $X_4$), strong and  dependent of $S$ ($X_5$ and $X_6$) and weak and  dependent of $S$ ($X_7$ and $X_8$). For strong (resp. weak) covariates, $|\beta_j| = 2.0$ (resp. $|\beta_j| = 0.2$).
% \begin{itemize}
%     \item $X_1$ and $X_2$ are strong and fair predictors ($\beta_1^* = \beta^*_2 = 1.0$ and $X_1,X_2 \perp S$)
%     \item $X_3$ and $X_4$ are weak and fair predictors ($\beta^*_3 = \beta^*_4 = 0.1$ and $X_1,X_2 \perp S$)
%     \item $X_5$ and $X_6$ are strong and unfair predictors ($\beta^*_5 = \beta^*_6 = 1.0$ and $X_1,X_2 \not \perp S$)
%         \item $X_7$ and $X_8$ are weak and unfair predictors ($\beta^*_7 = \beta^*_8 = 0.1$ and $X_1,X_2 \not \perp S$)
%     % $d_1$ which are independent of $S$ and with $\beta_i = 1.0$ (strong and fair predictors);
%     % \item $d_2$ which are independent of $S$ and with $\beta_i = 0.1$ (weak and fair predictors);
%     % \item $d_3$ which are dependent of $S$ and with $\beta_i = 1.0$ (strong and unfair predictors); 
%     % \item $d_4$ which are dependent of $S$ and with $\beta_i = 0.1$ (weak and unfair predictors).
% \end{itemize}
We sample the $S$-independent covariates i.i.d. from a standard normal while %for each unfair covariate $X_j$, 
we sample the other ones conditionally on $s_i$, i.e., $\mathbf X_{ij} \mid s_i\sim N( 1.5 s_i, 1)$.
%from the following distribution
% \begin{align*}
%     \mathbf X_{ij} \mid s_i\sim N( 1.5 s_i, 1), \quad i \in [n].
% \end{align*}
%with $\rho = 1.5$ \jack{confusing to use $\rho$ when it's not a correlation}. 
We then sample the outcomes $y_i$ as in \eqref{eq:glm} with $\alpha^* = 0$, thus the true model $f^*(x) = g((\beta^*)^Tx)$ is individually fair but not DP-fair. We consider a linear regression (Simulation 1) and a logistic regression (Simulation 2) settings. The results of Simulation 2 are reported in Appendix \ref{app:num_results} alongside additional experimental details.

\begin{figure*}[h!]
    \centering
\includegraphics[width=0.33\linewidth, trim = 0.5cm 0cm 1cm 1.35cm, clip]{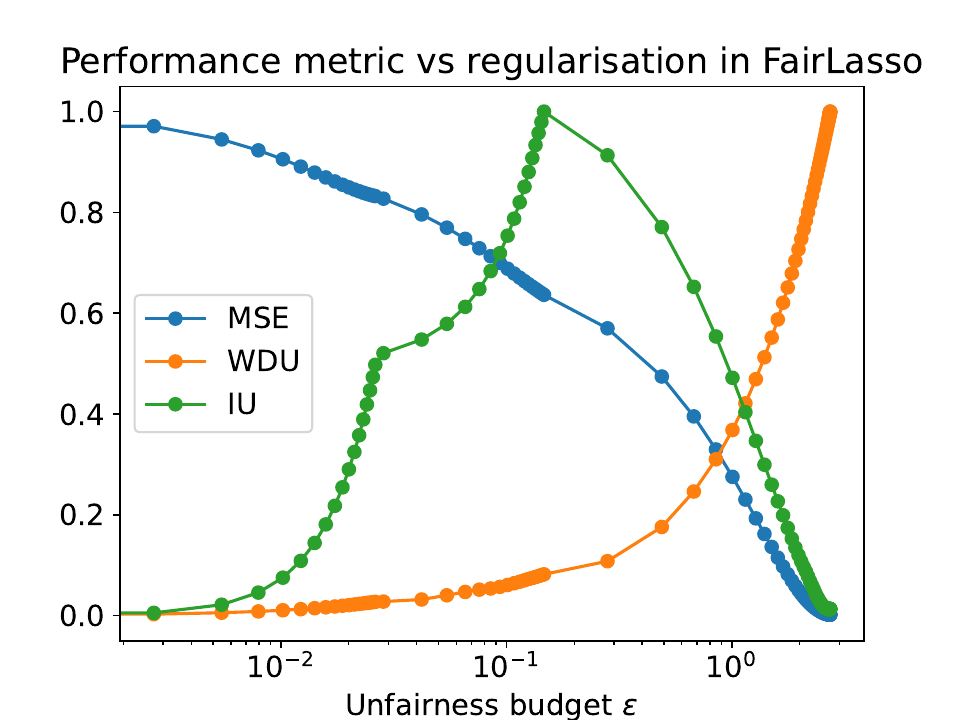}%
        \includegraphics[width=0.39\linewidth, trim = 1.0cm 0cm 1cm 1.5cm, clip]{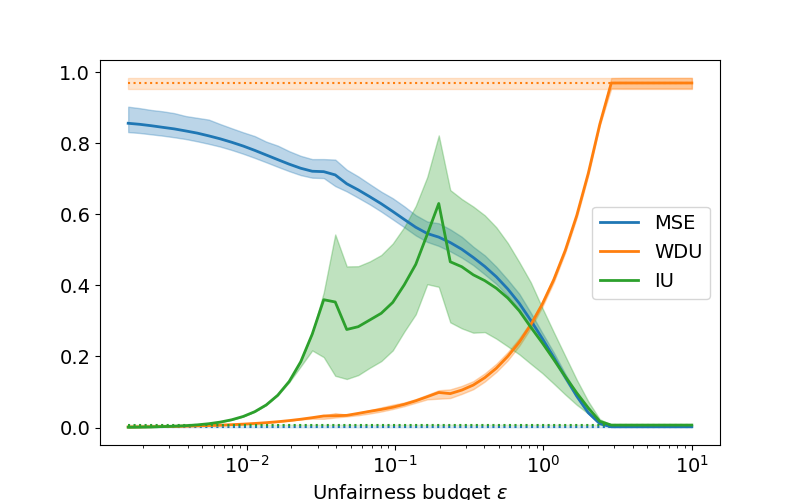}%
    \caption{
    %\jack{Change $\epsilon$ scale, whats important is the RHS, but i'm always drawn to the LHS as its bigger make $10^{-2}$ the smalest value}
    Left: Test performance (MSE, WDU, IU) of the Fair Lasso vs unfairness budget $\epsilon$ (in log scale). %\ds{plot 1/$\lambda$}. 
    Right: Test performance metrics of the Fair Posterior vs unfairness budget $\epsilon$; the bold lines correspond to the posterior averaged metric around which the 95\% pointwise credible bands are coloured. The horizontal dotted lines 
    %and colored areas 
    corresponds to the standard posterior.  %mean and 95\% credible bands respectively. 
    The metrics are normalised by their largest value in each method. 
    %IU is nonmonotonic (trade-off with WDU), CP when all unfair covariates are removed, the part after is not-interesting (just residual WDU) + add a dotted line at the CP
    %\jack{How do we get 0 MSE? This is probably why IU goes to 0 here, but can;t in the logistic case as there is still randomness even with the correct model (and hence BCE doesn't goto 0)}
    } 
    \label{fig:metrics-path}
\end{figure*}

In the linear regression setting, %$g(\mu) = \mu$, 
$v(\mu^*) = \sigma^2 = 0.2^2$. 
%Under this setting %$Y \perp S\mid X$, and therefore 
 % due to the dependence between $S$ and some of the covariates. 
%We will thus compare the ability of the different methods to reduce DP unfairness while assessing the relative impacts on statistical accuracy and IF, which are both expected to decrease.
%We set $d_1 = d_2 = d_3 = d_4 = 2$ and  
%
For all Bayesian methods we choose the base prior as Normal-Inverse-Gamma   \eqref{eq:normal-gamma} %the hyperparameters of the Normal-Gamma prior 
with $a=b=1$, %$\mu_0 = 0$, 
$\Sigma = I$. %$\phi = 0.5$. 
The training and test sample sizes are both equal to 200. %in linear regression.
% $n = 200$ and we evaluate the performance of the models % with the mean squared error 
% on a test set of size $n_{test} = 200$ samples generated from the same distribution as above. 
%and we set $\xi = 5$ in IU. % in the Individual Unfairness metric. 
% The correlation between each covariate and the sensitive is reported in Table \ref{tab:corr}.
% \begin{table}[h!]
%     \centering
%     \begin{tabular}{c|c|c|c|c|c|c|c}
%          $X_1$ & $X_2$ & $X_3$ &  $X_4$ & $X_5$ & $X_6$ & $X_7$ & $X_8$ \\ \hline
%          & & & & & & & 
%     \end{tabular}
%     \caption{Pearson correlation between each covariate $X_j$ and the sensitive attribute $S$ (averaged over 20 repeated experiments).}
%     \label{tab:corr}
% \end{table}
Figure \ref{fig:metrics-path} illustrates how Fair Lasso and Fair Posterior (FP)
%our penalised likelihood and Bayesian estimation methods 
allow for the trading off of DP violation (WDU) and  IF violation (IU), for different unfairness budgets $\epsilon$  %(or equivalently, different regularisation parameter $\lambda(\epsilon)$), 
and the induced effect %this has 
on MSE. %predictive mean-squared error (MSE).  
When $\epsilon$ is very small, %$\lambda(\epsilon)$ large, 
the estimated model is the null model (i.e., a constant intercept) for which $IU = 0$ and $WDU = 0$, but %such a model doesn't learn any relationship between $X$ and $y$ and so 
has large MSE. %On-the-other-hand, 
When $\epsilon$ is large, %$\lambda(\epsilon)$ small, 
the estimated model is the unconstrained least-squares estimate and close to the true model, which %is individually fair in this setting, has 
low MSE and IU, but high WDU due to the $S$-dependent predictors. 
Between these extreme budget values, MSE and WDU are respectively decreasing and increasing with $\epsilon$ %monotone functions of $\epsilon$ (respectively monotone decreasing and monotone increasing), 
while IU is not monotone. %This behaviour can be explained as follows.
As $\epsilon$ decreases from 3 to around 0.2, $S$-dependent covariates (weak then strong predictors) are removed which increases IU, traded-off to decrease WDU. However, after the last unfair covariate is removed from the model ($\epsilon \approx 0.2$), further decreasing $\epsilon$ only shrinks the coefficients corresponding to $S$-independent variables (for which $\widehat \rho_j$ is small but not 0) %(since $\widehat \rho_j$ is never 0 even if the population $\rho_j = 0$). This 
and eventually %tends to 
collapses to the null model ($IU=0$).
%where all predictions are the same constant value (i.e., the intercept) and IU decreases back to 0. 
This behaviour is qualitatively similar %the same 
for Fair Lasso and FP. % the Fair Posterior.
%The left and right hand sides of \Cref{fig:metrics-path} show qualitatively the same behaviour for the penalised likelihood and Bayesian estimation methods.

\begin{figure*}[h!]
    \centering
    \includegraphics[width=0.3\textwidth, trim = 0.5cm 0.cm 1cm 0cm, clip]{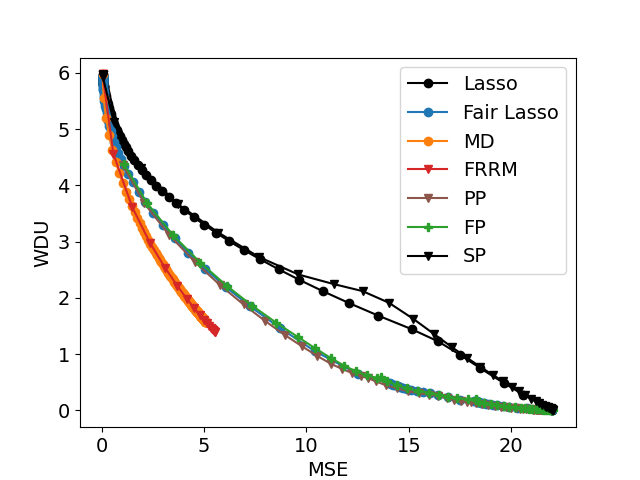}%
    \includegraphics[width=0.31\textwidth, trim = 0.5cm 0cm 1cm 0cm, clip]{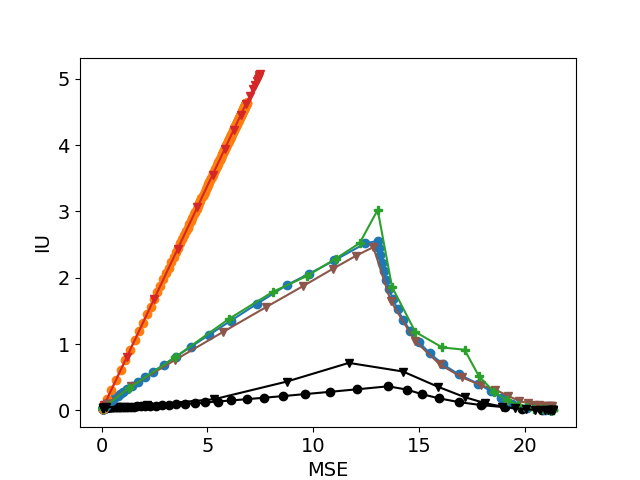}%
    \includegraphics[width=0.31\textwidth, trim = 0.5cm 0cm 1cm 0cm, clip]{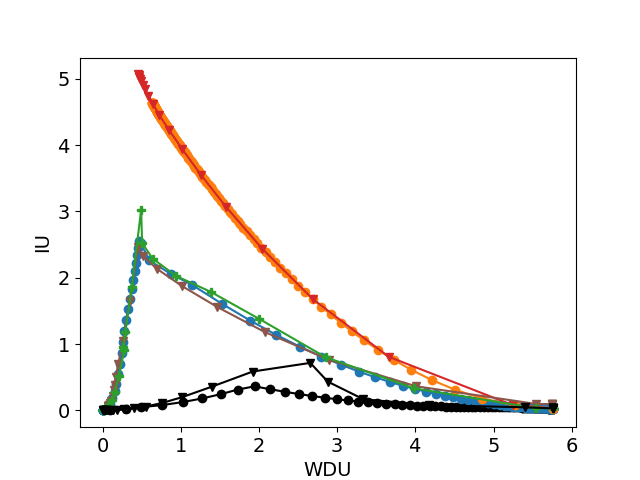}
    \includegraphics[width=0.31\textwidth, trim = 0.cm 0cm 0.5cm 0.5cm, clip]{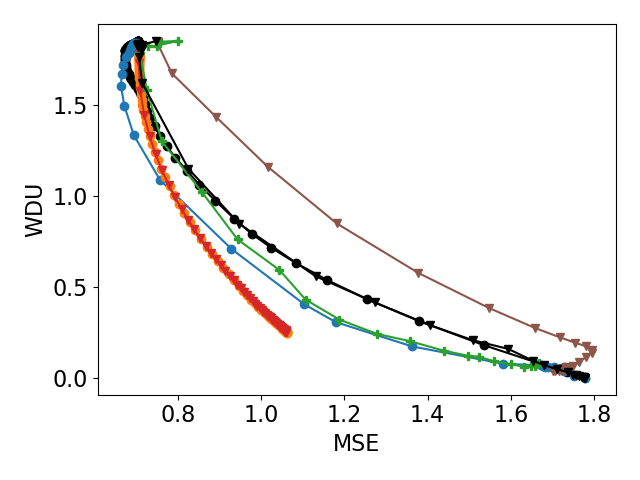}%
    \includegraphics[width=0.31\textwidth, trim = 0.cm 0cm 0.5cm 0.5cm, clip]{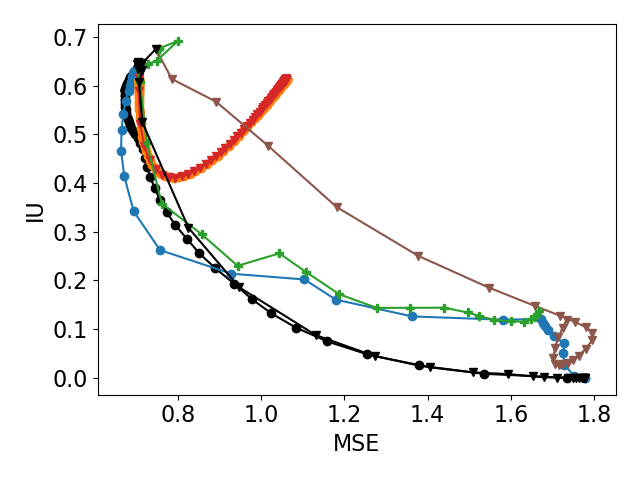}%
    \includegraphics[width=0.31\textwidth, trim = 0.cm 0cm 0.5cm 0.5cm, clip]{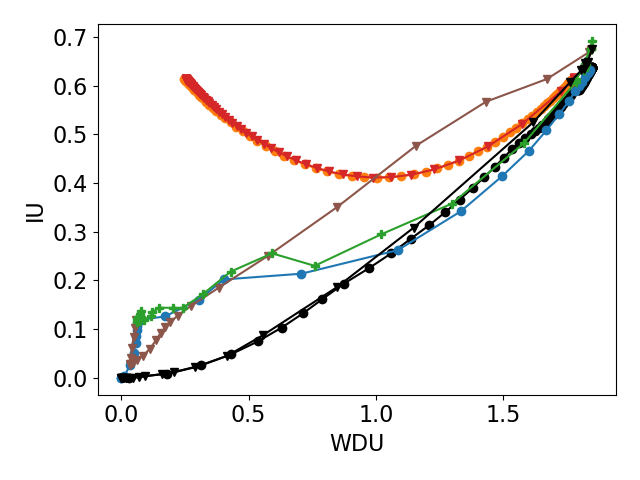}
    \caption{
    %\jack{increase font size slightly so we can see captions when 3 in a line}
    Pareto fronts of the proposed and baseline methods: Fair Lasso (blue), FP (green), PP (brown), MD (orange), FRRM (red) and the fairness-unaware methods Lasso and SP (black). Top row: linear regression simulation. Bottom row: Community dataset. %All metrics are evaluated on the test set. 
    For FP, PP and SP: metrics are averaged over the posterior samples for each value of the unfairness budget.  Left: WDU vs MSE. Middle: IU vs MSE. Right: IU vs WDU. }
    %Top: Simulated data, Bottom: Community data. %The performance metrics are normalised by their largest value across methods.
    %\jack{y-axis label cut}}
    \label{fig:reg-pareto-all}
\end{figure*}

Figure \ref{fig:reg-pareto-all} (top row) provides
%We now compare %these trade-offs to the ones achieved by the baselines. By varying the regularisation parameter, we estimate 
Pareto fronts of WDU, IU and MSE, for our methods and baselines when varying the unfairness budget or regularisation. %In Figure \ref{fig:reg-pareto-all} 
We observe 3 groups of methods: Fair Lasso, FP and PP; Lasso and SP; FRRM and MD.  %for all Pareto fronts, 
%Fair Lasso, FP %Fair Posterior (\textbf{FP}) 
%behave very similarly, as do Lasso and SP. %and the projected-posterior (\textbf{PP}) is slightly worse than these two.
%All these methods incorporate a penalisation parameter $\lambda$ and we can compute their Pareto fronts by computing the estimates for each value $\lambda$ in a suitable grid (see Figure \ref{fig:md-vs-fairlasso}). 
%\paragraph{Comparison to the MD estimate \citep{calders2013controlling}.} 
%In Figure \ref{fig:md-vs-fairlasso}, we compare the Pareto fronts of \textbf{FairLasso} and \textbf{MD} when considering the pairs of objectives: (a) DP unfairnesss and MSE; (b) Individual unfairness and MSE; (c) DP and Individual unfairness. 
The Pareto front WDU vs MSE (left plot) shows that the methods targeting DP (MD and FRRM) achieve better %MSE-WDU 
trade-offs than Fair Lasso and FP for moderate values of WDU, %when the unfairness is measured by DP Unfairness, 
however MD and FRRM cannot achieve very small WDU, because of their penalty %function  %This is due to their fairness penalty 
which can be null even without DP being achieved. Fair Lasso, FP, PP and Lasso can reach DP (WDU$=0$) at the null model %for which WDU 
%\jack{[WDU is a violation of DP]}
%is null and can 
but also achieve very low %levels of 
WDU once $S$-dependent covariates are removed. Moreover, for a given value of WDU, Fair Lasso and FP achieve much lower MSE than Lasso and SP.
%since Fair LASSO targets a compromise between DP and individual unfairness, its DP unfairness decreases less rapidly than MD and FRRM as the penalisation parameter increases, but more rapidly than standard LASSO which does not include any measure of fairness. 
%\jack{Think we need to make very obvious that decreasing WDU, implies increasing MSE for all methods, and therefore when we look at the right plot, increasing MSE corresponds to decreasing WDU}

The Pareto front IU vs MSE (middle plot) %in    Figure \ref{fig:reg-pareto-all}) 
shows that for Fair Lasso and FP, %lie in-between MD and FRRM on one hand, and Lasso and SP on the other hand. % methods and the fairness-unaware methods, i.e., its 
%This means for our methods, 
IU increases less rapidly than FRRM and MD %on one hand, 
but more rapidly than Lasso and SP 
%on the other hand, 
as the budget decreases (or regularisation increases). %We note that since 
Lasso and SP have the lowest IU across every MSE value  % - due to their targettting of the true model which is individually fair - 
however as we previously saw each MSE value (except %the 
extreme ones) corresponds to %much 
higher WDU than the other methods. Besides, %we remark that 
all methods achieve zero IU when the constraint or regularisation is inactive. %fairness penalty or regularisation is null. 
Finally, the Pareto front IU vs WDU (right plot) confirms that Fair Lasso, FP and PP achieve better trade-offs between these two unfairness criteria than FRRM and MD.

In summary, Fair Lasso and FP achieve better MSE for the same level of WDU  than fairness-unaware methods (Lasso and SP), and improved MSE for the same level of IU compared with DP methods (FRRM and MD). %Appendix \ref{app:num_results} contains analogous results for a simulated logistic regression experiment.

\subsection{Real-data experiments}

%\subsubsection{Community dataset}

The Communities and Crime dataset contains demographic, socio-economic and crime variables for 1,984 neighborhoods. Following previous work, we treat the normalised crime rate (between 0 and 1) as the outcome variable and the majority race (white or non-white) as the sensitive attribute. %Specifically, for each neighborhood $i$, $s_i=1$ if the neighborhood's population is mainly white  and $s_i=0$ otherwise. 
%White neighborhoods (about 80\%) and $S=0$ for non-White ones (about 20\%).  
%
%For pre-processing, 
%We standardize the predictors, remove the neighborhoods with  output values outside of the range $[0.01, 0.99]$, and transform the output via the logit function (not defined on 0 and 1). 
Figure \ref{fig:community} shows that the outcome distributions per group significantly differ. The correlation coefficient between $S$ and $Y$ %the outcome 
is $-0.459$, indicating that white neighborhoods have on average lower crime rates. After preprocessing (see Appendix \ref{app:num_results}), $n = 1940$  data samples remain and we split them into 80\% training and 20\% testing set.

The Pareto fronts %of our fair methods and the baselines are reported 
in Figure \ref{fig:reg-pareto-all} (bottom row) %and 
confirm previous observations: %the observations made in the simulated experiments. 
FRRM and MD can achieve better trade-offs between MSE and WDU compared to Fair Lasso and FP, but they display worse trade-offs between MSE and IU, and cannot achieve low levels of WDU nor IU. %The Pareto front of the Fair Lasso is sandwiched between that of the Lasso on one hand and those of FRRM and MD on the other hand, and therefore achieve the best trade-offs between IU, WDU and MSE.
To compare the resulting trade-offs between %all 3 measures of performance (
IU, WDU and MSE, %, we assume that 
we fix certain levels of WDU (resp. IU)  %is fixed (e.g., by a stake-holder or policy maker) 
and evaluate MSE and IU (resp. WDU) unfairness of the best predictive model (lowest MSE) for each method
%The relative advantages of the Fair Lasso can be more clearly evaluated by fixing a certain level of group (or individual) unfairness to be achieved and ranking the methods according to their individual (or group) unfairness and MSE.
%These results are reported in 
(Figure \ref{fig:community-barplot} and \ref{fig:community-barplot-iu} in Appendix \ref{app:num_results}).  %for % where methods are evaluated for
%3 increasing levels of WDU (resp. IU). %: WDU$\in \{0.05, 0.1, 0.5\}$ (resp. individual unfairness: IU$\in \{0.03, 0.05, 0.1\}$).
%we find the penalisation parameter for the Lasso, Fair Lasso, FRRM and MD with WDU lower or equal  than the fixed level and the lowest MSE, and we evaluate the IU of that model. 
%For the Bayesian methods, we select the model with the lowest median MSE and report the median metrics.
%\jack{I still think we should just look at the predictve values}. 
%These plots show on one hand, 
We observe on one hand that if WDU is fixed, %to a medium level, 
Fair Lasso and FP %can 
achieve lower levels of IU than FRRM and MD, and  lower MSE than Lasso and SP (particularly for small WDU). On the other hand, if IU is fixed %to a medium level 
then FP and Fair Lasso achieve lower MSE than FRRM and MD and lower WDU than Lasso and SP. 

This experiment thus confirms %experiment shows 
that Fair Lasso and FP %allow to 
can attain all levels of WDU and IU and generally %attain 
better trade-offs than the fairness-unaware and DP methods. Additional results %from applying logistic regression to 
on another dataset (German Credit Data) are reported in Appendix \ref{app:num_results}.

\section{Conclusion}

%In this work 
We introduced two frameworks for fair
%/legitimate 
variable selection, the Fair Lasso and the Fair Posterior induced by the Fair $\ell_1$-ball prior. 
%Posterior based, which leverage weighted $\ell_1$-ball penalty and prior respectively. 
Central to both methods is the specification of a set of $\epsilon$-fair GLM models, under unfairness budget $\epsilon$, %which 
using a measure of dependence between the covariates and $S$ %a sensitive attribute 
and the size of that feature's GLM coefficient.
%Both methods jointly select variables and coefficients under a certain unfairness budget $\epsilon$. 
To the best of our knowledge, the  Fair Posterior is the first fair Bayesian approach %, and since this posterior distribution is supported on the set of $\epsilon$-fair parameters, this 
which guarantees that any posterior point estimate and credible set is also $\epsilon$-fair. The selected variables can be used as the conditioning set in %the 
CDP %criterion 
when no policy or legal framework prescribes this choice. We  showed that %the 
corresponding sparse and CDP-fair GLMs achieves favourable trade-offs between  DP, IF and predictive performance. %group fairness and individual fairness, which are generally incompatible, and 
The simplification to a binary sensitive attribute is not essential for our methods. %was made in the exposition of our methods, however the latter can easily extend to more general categorical or continuous sensitive attribute.
%It can easily by extended to
Extensions to a single continuous sensitive attribute or multiple, categorical sensitive attributes is straightforward. %, by defining a group for all possible values of the sensitive attribute vector. 
%For multiple sensitive attributes with at least one continuous, further extension is needed.  
%While our method is exposed for the case of binary attribute, it easily extends to sensitive attributes taking more than two values. The correlation coefficient $\rho(X_i,S)$ can be used even if $S$ takes values in a infinite or continuous set. Other measures of dependence like the Wasserstein distance between conditional distributions \eqref{eq:dis-distance} can easily be adapted when $S$ takes a finite number of values (e.g., via a Wasserstein barycenter). If there are more than one sensitive attribute, e.g., $S_1, S_2, \dots, S_K$, one could consider an additive measure of dependence $\rho_i = \rho(\mathbf X_i, (S_k)_k) = \sum_{k=1}^K \rho(\mathbf X_i, S_k)$ and follows the same methodology. More elaborate methods for dealing with interdependence (intersectionality) is left for future work.
%Finally, we stress an important limitation of our %fair variable selection 
Limitations of our methods include that  we do not account for estimation error of $\widehat \rho$, %the dependence between each covariate and $S$, %feature and the sensitive attribute, 
that we measure pairwise dependence %between features and the sensitive attribute 
rather than mutual dependence, and that we do not the estimate the partial dependence of $S$ %the sensitive attribute 
on the observed outcome, %and so cannot account and correct for this type of bias or measurement error, 
which could %well %have an indirect 
impact %on 
the selected set. % of selected covariates. 
These %issue 
will be addressed in future work.
%Further, in higher dimensional settings, the $\ell_1$-ball prior %sampling of non-sparse latent variables 
%becomes computationally challenging and spike-and-slab methods priors that only sample non-zeros would be preferable.
%it is important to recognise what CDP and our method cannot achieve. First it does not provide a solution for data biases such as systematic measurement error or direct causal effect of the sensitive attribute onto the outcome. In contrast, DP partly solves these problems, but specific methodologies that deal with them could also be combined with our approach. 

%\jack{Note about scalability of $\ell_1$-ball for large $d$ and exact 0's being better}

% \section{Supplementary Material}

% If you need to include additional appendices during submission, you can include them at the end of the main submission file (here) OR submit them separately as part of the supplementary material.
% When submitting a separate supplementary material file, it may be either a PDF file (such as proof details) or a ZIP file for other formats/more files (such as code or videos).
% Note that reviewers are under no obligation to examine your supplementary material.
% If you have only one supplementary PDF file, please upload it as is; otherwise, gather everything to the single ZIP file.

% For the separate supplementary material file must use \texttt{aistats2027.sty} as a style file and follow the same formatting instructions as in the main paper.
% The only difference is that it must be in a \emph{single-column} format.

\subsection*{AI use statement}

In this work, we have used generative AI tools to generate the code of the projection function used to construct the projected fair prior in \Cref{sec:fair-bayes} and feedback on a earlier version of this draft (LLM-powered feedback through Google's Paper Assistant Tool).
We have not used generative AI tools for generating the research idea, the methodology and writing the paper.
%[other tasks with required disclosure], and [the rest of the required disclosure tasks] are not applicable to this work.
Additionally, we used generative AI tools for checking if the methodology had been previously proposed in the literature.
% [tasks with recommended disclosure].
% We have reviewed all AI-assisted work.
% [Elaborate. For example, ``we checked LLM-generated research ideas for potential plagiarism through a manual literature survey'', ``LLM-generated code was verified and tested for correctness by two authors'', ``all LLM-assisted proofs were checked line by line by the authors'', etc.].
We take responsibility for the final content of this work, including text, claims, or artifacts produced with the aid of generative AI.

% See the AISTATS 2027 AI Policy for Authors for more details. This statement should not be more than 1 page.

% \subsubsection*{Acknowledgements}
% All acknowledgements go at the end of the paper, including thanks to reviewers who gave useful comments, to colleagues who contributed to the ideas, and to funding agencies and corporate sponsors that provided financial support.
% To preserve the anonymity, please include acknowledgements \emph{only} in the camera-ready papers. The acknowledgements do not count against the 9-page page limit in the camera-ready.

%\subsubsection*{References}

\bibliography{bib}

% References follow the acknowledgements. Use an unnumbered third level
% heading for the references section.  Please use the same font
% size for references as for the body of the paper---remember that
% references do not count against your page length total.

% \begin{thebibliography}{}
% \setlength{\itemindent}{-\leftmargin}
% \makeatletter\renewcommand{\@biblabel}[1]{}\makeatother
% \bibitem{} J.~Alspector, B.~Gupta, and R.~B.~Allen (1989).
%     \newblock Performance of a stochastic learning microchip.
%     \newblock In D. S. Touretzky (ed.),
%     \textit{Advances in Neural Information Processing Systems 1}, 748--760.
%     San Mateo, Calif.: Morgan Kaufmann.

% \bibitem{} F.~Rosenblatt (1962).
%     \newblock \textit{Principles of Neurodynamics.}
%     \newblock Washington, D.C.: Spartan Books.

% \bibitem{} G.~Tesauro (1989).
%     \newblock Neurogammon wins computer Olympiad.
%     \newblock \textit{Neural Computation} \textbf{1}(3):321--323.
% \end{thebibliography}

%%%%%%%%%%%%%%%%%%%%%%%%%%%%%%%%%%%%%%%%%%%%%%%%%%%%%%%%%%%%

\section*{Checklist}

% %%% BEGIN INSTRUCTIONS %%%
%The checklist follows the references. For each question, choose your answer from the three possible options: Yes, No, Not Applicable.  You are encouraged to include a justification to your answer, either by referencing the appropriate section of your paper or providing a brief inline description (1-2 sentences).
%Please do not modify the questions.  Note that the Checklist section does not count towards the page limit. Not including the checklist in the first submission won't result in desk rejection, although in such case we will ask you to upload it during the author response period and include it in camera ready (if accepted).

%\textbf{In your paper, please delete this instructions block and only keep the Checklist section heading above along with the questions/answers below.}
% %%% END INSTRUCTIONS %%%

\begin{enumerate}

  \item For all models and algorithms presented, check if you include:
  \begin{enumerate}
    \item A clear description of the mathematical setting, assumptions, algorithm, and/or model. [Yes]
    \item An analysis of the properties and complexity (time, space, sample size) of any algorithm. [Not Applicable]
    \item (Optional) Anonymized source code, with specification of all dependencies, including external libraries. [Yes]
  \end{enumerate}

  \item For any theoretical claim, check if you include:
  \begin{enumerate}
    \item Statements of the full set of assumptions of all theoretical results. [Not Applicable]
    \item Complete proofs of all theoretical results. [Not Applicable]
    \item Clear explanations of any assumptions. [Not Applicable]
  \end{enumerate}

  \item For all figures and tables that present empirical results, check if you include:
  \begin{enumerate}
    \item The code, data, and instructions needed to reproduce the main experimental results (either in the supplemental material or as a URL). [Yes]
    \item All the training details (e.g., data splits, hyperparameters, how they were chosen). [Yes]
    \item A clear definition of the specific measure or statistics and error bars (e.g., with respect to the random seed after running experiments multiple times). [Yes]
    \item A description of the computing infrastructure used. (e.g., type of GPUs, internal cluster, or cloud provider). [Not Applicable]
  \end{enumerate}

  \item If you are using existing assets (e.g., code, data, models) or curating/releasing new assets, check if you include:
  \begin{enumerate}
    \item Citations of the creator If your work uses existing assets. [Yes]
    \item The license information of the assets, if applicable. [Not Applicable]
    \item New assets either in the supplemental material or as a URL, if applicable. [Yes]
    \item Information about consent from data providers/curators. [Not Applicable]
    \item Discussion of sensible content if applicable, e.g., personally identifiable information or offensive content. [Not Applicable]
  \end{enumerate}

  \item If you used crowdsourcing or conducted research with human subjects, check if you include:
  \begin{enumerate}
    \item The full text of instructions given to participants and screenshots. [Not Applicable]
    \item Descriptions of potential participant risks, with links to Institutional Review Board (IRB) approvals if applicable. [Not Applicable]
    \item The estimated hourly wage paid to participants and the total amount spent on participant compensation. [Not Applicable]
  \end{enumerate}

\end{enumerate}

\clearpage
\appendix
\thispagestyle{empty}

% Supplementary material: To improve readability, you must use a single-column format for the supplementary material.
\onecolumn
\aistatstitle{%Instructions for Paper Submissions to AISTATS 2027: \\
Supplementary Materials}

% Reset the figure counter to 0
\setcounter{figure}{0} 
% Redefine the figure numbering format to "A.1", "A.2", etc.
\renewcommand{\thefigure}{S\arabic{figure}} 

% Note: You can choose whether to include you appendices as part of the main submission file (here) OR submit them separately as part of the supplementary material. The preferred option is to include appendices as part of the main submission file. It is the authors' responsibility that any supplementary material does not conflict in content with the main paper (e.g., the separately uploaded additional material is not an updated version of the one appended to the manuscript).

\section{Fair Projected-Posterior}\label{app:projected-post}

In this section we present an alternative approach that also uses projections to construct a fair posterior supported on the set of fair parameters $\widehat B(\epsilon)$.
%and selecting a set of fair variables.  
%legitimate covariates in a Bayesian fashion and designing a CDP-fair posterior. 
For this we leverage ideas from the projected (or, immersion) posterior \citep{pal2025bayesianhighdimensionallinearregression} in sparse linear regression, where a standard (unconstrained) prior is used and the induced posterior is post-processed by projecting its samples onto the constrained space (a standard $\ell_1$-ball in \citep{pal2025bayesianhighdimensionallinearregression}). For simplicity here we consider the linear regression setting, however the following derivations also apply to GLMs.
%are computed  and apply the problem's constraints (such as sparsity) as a post-processing step of the standard posterior. 

%In this way, we can design a method complying with the fairness budget constraint while retaining the main advantages of the Bayesian approach - incorporation of a-priori information and uncertainty quantification. Broadly speaking, since the Bayesian predictive distribution can be seen as an aggregation of multiple models, we require that each model satisfy the constraint by projecting the posterior distribution's samples onto the space of fair models. This post-processing technique of the posterior allows to use standard (e.g., conjugate) prior distributions, which is often computationally convenient. 

%Here the constraint on our posterior is the unfairness budget, i.e., its support should be the space of $\epsilon$-fair samples defined as
Recall that our space of  $\epsilon$-fair GLM coefficients is
\begin{align*}
    \widehat B(\epsilon):= \{\beta \in \mathbb R^d :  \widehat{ \mathcal{U}}(\beta) = \|\widehat \rho \circ \beta\|_1\leq \epsilon \}.
\end{align*}
Given a standard (possibly conjugate) prior $\pi(\beta_0, \Tilde\beta, \sigma^2
)$ (with $\sigma^2$ the noise variance) and the standard posterior
\begin{align*}
    \pi(\beta_0,\Tilde\beta, \sigma^2 | \mathbf X, \mathbf y) \propto \exp\left\{-\mathcal{L}(\mathcal{D}_n; \beta_0, \Tilde\beta, \sigma^2)\right\} \pi(\beta_0, \Tilde\beta, \sigma^2),
\end{align*}
we can construct a $\epsilon$-fair projected-posterior distribution $\Tilde \pi_\epsilon(\beta, \beta_0, \sigma^2 |\mathbf X, \mathbf y)$ via a projection step onto $ \widehat B(\epsilon)$:
\begin{align}\label{eq:projected_post}
    \Tilde \pi_\epsilon(\beta_0 \in A, \beta \in A', \sigma^2 \in A'' |\mathbf X, \mathbf y) =  \int \mathds{1}_{P_{ \widehat B(\epsilon)}(\Tilde\beta) \in A'} \mathds{1}_{\beta_0 \in A} \mathds{1}_{\sigma^2 \in A''} \pi(\beta_0, \Tilde\beta, \sigma^2 | \mathbf X, \mathbf y) d \beta_0 d \Tilde\beta d\sigma^2, \quad
\end{align}
for $A \times A' \times A'' \subseteq \mathbb R \times \mathbb R^d \times \mathbb R^+$, 
% \begin{align*}
%     \Tilde \pi( A \times A' |\mathbf X, \mathbf y) =  \int \mathds{1}_{P_{ \widehat B(\epsilon)}(\beta) \in A} \mathds{1}_{\sigma^2 \in A'} \pi(\beta, \sigma^2 | \mathbf X, \mathbf y) d \beta d\sigma^2, \qquad A \times A' \subseteq \mathbb R^d \times \mathbb R^+
% \end{align*}
with $P_{ \widehat B(\epsilon)}$ the projection operator from $\mathbb R^d$ to  $\widehat B(\epsilon)$.

In practice,  sampling from \eqref{eq:projected_post} requires sampling from $\pi(\beta_0, \Tilde\beta, \sigma^2 |\mathbf X, \mathbf y)$ and applying the projection. Precisely, a sample $(\beta_0, \beta, \sigma^2)$ from $\Tilde \pi_\epsilon$ is obtained in 2 steps:
\begin{enumerate}
    \item Sample $(\beta_0, \Tilde\beta, \sigma^2) \sim \pi(\beta_0, \Tilde\beta, \sigma^2 |\mathbf X, \mathbf y)$
    \item Compute $\beta = P_{ \widehat B(\epsilon)}(\Tilde\beta)$ (see Appendix \ref{app:weighted_proj}). % and set $\beta_0 = \beta_0$ and $\sigma^2 = \sigma^2$.
    % i.e.,
    %     \begin{align*}
    %     \tilde \beta = \argmin_{\beta' \in \R^d : \mathcal{\hat U}(\beta') \leq \epsilon} \| \beta' - \beta \|_2^2,
    % \end{align*}
\end{enumerate}

We briefly comment here the differences between the fair projected-posterior $\Tilde \pi_\epsilon$ and the Fair Posterior $\pi_\epsilon$ induced by the fair $\ell_1$-ball prior (see Section \ref{sec:fair-bayes}). Firstly, rephrasing \cite{xu2020bayesian}, post-processing the posterior distribution is not a fully Bayesian approach and does not provide valid interpretation of the credible sets, contrary to the projected-prior approach. Secondly, a main difference between the two approaches is that in the projected-posterior, the un-processed samples enter the likelihood
\begin{align*}
d\Tilde \pi_{\epsilon}(\beta_0,  \beta, \sigma^2 
    \mid \mathbf X, \mathbf y) \propto \mathds{1}_{\beta = P_{ \widehat B(\epsilon)}(\Tilde\beta)} \exp\left\{-\mathcal{L}(\mathcal{D}_n; \beta_0, \Tilde\beta, \sigma^2)\right\} d\pi(\beta_0, \Tilde\beta, \sigma^2),
\end{align*}
while for the projected-prior it is the projected samples:
\begin{align*}
    d\pi_{\epsilon}(\beta_0, \beta, \sigma^2
    \mid \mathbf X, \mathbf y) \propto \exp\left\{-\mathcal{L}(\mathcal{D}_n; \beta_0, \beta \right\} d\pi_\epsilon(\beta_0, \beta, \sigma^2).
\end{align*}
This implies that the projected-prior method is more likely to higher posterior mass to the best fitting regions of $\widehat B(\epsilon)$ compared to the  projected-posterior. Our simulated experiments in logistic regression confirm this intuition. 
%that better inference is achieved by the projected-prior. This can be understood intuitively since the post-processing projection ignores the likelihood and effectively projects posterior samples to a region of 0 posterior mass (unless the posterior is itself sparse). In contrast, the projected-prior leads to a valid posterior distribution where well-fitting regions within the support of the prior are given higher mass than underfitting ones.}

%One can naturally wonder how do the two projections, of the posterior and of the prior, compare. Intuitively, we expect the projection of the prior to lead to more accurate inference, since the likelihood shifts the projected prior towards  parameter that better fits the data, while the projection of the standard posterior ignores the data in the projection step (some underfitting parameter may be highly weighted under this projected posterior). 

Nonetheless, the projected-posterior can be computationally advantageous in some settings. For instance,  when using conjugate priors running MCMC algorithms can be avoided in the first step. For instance, in linear regression, with the conjugate Normal-Inverse-Gamma prior 
\begin{align}\label{eq:normal-gamma}
    &\sigma^2 \sim IG(a,b)\\
    &\beta \mid \sigma^2 \sim N(0,\sigma^2\Sigma)
\end{align}
where $a, b > 0$ %are hyperparameters which  have generally little influence on the performance, 
and $\Sigma \succ 0$ is a positive-definite covariance matrix, e.g., $\Sigma  = \phi^{-1} I_d$, $\phi > 0$,
%Together with the generating model \eqref{eq:true-model}, this prior induces 
the  posterior is also a Normal-Inverse-Gamma distribution
\begin{align*}
    &\sigma^2  \mid  \mathbf X, \mathbf  y \sim IG \left(a + \frac{n}{2},b + \frac{\mathbf y^T \mathbf y - \mathbf y^T  \mathbf X ( \mathbf X^T  \mathbf X + \Sigma^{-1})^{-1} \mathbf X^T \mathbf y}{2} \right) \\
    &\beta \mid \sigma^2,  \mathbf X, \mathbf  y \sim N \big( ( \mathbf X^T  \mathbf X + \Sigma^{-1})^{-1} \mathbf X^T \mathbf  y, \sigma^2 ( \mathbf X^T  \mathbf X+ \Sigma^{-1})^{-1} \big).
\end{align*}
% Additionally, the marginal posterior on $\beta$ is %also available in analytical form: $\pi(\beta |  \mathbf X, y)=MVT_{\nu^*}(\mu^*, \Sigma^*)$,  
% a multivariate t-density with $\nu^\star = 2a + n$ degrees of freedom and parameters
% \begin{align*}
%     &\mu^* = (\mathbf X^T\mathbf X + \Sigma^{-1})^{-1}\mathbf X^T y\\
%     &\Sigma^* =\frac{2b + y^Ty - y^T \mathbf X (\mathbf X^T \mathbf X + \Sigma^{-1})^{-1}\mathbf X^T y}{2a + n}  (\mathbf X^T \mathbf X + \Sigma^{-1})^{-1},
% \end{align*}
% which means that one can directly sample from the posterior $\pi(\beta |\mathbf X, y)$ without the need of an approximation method.
%which bypasses the need for MCMC (which is required with the $\ell_1$-ball prior). 
Moreover, %even if the unprocessed posterior samples are obtained via MCMC (e.g., if the chosen prior is heavy-tailed or sparse), 
one can compute the projected-posterior for different unfairness budgets $\epsilon$ in Step 2 without needing to re-sample in Step 1. %the need to re-run the MCMC algorithm. 
In contrast, the projected-prior requires posterior sampling for each unfairness budget. %the MCMC to sample from the fair posterior needs to be run for each unfairness budget.

%Nonetheless, this method can be used in combination with any other prior such as heavy-tailed priors and sparse priors, as long as samples from the standard posterior can be obtained.

% \jack{[Add soemthing on sampling]}

% \begin{align}\label{eq:fair-post}
%     %\pi(\beta_0, \beta,  \sigma^2 \mid \mathbf X, \mathbf y) \propto \prod_{i=1}^n \ell (y_i ; \beta_0, \beta, \sigma^2, x_i) \tilde \pi(\beta) \pi_0(\beta_0)\pi_\sigma(\sigma^2),
%     &\pi_{\epsilon}(\beta_0, \beta  %\theta 
%     \mid \mathbf X, \mathbf y) \propto \exp\left\{-\mathcal{L}(\mathcal{D}_n; \beta_0, \beta)\right\} \pi_{\epsilon}(\beta) \pi_0(\beta_0).\\%\pi_\theta(\theta),
%     &\pi_{\epsilon}(\beta_0, \tilde{\beta}  %\theta 
%     \mid \mathbf X, \mathbf y) \propto \exp\left\{-\mathcal{L}(\mathcal{D}_n; \beta_0, \beta = P_{\widehat B(\epsilon)}(\Tilde\beta))\right\} \pi(\tilde{\beta}) \pi_0(\beta_0).
% \end{align}

\section{Weighted $\ell_1$-ball projection}\label{app:weighted_proj}

In this section we show how to compute weighted $\ell_1$-ball projections using a  modification of the algorithm in \cite{duchi2008efficient}.

The weighted $\ell_1$-ball projection operator is defined as:
\begin{align*}
    P_1(x) = \argmin_{y \in B_1(w, r)} \frac{1}{2} \|y-x\|^2
\end{align*}
with $B_1(w,r) := \{ x  : \sum w_i |x_i| \leq r \}$ with $w$ a weight vector and $r > 0$ a radius. This is a convex problem and we can use the KKT conditions to find the solution $P_1(x)$. Without loss of generality we can assume $x_i \geq 0$ (otherwise we just need to flip the sign of the $[P_1(x)]_i$ for which $x_i < 0$). We also assume that $\sum w_i x_i > r$ (otherwise $P_1(x) = x$). Defining the Lagrangian function
\begin{align*}
    \mathcal{L}_g(y, \lambda, \nu) = \frac{1}{2} \|y-x\|^2 + \lambda (\sum w_i y_i - r) - \sum \nu_i y_i, \qquad \lambda, \nu \in \mathbb R,
\end{align*}
the KKT conditions are
\begin{align*}
     &\nabla_y \mathcal{L}_g(y^*, \lambda^*, \nu^*) = 0 \\
     &\nu_i^* y_i^* = 0, \forall i \\
     &\lambda^*  (\sum w_i y_i^* - r) = 0 \\
     &\nu_i^* \geq 0, \forall i \\
     &\lambda^* \geq 0
\end{align*}
First condition implies:
\begin{align*}
    (y^* - x) + \lambda w - \nu = 0 \iff  y^* = \nu + x - \lambda w
\end{align*}
Second and fourth condition implies that 
\begin{align*}
    &y_i^* > 0 \implies \nu_i^* = 0 \iff y_i^* = x_i - \lambda^* w_i \\
    &y_i^* = 0 \implies \nu_i^* = \lambda^* w_i - x_i \geq 0
\end{align*}
thus
\begin{align*}
    y_i^* = \max ( x_i - \lambda^* w_i, 0).
\end{align*}
and $\lambda^* > 0$ since $x \not \in B_1(w,r)$. Third condition implies:
\begin{align*}
    \sum w_i y_i^* - r = 0 \iff \sum_i w_i^2 \max ( \frac{x_i}{w_i} - \lambda^* , 0) = r.
\end{align*}
Let
\begin{align*}
    f(\lambda) = \sum_i w_i^2 \max ( \frac{x_i}{w_i} - \lambda, 0)
\end{align*}
The function $f$ is piece-wise linear (in particular, continuous), non-increasing, and its image is $[0,+\infty)$ thus there exists $\lambda^*$ such that $f(\lambda^*) = r$. Let $\theta_i = \frac{x_i}{w_i}$ and $(\theta_{\sigma(i)})_{i}$ the ordered sequence (by decreasing value). Then $f(\lambda)$ is linear on each interval $I_i = [\theta_{\sigma(i+1)}, \theta_{\sigma(i)}]$ since on this interval
\begin{align*}
     f(\lambda) = \sum_{j \leq i} w_{\sigma(j)}^2 ( \theta_{\sigma(j)} - \lambda) =  \sum_{j\leq i} w_{\sigma(j)}^2 \theta_{\sigma(j)} - \lambda  \sum_{j\leq i} w_{\sigma(j)}^2
\end{align*}
Therefore the image of $f$ on $I_i$ is
\begin{align*}
 f(I_i) =  \left[ \sum_{j\leq i} w_{\sigma(j)}^2 \theta_{\sigma(j)}  -\theta_{\sigma(i)} \sum_{j\leq i} w_{\sigma(j)}^2, \sum_{j\leq i} w_{\sigma(j)}^2 \theta_{\sigma(j)} - \theta_{\sigma(i+1)}  \sum_{j\leq i} w_{\sigma(j)}^2 \right]
\end{align*}
and we need to find $i$ such that $r \in f(I_i)$ and then
\begin{align*}
    \lambda^* =  \frac{\sum_{j\leq i} w_{\sigma(j)}^2 \theta_{\sigma(j)} - r}{ \sum_{j\leq i} w_{\sigma(j)}^2} =  \frac{\sum_{j\leq i} w_{\sigma(j)} x_{\sigma(j)} - r}{ \sum_{j\leq i} w_{\sigma(j)}^2}
\end{align*}
Thus the projection is
\begin{align*}
    [P_1(x)]_i = x_i  \max ( x_i - \lambda^* w_i, 0).
\end{align*}
and for general $x \in \mathbb R^d$,
\begin{align*}
    [P_1(x)]_i = sign(x_i)  \max ( |x_i| - \lambda^* w_i, 0).
\end{align*}
The pseudo-code of the algorithm for computing the projection $P_1(x)$ of a vector $x \in \mathbb R^d$ onto a weighted $\ell_1$-ball with weights $w$ and radius $r > 0$, is reported in Algorithm \ref{alg:l1projection} for completeness.

\begin{algorithm}[t]
\caption{Projection onto the weighted $\ell_1$ ball}
\label{alg:weighted-l1-projection}
\begin{algorithmic}[1]
\Require input vector $x \in \mathbb{R}^d$, vector of weights $w \in \mathbb{R}_+^d$, radius $r > 0$
\Ensure $y = P_{1}(x)$
\State $y \gets x$ 
\State $\mathcal{P} \gets \{ i : w_i > 0 \}$
\State $m \gets |\mathcal{P}|$
\If{$\sum_{i \in \mathcal{P}} w_i |x_i| \le r$}
    \State \Return $y$ 
\EndIf
\State $u_i \gets |x_i|$, \; $t_i \gets x_i / w_i$ \; for all $i \in \mathcal{P}$
\State Order $\mathcal{P}$ so that $t_{(1)} \ge t_{(2)} \ge \dots \ge t_{(m)}$
\For{$k = 1, \dots, m$} 
    \State $\lambda_k \gets \Big( \sum_{j=1}^{k} w_{(j)} u_{(j)} - r \Big) \Big/ \sum_{j=1}^{k} w_{(j)}^2$
\EndFor
\State $K \gets \max\{ k : \lambda_k < t_{(k)} \}$
\State $\lambda^\star \gets \lambda_K$
\For{$i \in \mathcal{P}$}
    \State $y_i \gets \operatorname{sign}(x_i)\, \max\big(u_i - \lambda^\star w_i,\, 0\big)$
\EndFor
\State \Return $y$
\end{algorithmic}
\label{alg:l1projection}
\end{algorithm}

% Defining:
% \begin{align*}
%     &s_i = sign(x_i) \\
%     &t_i = |x_i| - \lambda^* w_i \\
%     &\lambda =  \frac{\sum_{j\leq i} w_{\sigma(j)} x_{\sigma(j)} - r}{ \sum_{j\leq i} w_{\sigma(j)}^2}
% \end{align*}
% and $(t,s, \lambda) = g(x)$ and we can compute the Jacobian of $g^{-1}$ as follows: since $x_i = s_i (t_i + \lambda^* w_i)$,
% \begin{align*}
%     &\frac{\partial x_i}{\partial t_i} = s_i  \\
%     &\frac{\partial x_i}{\partial s_i} = t_i  \\
%     &\frac{\partial x_i}{\partial \lambda} = s_i w_i
% \end{align*}

\section{Description of baseline methods}\label{app:baselines}

Here we list and briefly describe the baseline methods we compare to in our numerical experiments:
\begin{enumerate}
    \item \textbf{Standard LASSO} \citep{tibshirani1996regression} $\hat \beta_{L}$ %(see \eqref{eq:lasso}),
    \begin{align}\label{eq:lasso}
    % (\widehat \beta_{L0}, \hat \beta_{L}) = \argmin_{(\beta_0, \beta) \in \mathbb R \times \R^d}  \frac{1}{n} \|\mathbf X \beta + \beta_0 \mathds{1}_n -  \mathbf y \|_2^2 + \lambda \|\beta \|_1,\\
    (\widehat \beta_{L0}, \hat \beta_{L}) = \argmin_{(\beta_0, \beta) \in \mathbb \R^{d+1}}  \frac{1}{n} \mathcal{L}( \mathcal{D}_n; \beta_0, \beta) + \lambda \|\beta \|_1,
\end{align}
    % the standard LASSO  estimate is \ds{remove expression}
    % \begin{align*}
    %         \hat \beta_{L} = \argmin_{\beta \in \R^d}  \|\mathbf X \beta -  \mathbf y \|_2^2 + \lambda \| \beta \|_1,
    % \end{align*}
    %the penalisation function is $p(\beta ; \lambda) = \lambda \|\beta\|_1$ where  $\lambda > 0$ is the regularisation parameter.
    %The LASSO penalises all coefficients equally, regardless of their (un)fairness. In our experiments, 
    %We compute $\hat \beta_{L}$ using %the implementation in 
    computed using the Python package \textbf{scikit-learn}. %This method is blind to any fairness constraint. 
    \item \textbf{Mean-Difference estimate (MD)} \citep{calders2013controlling}: this estimate is only defined in linear regression and is the solution of 
    \begin{align*}
            (\widehat \beta_{0MD}, \widehat \beta_{MD}) = \argmin_{(\beta_0, \beta) \in \R \times \R^d}  \|\mathbf X \beta + \beta_0 \mathds{1}_n -  \mathbf y \|_2^2 +  \lambda ( \beta^T m )^2,
    \end{align*}
    %\jack{$\mathcal{L}$?}
    %optimiser of a penalised least-square problem where the penalisation function is
    where $m =  \frac{1}{n_1} \sum_{i: s_i=1} x_{i} - \frac{1}{n_0} \sum_{j: s_j=0} x_j$,
    % \begin{align*}
    %     &p(\beta ; \lambda) = \lambda ( \beta^T d )^2 \\
    %     &d =  \frac{1}{n_1} \sum_{i: s_i=1} x_{i} - \frac{1}{n_0} \sum_{j: s_i=0} x_j
    % \end{align*}
    $n_1,n_0$ are the size of the groups of observations $G_k :=\{i: s_i=k\}$, $k \in \{0,1\}$. Here the penalty is the squared difference between the average predicted outcomes in the groups $G_0$ and $G_1$. This penalty can be interpreted as enforcing a relaxed version of DP, where only the conditional mean outcome $\mathbb E[f(X,S) \mid S ]$ needs to be independent of $S$.  %\jack{have we said this?}

    \item \textbf{Fair Ridge Regression Model (FRRM)} \citep{scutari2022achieving}\footnote{Implementation in the R package \texttt{fairml}  \url{https://cran.r-project.org/web/packages/fairml/index.html}}: this estimation method targets DP by solving a penalised objective where the penalisation function is the proportion of outcome variance explained by the sensitive attribute $S$. We note that in this method, the outcome $Y$ is also regressed on $S$. 

    \item \textbf{Fair Feature Selection (FFS)} \citep{belitz2021automating}\footnote{Code available at \url{https://github.com/pnb/fairfs/releases/tag/v0.1-aies} and adapted using Claude.}: this is a greedy algorithm that selects features in a sequential manner by minimising a fairness-penalised objective, in a classification setting. To align with other methods, we use a logistic regression model and set the performance metrics as the BCE and the unfairness metrics as a measure of DP violation. For each value of the regularisation hyperparameter, the selected set of features is obtained my minimising the penalised objective on the cross-validated set.

    \item \textbf{Standard Bayes posterior (Bayes)}: this posterior distribution is obtained from the conjugate Normal-Gamma prior in the linear regression setting and a standard Normal prior in logistic regression (see Appendix \ref{app:projected-post}). It is also the base posterior for the projected-posterior (see below).
    %\jack{Do we change the prior variance?}
    
    \item \textbf{Sparse posterior (SP)}: this is the posterior distribution obtained from the standard $\ell_1$-ball prior \citep{xu2020bayesian}. In this method the base prior is also the conjugate Normal-Gamma prior in the linear regression setting and a standard Normal prior in logistic regression. 
    %Contrary to the original method where the radius of the $\ell_1$-ball is given an exponential prior, here 
    We compute the posteriors obtained for a grid of values of the radius. %and compare the sequence of posteriors to the sequence of fair posteriors under different unfairness budgets.}
    %\jack{and the reccomended expnential priro on the radius}\ds{I actually vary it, explain why it is varied}

        \item \textbf{Projected Posterior (\textbf{PP})}: this is a posterior distribution constructed from projecting the standard Bayes posterior above onto $\widehat B(\epsilon)$  (see Appendix \ref{app:projected-post}).

    % \item \ds{\textbf{Fair Empirical Bayes posterior (FEB)} \citep{carrizosa2024empirical}: this method constructs a posterior distribution which hyperparameters are tuned so that the posterior mean $\hat \beta$ satisfy the mean difference constraint $\hat \beta^T d = 0$. Code in Matlab and difficult to adapt to a new data set.}
    % \item \ds{\textbf{DP-fair linear model}: we use the methodology of \cite{chzhen2022minimax} to build an approximately DP-fair linear model.}
    % \item \ds{\textbf{CDP-fair model} of \cite{ghassemi2025auditing}?}
\end{enumerate}
%\jack{Can we defer all prior setups explanation fo other methods to an appendix section}

\section{Additional material and results from the  numerical experiments}\label{app:num_results}

%In this section we report additional results of the simulated experiments.

The code to reproduce all our numerical experiments is available on Github.\footnote{ Anonymous Github repository \url{https://anonymous.4open.science/r/FairVariableSelection-1A87/README.md}}

In all experiments, we compute $\widehat \rho$ using the training set. Moreover, for the Bayesian methods, we take 2000 samples from the posterior. For the non-conjugate posterior distribution, we run the NUTS algorithm via the NumPyro implementation with 1000 warm-up iterations.

\subsection{Simulation 1: Linear regression}

In the linear regression setting, the intercept $\beta_0 = 0$, the regression coefficient is
\begin{align}\label{eq:beta_star}
    \beta^* = (2, 2, 0.2, 0.2, -2, -2, -0.2, -0.2),
\end{align}
and we recall that $X_1, \dots, X_4$ are independent of $S$ while $X_5, \dots, X_8$ are dependent of $S$.
Moreover the training and test sample size is 200. We set $\widehat \rho$ as the empirical correlation coefficient and $\xi = 5$  in the IU metric. To plot the Pareto fronts of the Bayesian methods, for each posterior sample $\beta$, we compute the expected outcomes $\widehat y_i = \beta^T x_i$ in the test set and the corresponding performance metrics, and take the average of these metrics over posterior samples. %\jack{[Is this final bit still true?]}. 

\begin{figure}[h!]
    \centering
    \includegraphics[width=0.45\linewidth, trim = 1cm 0cm 1cm 1.35cm, clip]{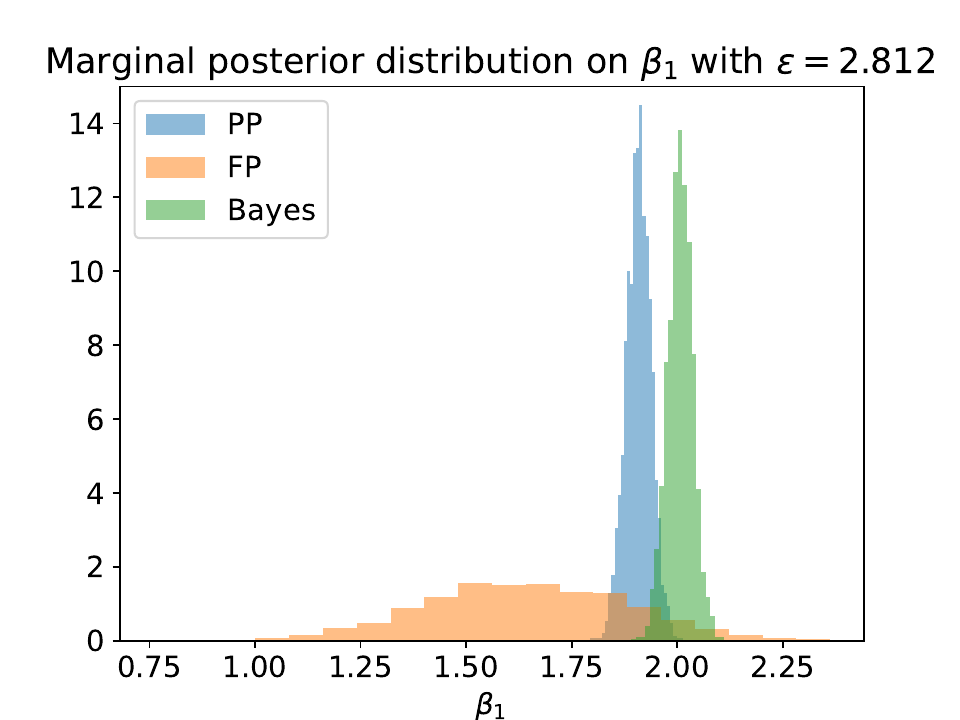}%
    \includegraphics[width=0.45\linewidth, trim = 1cm 0cm 1cm 1.35cm, clip]{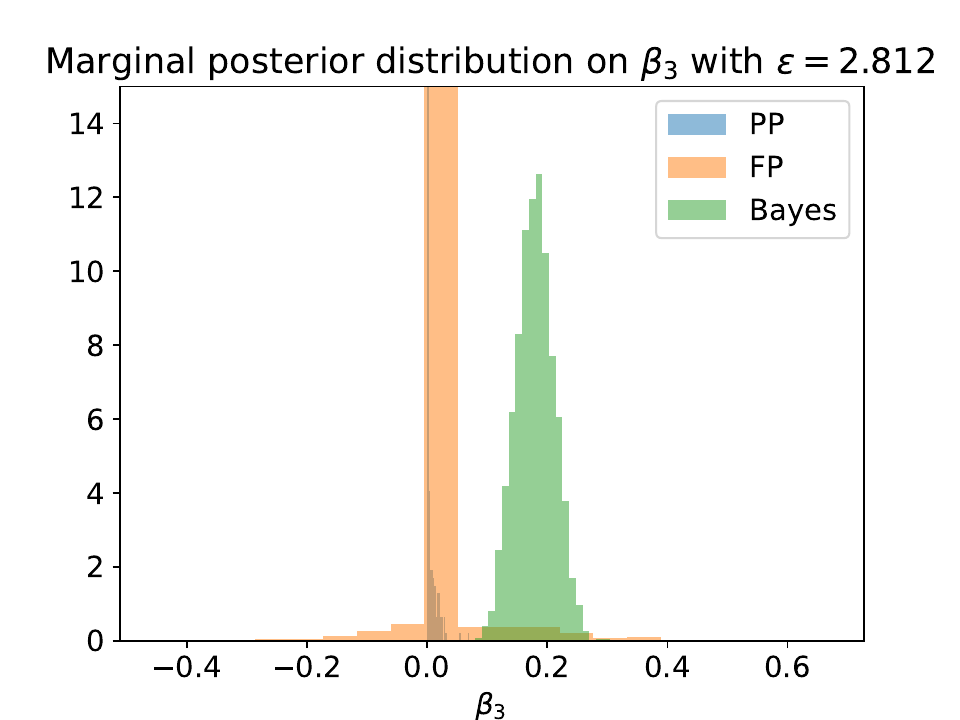}
    \includegraphics[width=0.45\linewidth, trim = 1cm 0cm 1cm 1.35cm, clip]{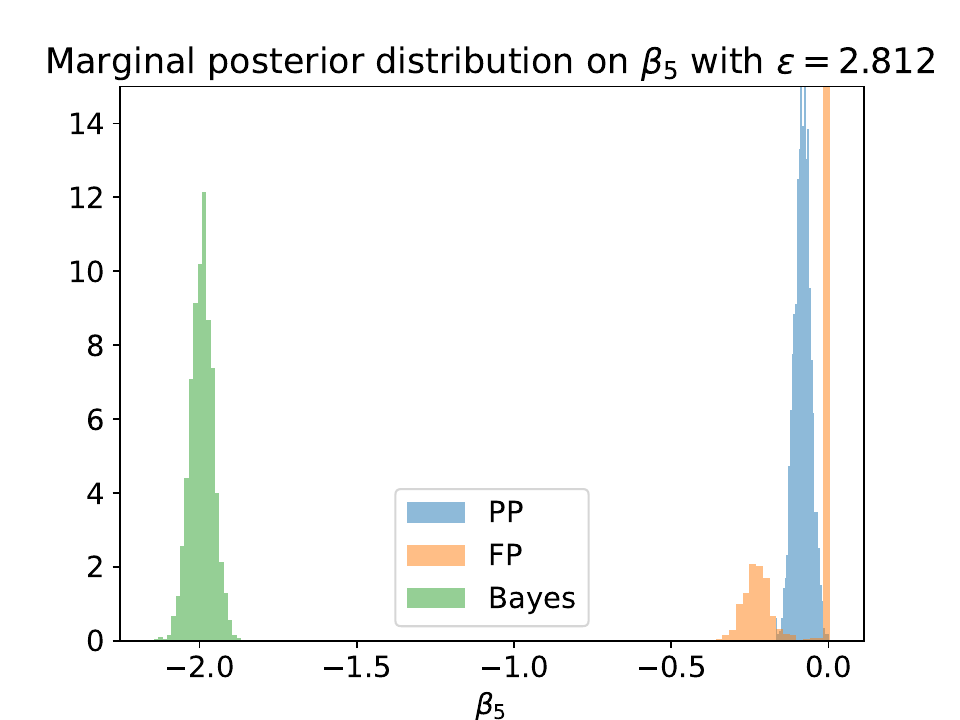}%
    \includegraphics[width=0.45\linewidth, trim = 1cm 0cm 1cm 1.35cm, clip]{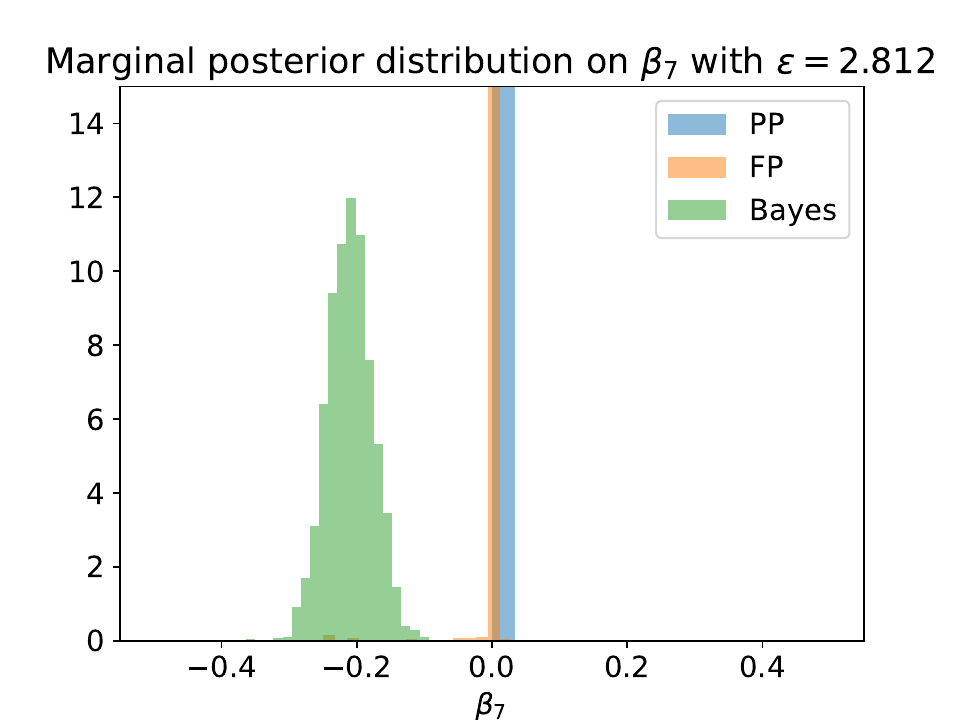}
    \caption{Marginal posterior distribution on $\beta_1$ (top left, fair and strong), $\beta_3$ (top right, fair and weak), $\beta_5$ (bottom left, unfair and strong) and $\beta_7$ (bottom right, unfair and weak) of the Fair Posterior (FP), Projected Posterior (PP) and the conjugate standard posterior (Bayes) for an unfairness budget $\epsilon = 2.812$.}
    \label{fig:marg-post}
\end{figure}

\begin{figure}
    \centering
    \includegraphics[width=0.45\linewidth, trim = 0cm 0cm 1cm 1.35cm, clip]{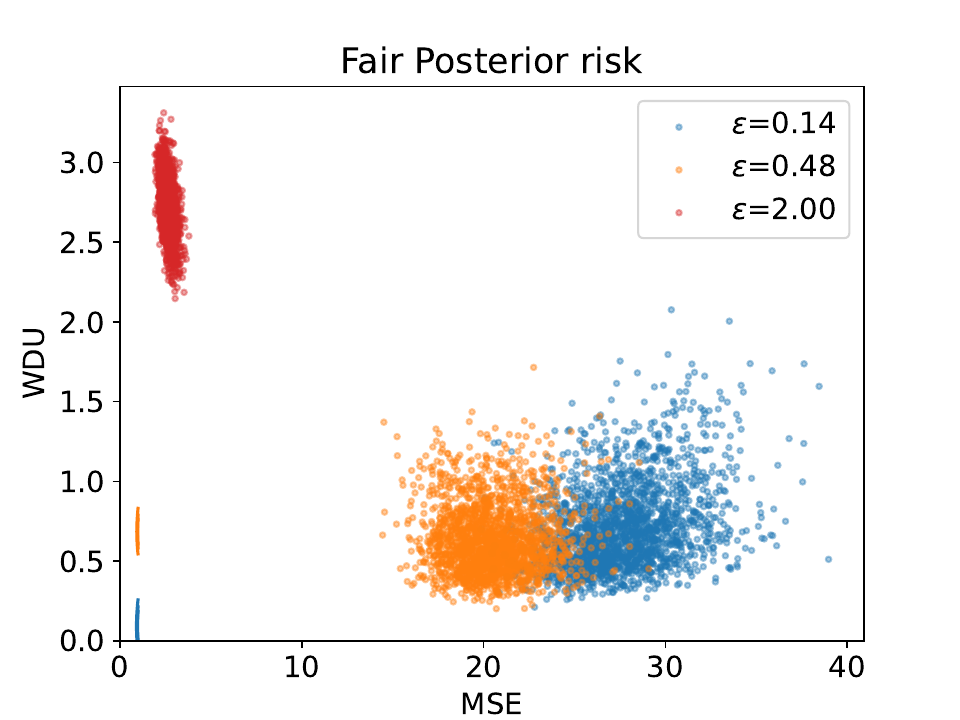}%
    \includegraphics[width=0.45\linewidth, , trim = 0cm 0cm 1cm 1.35cm, clip]{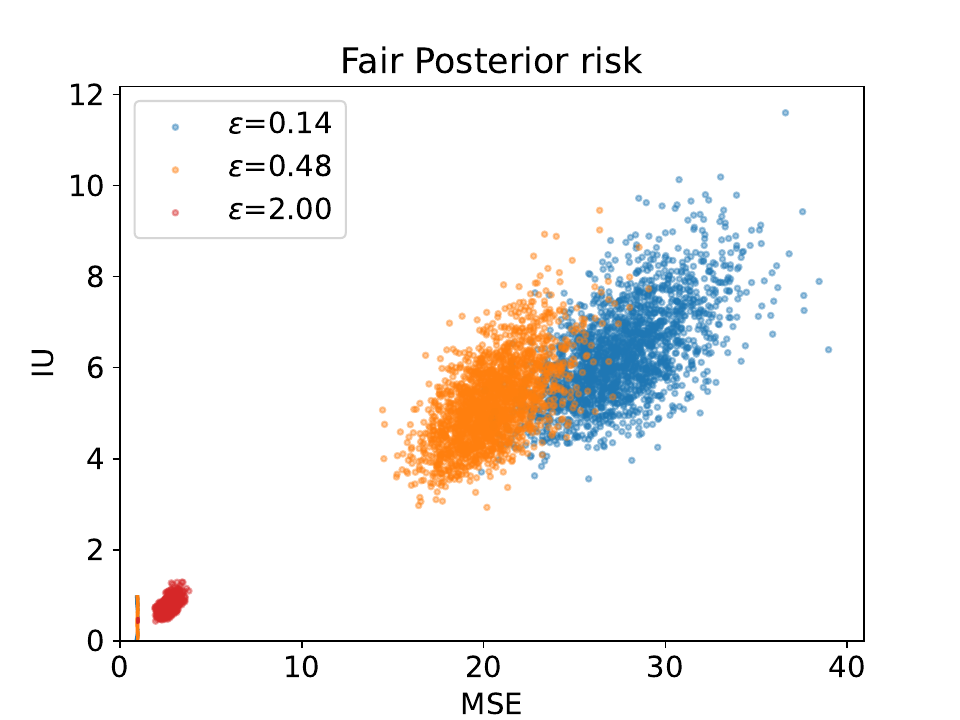}
    \caption{Performance of each posterior sample for 3 values of the unfairness budget $\epsilon \in \{0.39, 0.96, 2\}$. Left: WDU vs MSE. Right: IU vs MSE.}
    \label{fig:post-risk-ppp}
\end{figure}

Below, we provide a comparison of the marginal posterior distributions of FP and PP (\Cref{fig:marg-post}) and a visualisation of the performance of FP across different unfairness budgets (\Cref{fig:post-risk-ppp}).
%
%Figure \ref{fig:marg-post} represents the marginal posterior distributions on some coefficients $\beta_i$ for a budget $\epsilon = 0.155$. 
%
%
While the Pareto plots in \Cref{fig:reg-pareto-all} (top row) show very similar performance of the Fair Posterior (FP) and the projected posterior (PP), Figure \ref{fig:marg-post} demonstrates that they are nonetheless significantly different posterior distributions. Figure \ref{fig:marg-post} 
%in Appendix \ref{app:num_results} 
 represents the marginal posterior distributions on four coefficients for a fixed unfairness budget ($\epsilon = 2.812$). We observe that FP generally lead to higher shrinkage and has a higher variance %the fair coefficients and less the unfair ones 
than PP. However for both posteriors we observe that coefficients corresponding to unfair and/or weak covariates %($\beta_5$ and $\beta_7$ 
have a non-zero posterior probability of being 0, thus of being excluded.

%\jack{[Note for later to add non-fair l1-ball here]}

In Figure \ref{fig:post-risk-ppp}, we visualise the performance of 1000 posterior samples from the Fair Posterior (FP) for different unfairness budgets in two dimensions: WDU vs MSE (left) and IU vs MSE (right). This figure shows how, as the unfairness budget increases, posterior samples become more DP-unfair, less individually unfair and lead to better fit, and the variance decreases.

\subsection{Simulation 2: Logistic regression}

In the logistic regression setting, we sample the sensitive attributes and the design matrix as in linear regression in Simulation 1, and %we sample 
the outcomes $(y_i)_i$ as $y_i \sim Ber( \mu_i)$ with
\begin{align*}
    \mu_i = (1 + e^{-(\beta_0 + \beta^T x_i)} )^{-1},
\end{align*}
with $\beta^*$ as in \eqref{eq:beta_star} and a null intercept.
\ds{The training and test sample size is 500. We set $\widehat \rho$ as the empirical correlation coefficient and $\xi = 10$  in the IU metric. To plot the Pareto fronts of the Bayesian methods, we average the performance metrics across the posterior predictive distribution, e.g., for the BCE we compute
\begin{align*}
  \frac{1}{n_{test}} \sum_i - y_i \log p(\hat y_i=1 \mid x_i) - (1-y_i) \log ( 1- p(\hat y_i=1 \mid x_i)),
\end{align*}
where $p(\hat y_i=1 \mid x_i) = \int p(\hat y_i=1 \mid x_i, \beta_0, \beta)d\pi(\beta_0, \beta \mid \mathbf{X}, \mathbf y)$ is the posterior predictive, approximated via Monte-Carlo.}

\begin{figure}
    \centering
        \includegraphics[width=0.4\linewidth, trim = 0.5cm 0cm 1cm 1.4cm, clip]{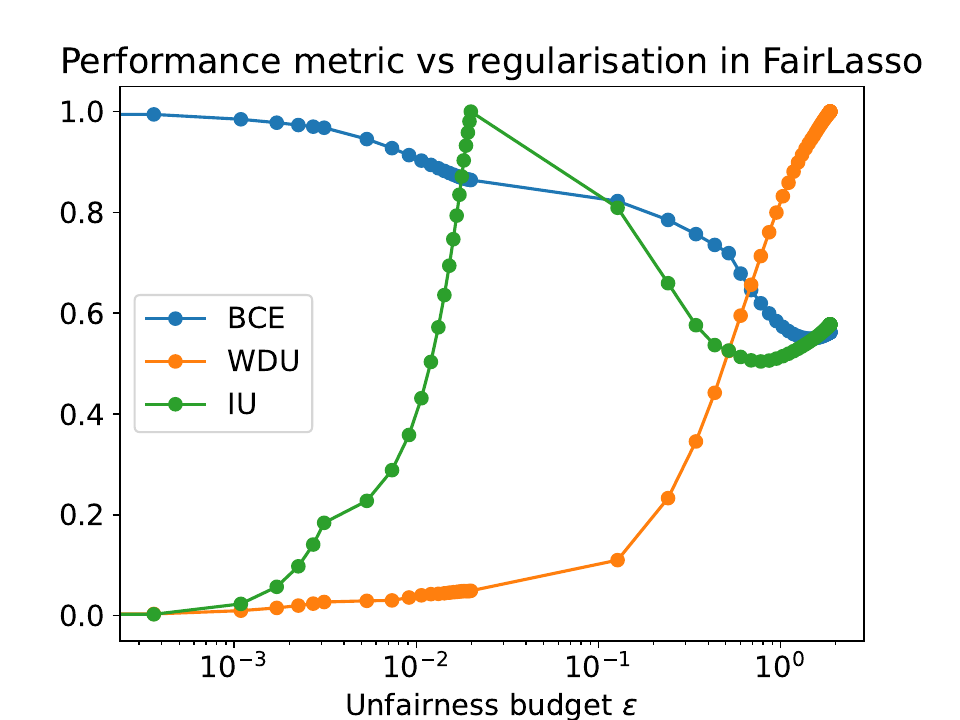}%
        \includegraphics[width=0.48\linewidth, trim = 0.5cm 0cm 2cm 0cm, clip]{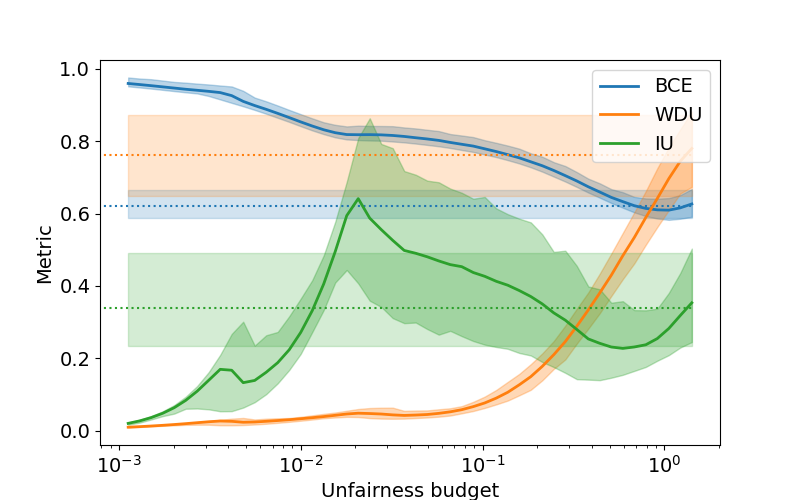}%
    \caption{
    Left: Test performance (BCE, WDU, IU) of the Fair Lasso vs unfairness budget $\epsilon$ (in log scale) in the logistic regression setting. Right: Test performance metrics of the Fair Posterior vs unfairness budget $\epsilon$; the bold lines correspond to the posterior averaged metric around which the 95\% pointwise credible bands are coloured. The horizontal dotted lines and colored areas correspond to the conjugate posterior mean and 95\% pointwise credible bands respectively. The metrics are normalised by their largest value in each method. 
    %\ds{look at path and for interpretation. Ask Marco for explanation for penalty in logistic regression
    }
    %Performance (AUC, Parity Gap, and Individual Fairness) vs unfairness budget $\epsilon$ of PPP (left) and PP (right). % and SP (bottom plot). 
    %The bold lines correspond to the posterior averaged metric around which the 95\% pointwise credible bands are coloured. The horizontal dotted lines and colored areas correspond to the conjugate posterior mean and 95\% credible bands respectively.
    \label{fig:class-comp-cp}
\end{figure}

We first evaluate the variation of performance (in terms of IU, WDU, BCE) of the Fair Lasso and Fair Posterior as the unfairness budget decreases. In \Cref{fig:class-comp-cp} we observe similar trends as in the linear regression simulation: BCE (resp. WDU) monotonically increases (resp. decreases) as $\epsilon$ decreases while IU first increases (to trade-off some DP unfairness as variables dependent on $S$ are excluded) then decreases as coefficients corresponding to $S$-independent variables are shrunk.
%However, contrary to the linear regression simulation where the residual error (MSE) and IU were very small for high unfairness budgets, here these two metrics remain around 0.6 in the unconstrained model due to the relatively small sample size.

%IU is not null, although the true model does not have any direct effect of $S$ onto the outcome probability, which \ds{is probably due to the fact that the unconstrained model has residual error on individual probabilities.

%\jack{`which may be' is too weak}
%\jack{We need to be more concrete than this, see comment on MSE = 0 vs BCE not 0}
% the unconstrained model is not individually nor DP fair, thus here all performance metrics  globally decrease as $\lambda$ increases .

\begin{figure}
    \centering
    \includegraphics[width=0.33\linewidth, trim = 0.cm 0cm 1cm 0cm, clip]{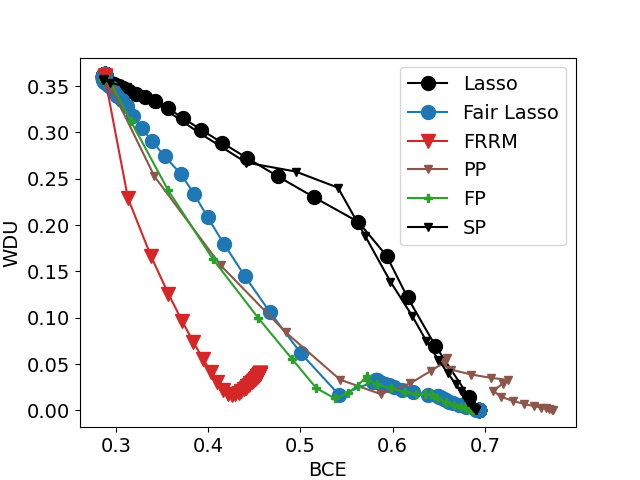}%
    \includegraphics[width=0.33\linewidth, trim = 0.cm 0cm 1cm 0cm, clip]{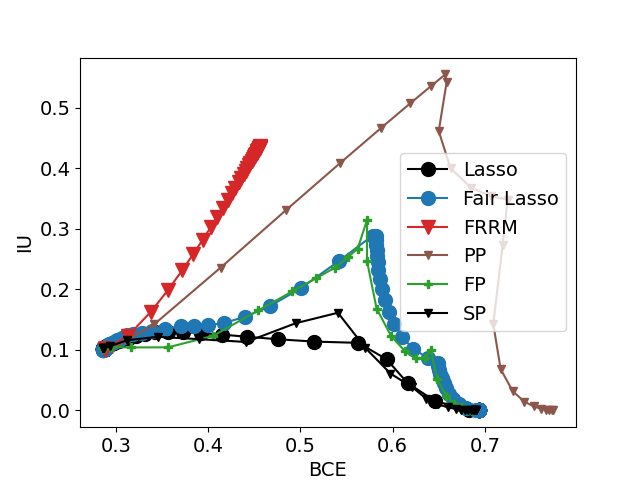}
    \includegraphics[width=0.33\linewidth, trim = 0.cm 0cm 1cm 0cm, clip]{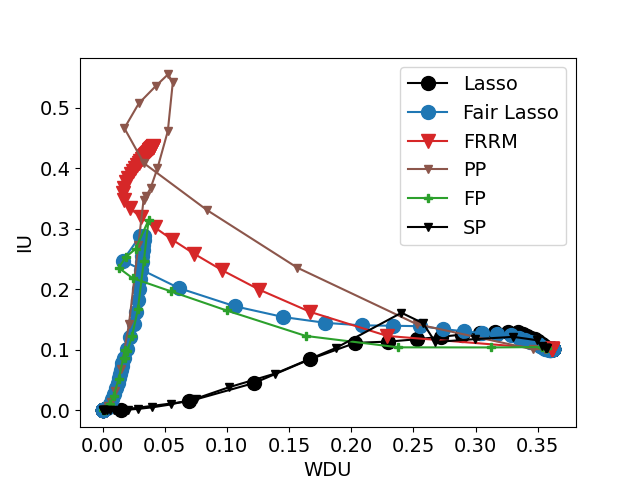}
    \caption{
    Pareto fronts of the proposed and baseline methods: Fair Lasso (blue), FP (green), PP (brown), FRRM (red) and the fairness-unaware methods Lasso and SP (black), in the logistic regression simulation. All metrics are evaluated on the test set. For FP, PP and SP: metrics are averaged over the posterior predictive distribution for each unfairness budget.  Left: WDU vs BCE. Middle: IU vs BCE. Right: IU vs WDU.}
    \label{fig:class-pareto-all}
\end{figure}

We then compare the Pareto fronts of Fair Lasso and FP to the baseline methods. In Figure \ref{fig:class-pareto-all} we observe that while FRRM and MD achieve better WDU-BCE trade-offs when WDU is moderately large (between 0.03 and 0.15), it cannot achieve very small values of WDU as well as of IU. In comparison, for a fixed level of WDU, Fair Lasso and FP have lower BCE than Lasso and SP and lower IU than FRRM and MD. These plots thus show once again that Fair Lasso and FP can achieve better performance trade-offs between BCE, IU and WDU.
%The performance trade-offs of PP and FP as $\epsilon$ varies are represented in Figure \ref{fig:class-comp-cp}. In Figure \ref{fig:class-pareto-all} we represent the  Pareto plots of our proposed and baseline methods.
Besides, this simulation clearly shows that FP outperforms PP for almost all unfairness budgets, confirming our intuition that our Fair Posterior based on projecting a prior leads to better predictive performance and thus better trade-offs with fairness than a projected posterior.

\begin{figure}
    \centering 
    \includegraphics[width=0.33\linewidth, trim = 0.cm 0cm 1cm 0cm, clip]{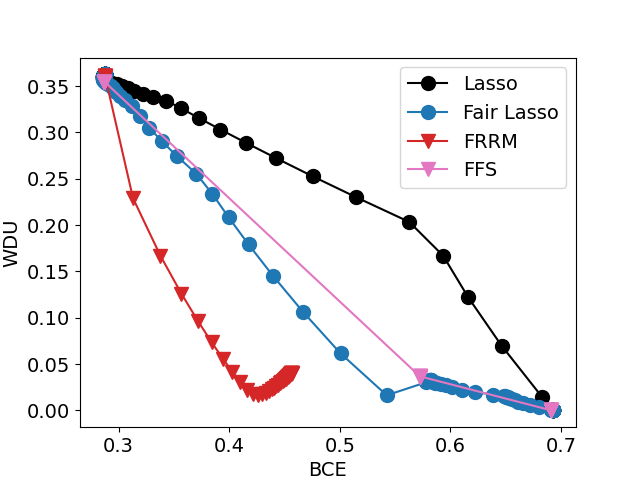}%
    \includegraphics[width=0.33\linewidth, trim = 0.cm 0cm 1cm 0cm, clip]{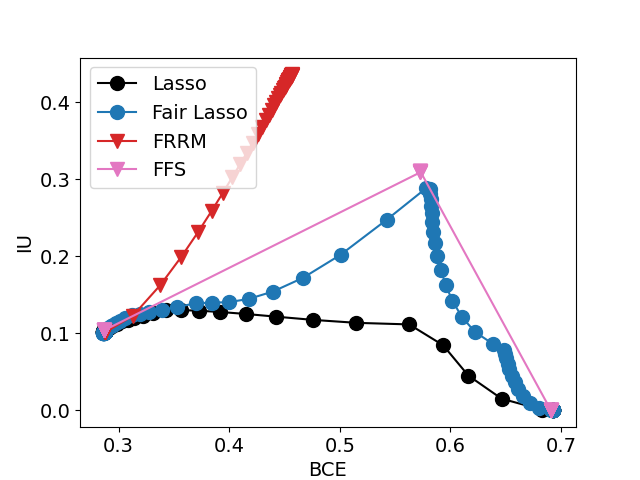}%
    \includegraphics[width=0.33\linewidth, trim = 0.cm 0cm 1cm 0cm, clip]{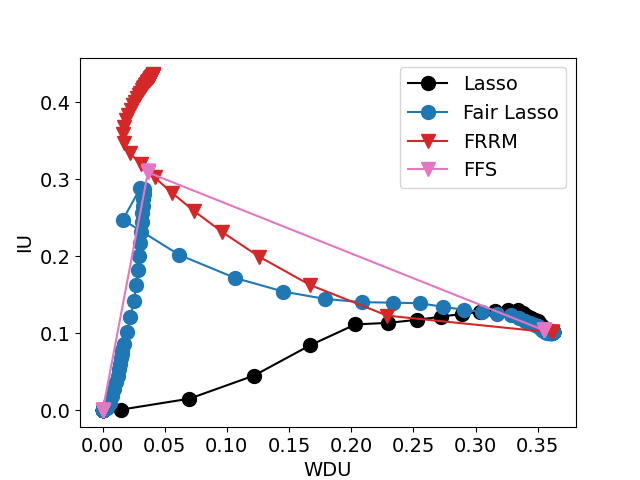}
    \caption{
    Pareto fronts of the Fair Lasso (blue), FRRM (red), Lasso (black) and FFS (pink), in the logistic regression simulation. All metrics are evaluated on the test set. Left: WDU vs BCE. Middle: IU vs BCE. Right: IU vs WDU.}
    \label{fig:class-pareto-pen}
\end{figure}

Finally, we report here an additional comparison of the Fair Lasso to the FFS method \citep{belitz2021automating}. \Cref{fig:class-pareto-pen} represents the Pareto front of all fairness-penalised methods (Fair Lasso, FRRM, Lasso and FFS). We observe that while FFS can achieve similar WDU-BCE trade-offs for high and low values of WDU (left), it cannot achieve intermediate values of WDU (since no shrinkage is applied to the parameters) and lead to worse trade-offs with IU (middle and right plots). %This therefore shows that the greedy algorithm is not able to optimally select the set of featres in this setting  .

\subsection{Communities and Crime data set}

\begin{figure}[h!]
    \centering
    \includegraphics[width=0.49\linewidth, trim = 0.5cm 0cm 1cm 1.4cm, clip]{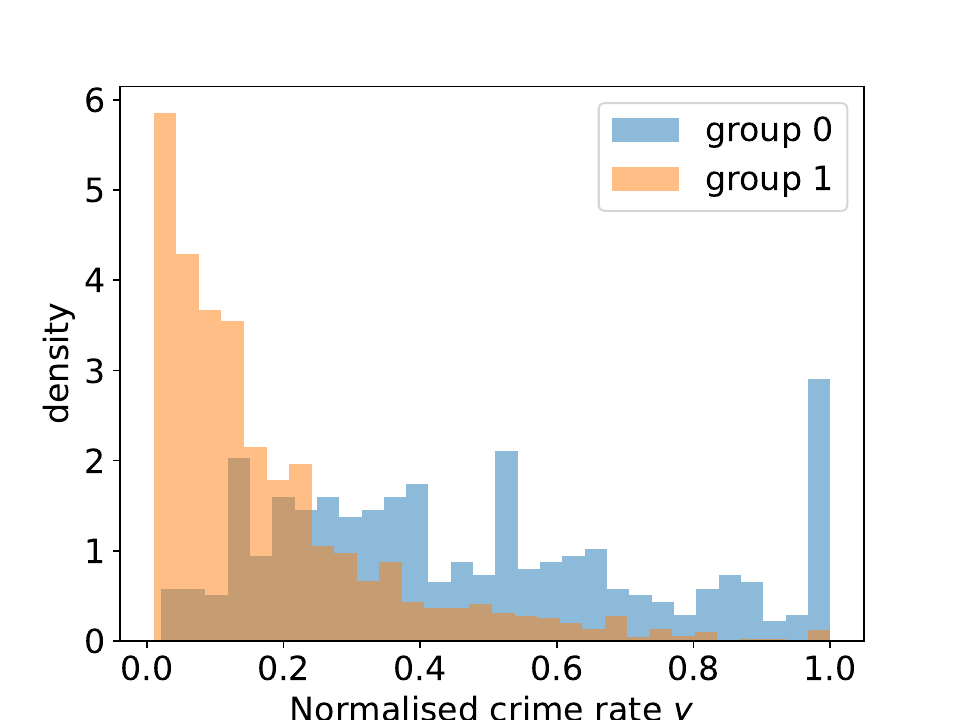}%
    \includegraphics[width=0.49\linewidth, trim = 0.cm 0cm 1cm 1.4cm, clip]{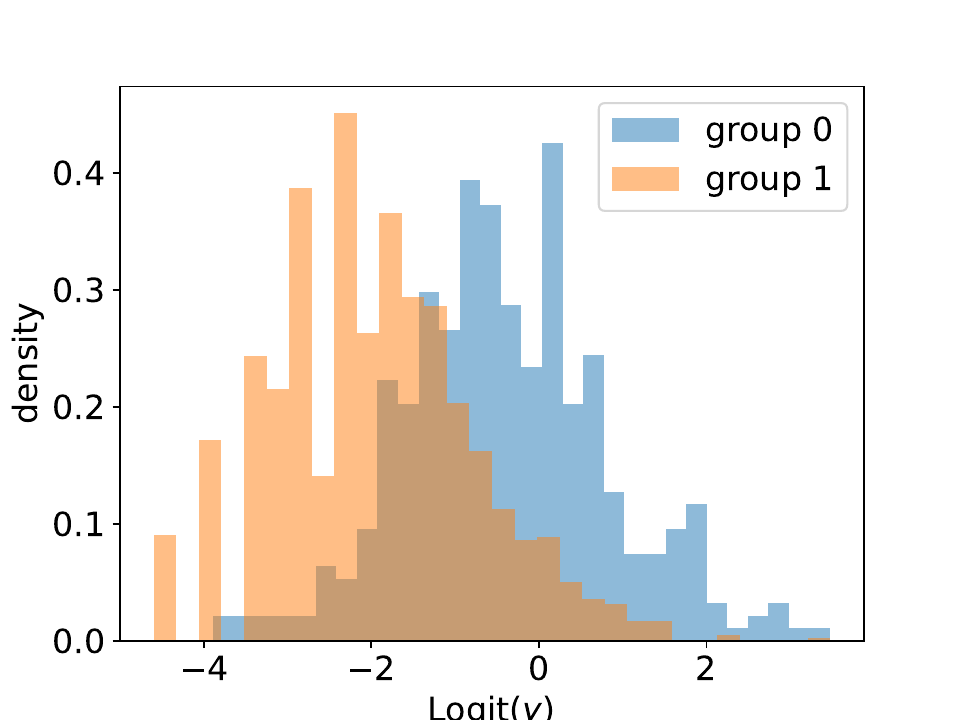}
    \caption{Crime rate distribution in the Community data before pre-processing (left) and after pre-processing (right).  Group 1 (resp. group 0) gathers neighborhoods with a majority of white (resp. non-white) population.
    % \jack{Needs x and y axis labels - will help identify the differences}\ds{density vs y or log(y)}
    % \jack{I think this one can go in the appendix}
    }
    \label{fig:community}
\end{figure}

\ds{This dataset is one of the benchmark datasets in fair machine learning and is available on the UCI Machine Learning Repository.\footnote{https://archive.ics.uci.edu/dataset/183/communities+and+crime} Following previous analysis, we treat race as the binary and sensitive attributes ($s_i=1$ for neighborhood for mainly white population and 0 otherwise). We apply standard pre-processing of the variables, namely a standardisation of the design matrix (statistics computed from the training set and also applied to both the training and test sets) and a logit transformation of the target outcome (the crime rate, between 0 and 1), after removing the samples with  output values outside of the range $[0.01, 0.99]$ (since the logit function is not defined on 0 and 1). The crime rate distributions in the whole dataset before and after pre-processing are plotted in Figure \ref{fig:community}. The DP unfairness of the observed outcome is $WDU=1.578$. The training and test (80\%/20\%) sample sizes are respectively 1552 and 388.}

\begin{figure}
    \centering
    \includegraphics[width=0.55\linewidth, trim = 0.5cm 0cm 0.3cm 1.2cm, clip]{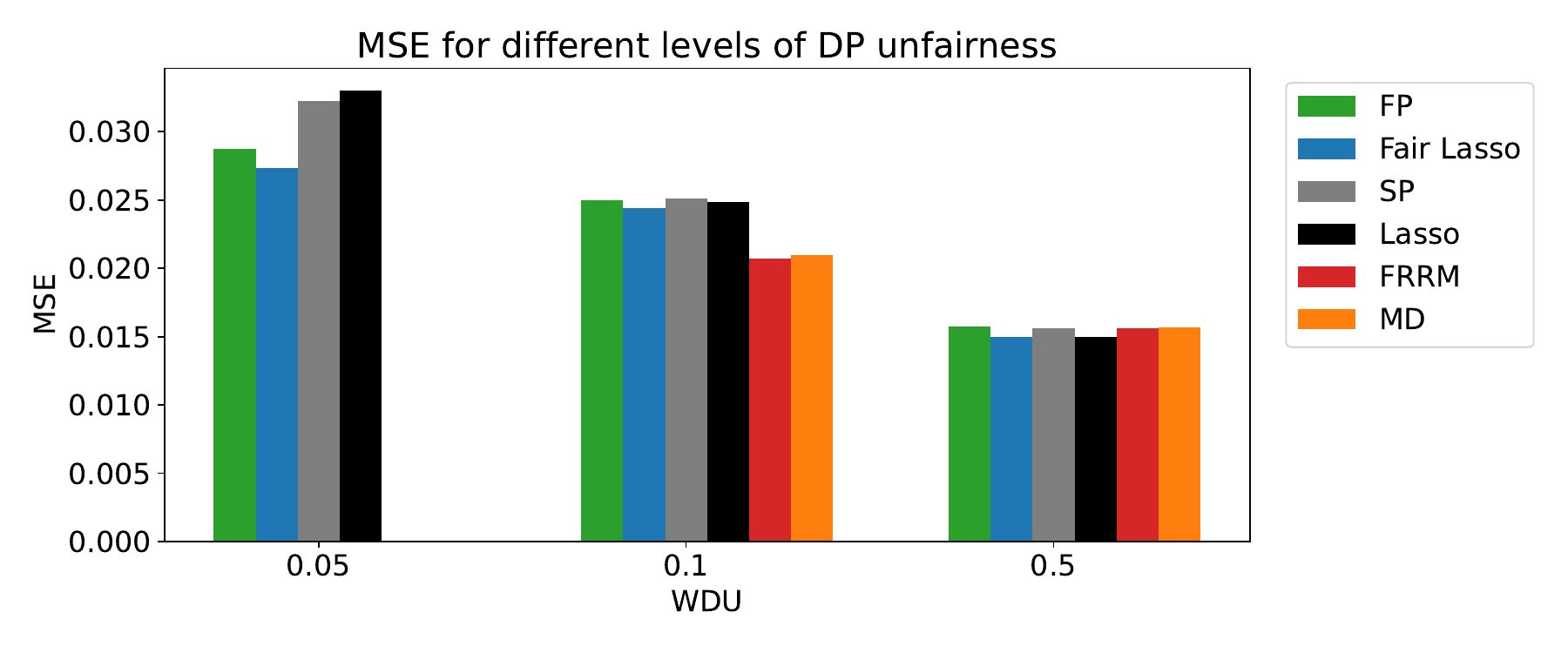}%
    \includegraphics[width=0.45\linewidth, trim = 0.5cm 0cm 6cm 1.2cm, clip]{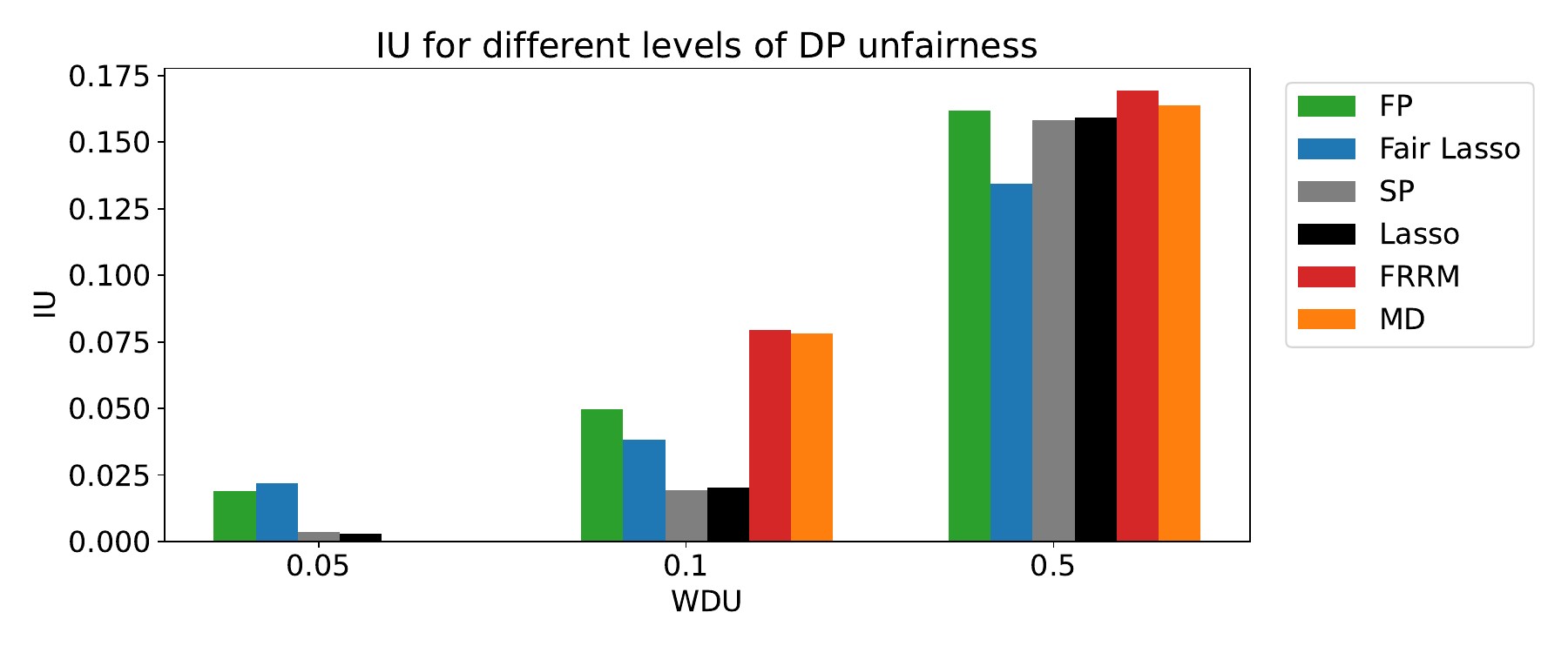}
    \caption{MSE (left) and IU (right) of the different methods at 3 levels of DP unfairness (WDU $= 0.05, 0.1, 0.5$). 
    %\jack{Can we fix the colours across plots}
    %\jack{So for fixed WDU, we have better or equal mse than LASSO and SP, and better IU than FRMM and MD}
    }
    \label{fig:community-barplot}
\end{figure}

\begin{figure}
    \centering
    \includegraphics[width=0.55\linewidth, trim = 0.5cm 0cm 0.3cm 1.2cm, clip]{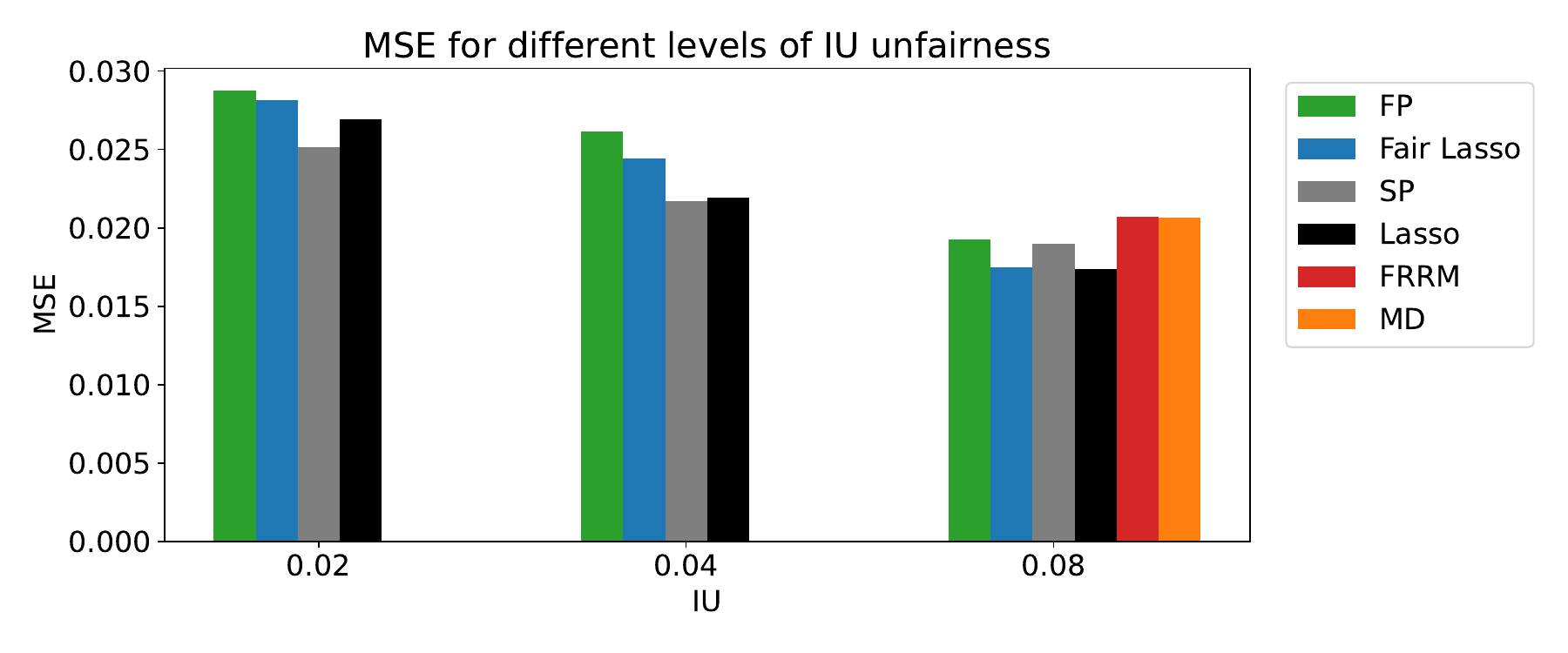}%
    \includegraphics[width=0.45\linewidth, trim = 0.5cm 0cm 6cm 1.1cm, clip]{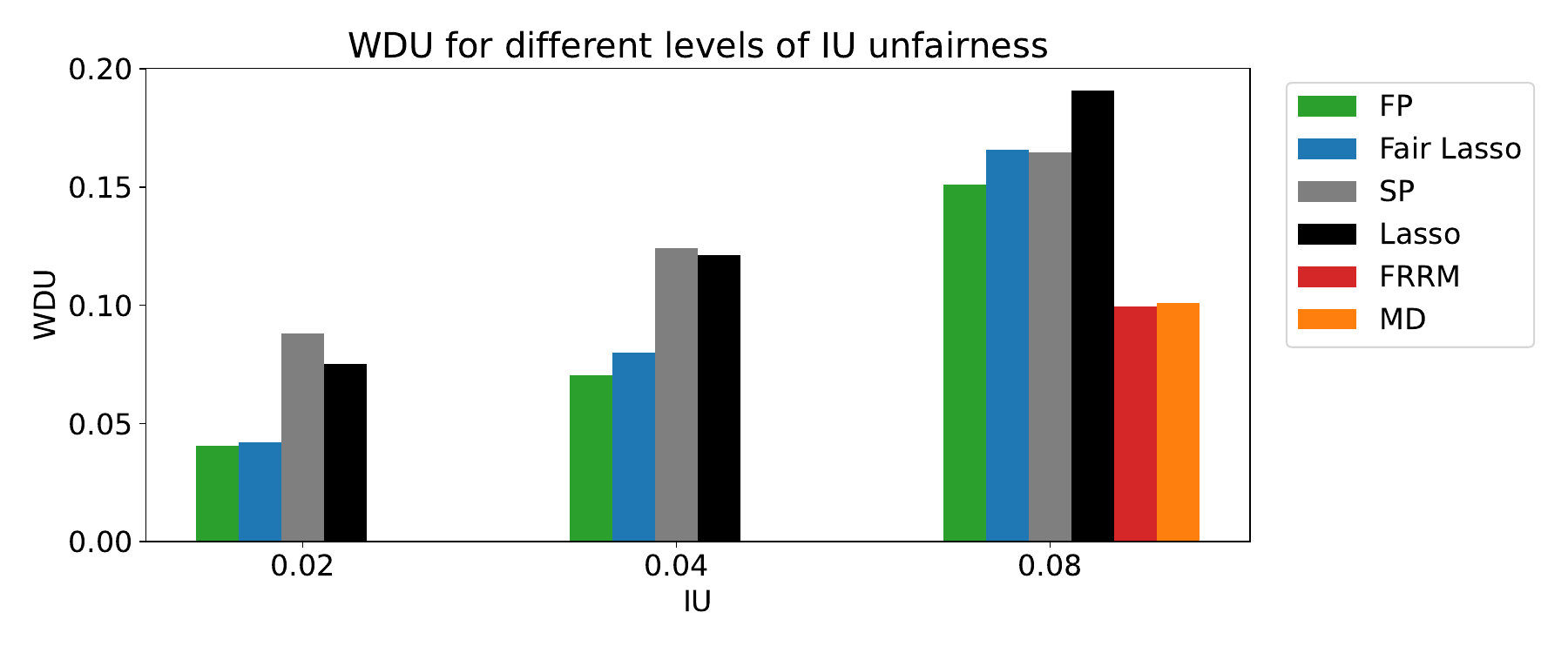}
    \caption{MSE (left) and WDU  (right) of the different methods at 3 levels of individual unfairness (IU $= 0.03, 0.05, 0.1$). 
    %For fixed IU, lower MSE than FRRM and MD and lower WDU than Lasso
    }
    \label{fig:community-barplot-iu}
\end{figure}

\ds{We set $\widehat \rho$ as the correlation coefficient and $\xi=2$ in IU. The Pareto fronts in Figure \ref{fig:reg-pareto-all} (bottom row) are computed as in Simulation 1. Note that contrary to our simulation settings, here both WDU and IU decreases as MSE increases (i.e., the unfairness budget decreases) because the unconstrained model (estimated for high $\epsilon$) is not individually fair. To better evaluate fairness and predictive performance trade-offs, we evaluate MSE and IU (resp. MSE and WDU) for fixed levels of WDU (resp. IU) of the best predictive (lowest MSE) of each method method. For the Bayesian methods, we select the model with lowest median MSE and report the median metrics across posterior samples %\jack{[still true?]}
. These results are plotted in Figures \ref{fig:community-barplot} and Figure \ref{fig:community-barplot-iu}).}

\subsection{German credit data}

\begin{figure}[h!]
    \centering
    \includegraphics[width=0.5\linewidth, trim = 0.5cm 0cm 1cm 1.4cm, clip]{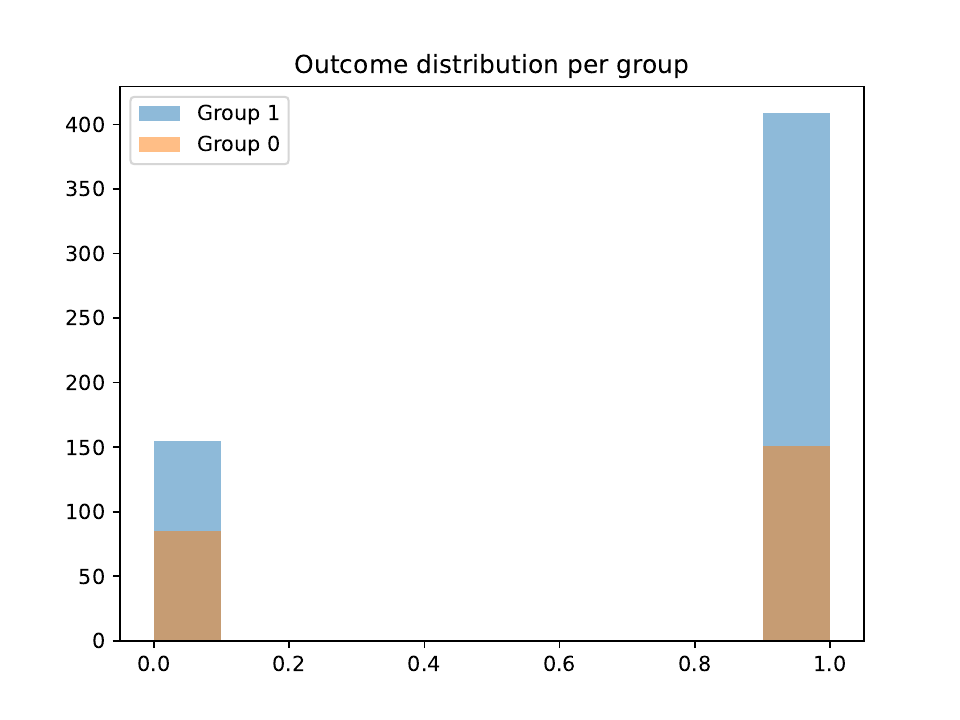}%
    \caption{
    %\jack{Appendix}
    Outcome (credit risk) distribution in the German credit data.  Group 1 corresponds to men while Group 0 corresponds to women.}
    \label{fig:credit}
\end{figure}

This is another benchmark dataset, available via the package AIF360\footnote{https://github.com/Trusted-AI/AIF360}. It contains attributes and the credit risk (1 for good and 0 for bad) of 1,000 credit applicants. Here we consider sex as the sensitive attribute, i.e., $s_i=1$ for men (around 71\% of total samples) and 0 for women (around 29\% of total samples). % and the outcome variable is the credit risk .
The credit risk distribution across mean and women is plotted in Figure \ref{fig:credit} and shows that a larger proportion of male applicants have good credit risk (73\% vs 64\% for women). In fact, the Parity Gap, defined as the absolute difference between the frequency of positive outcome for men and women, in the training set (80\% of total samples) is around $0.085$.
%\jack{This is the first time we define this!}. 

\ds{For pre-processing, categorical variables are encoded as one-hot vectors and continuous  variables are standardised. We set $\widehat \rho$ as the correlation coefficient for continuous variables and the Cramer's V association coefficient for the categorical ones.} 

\begin{figure*}[h!]
    \centering
    \includegraphics[width=0.33\linewidth, trim = 0.cm 0cm 1cm 0cm, clip]{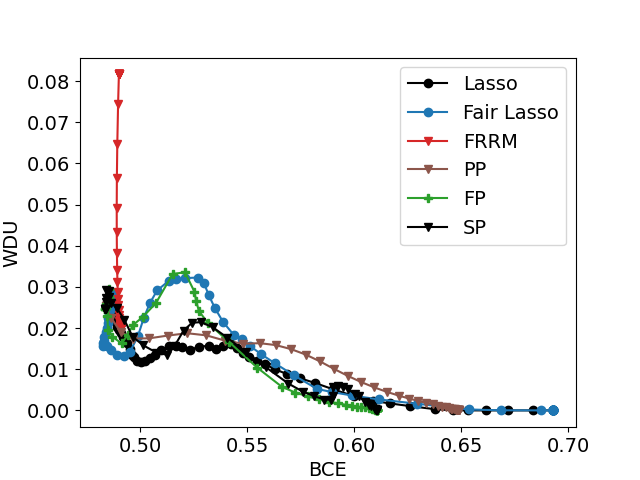}%
    \includegraphics[width=0.33\linewidth, trim = 0.cm 0cm 1cm 0cm, clip]{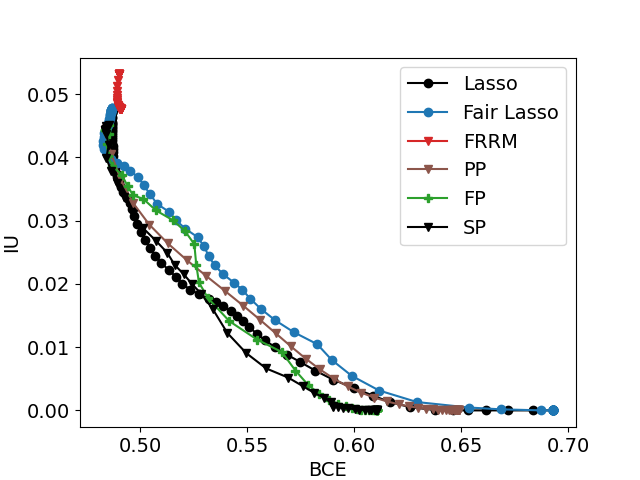}
    \includegraphics[width=0.33\linewidth, trim = 0.cm 0cm 1cm 0cm, clip]{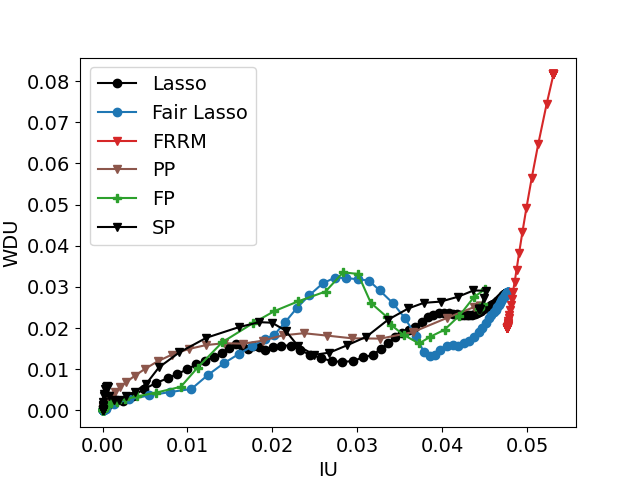}
    \caption{Pareto fronts of the Fair Lasso, Lasso, FRRM, FP, PP and SP on the test set of the German Credit dataset. Top left: WDU vs BCE. Top right: IU vs BCE. Bottom: IU vs WDU.} % The performance metrics are normalised by their largest value across methods.}
    \label{fig:credit-pareto-all}
\end{figure*}

\begin{figure}[h!]
    \centering
    \includegraphics[width=0.55\linewidth, trim = 0.cm 0cm 0.3cm 0.7cm, clip]{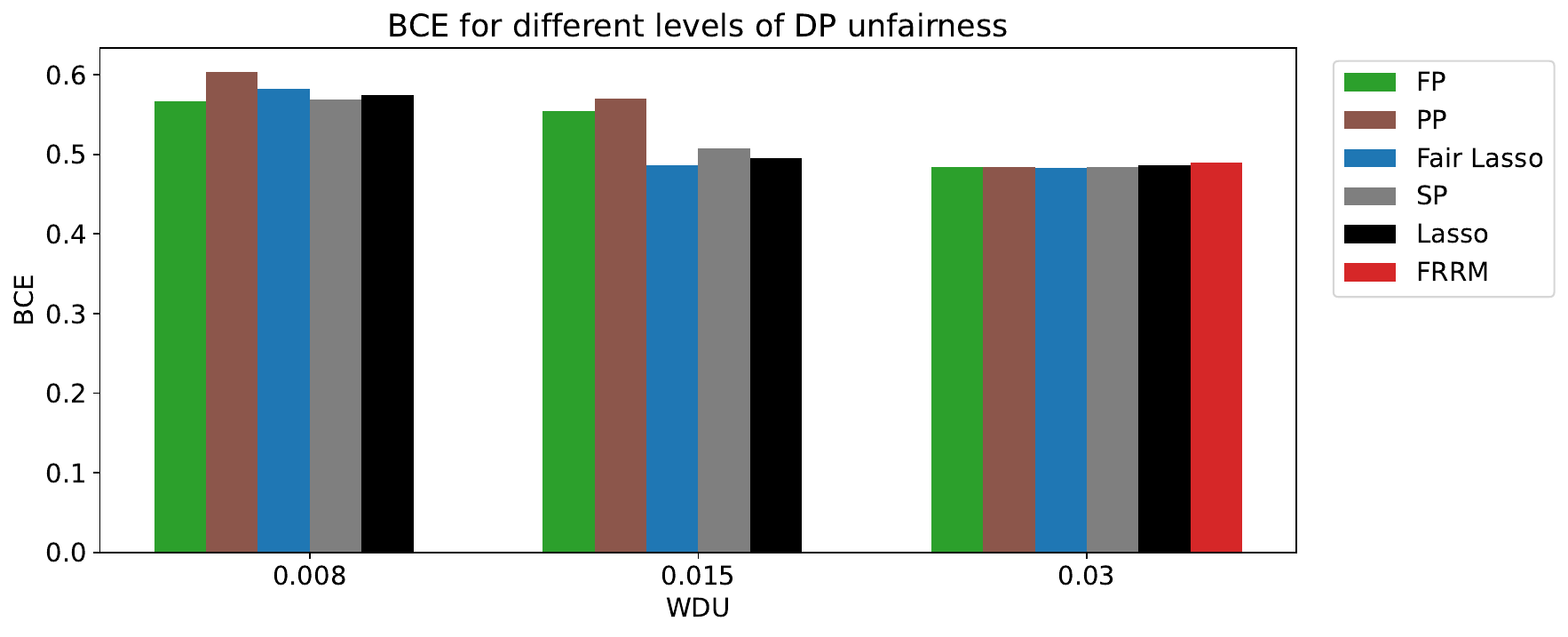}%
    \includegraphics[width=0.45\linewidth, trim = 0.cm 0cm 5cm 0.7cm, clip]{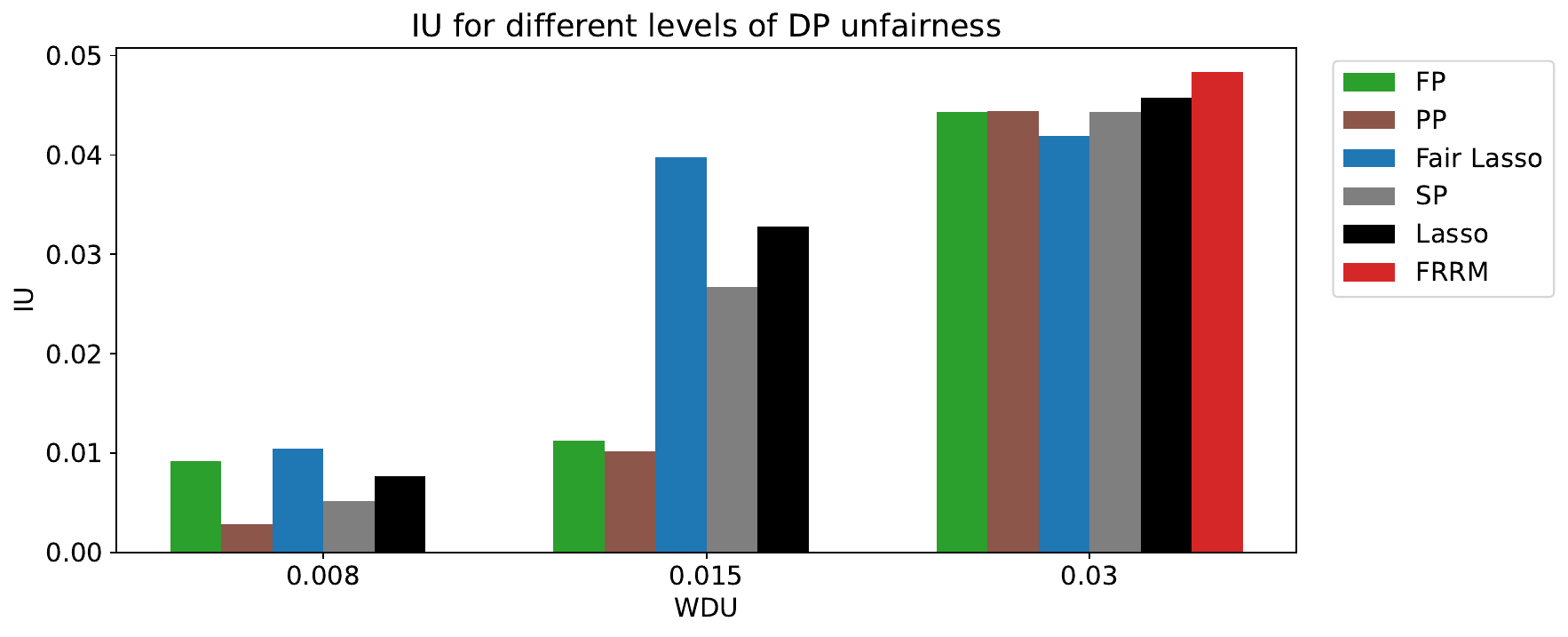}
    \caption{BCE (left) and IU (right) of the different methods at 3 levels of WDU ($0.02, 0.04, 0.07$) on the test set of the German Credit dataset. %PP is excluded from this graph as it is always worse than the other methods.
    }
    \label{fig:credit-barplot}
\end{figure}

\begin{figure}
    \centering
    \includegraphics[width=0.55\linewidth, trim = 0.cm 0cm 0.3cm 0.7cm, clip]{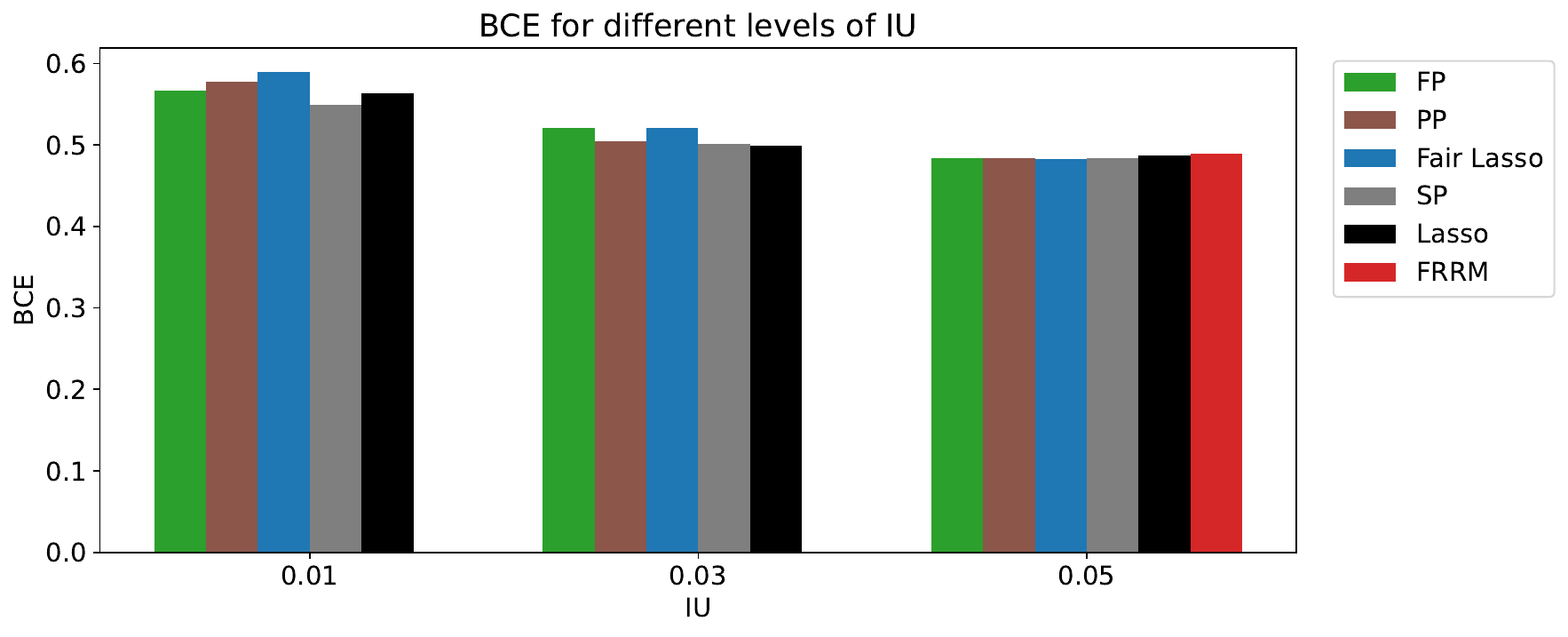}%
    \includegraphics[width=0.45\linewidth, trim = 0.cm 0cm 5cm 0.7cm, clip]{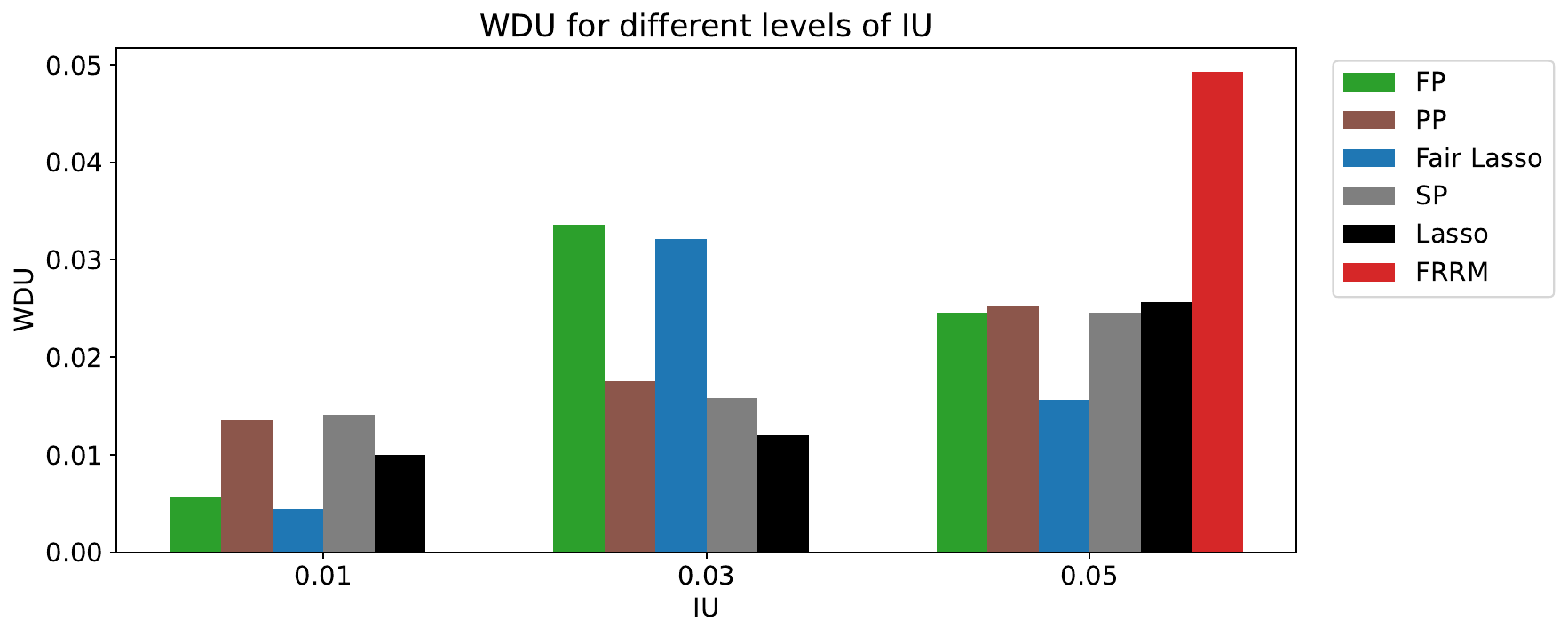}
    \caption{BCE (left) and WDU (right) of the different methods at 3 levels of individual unfairness ($0.015, 0.03, 0.05$) on the test set of the German Credit dataset. %PP is excluded from this graph as it is always worse than the other methods.
    }
    \label{fig:credit-barplot-iu}
\end{figure}

\ds{Figure \ref{fig:credit-pareto-all} represents the %2D
Pareto fronts of all methods (the FFS estimate was excluded because of its much worse performance than any other method in simulation). We observe here that all methods except FRRM (which is always worse) have very close performance. We make similar observation on the bar plots in Figures \ref{fig:credit-barplot} and \ref{fig:credit-barplot-iu}, where BCE and IU (resp. BCE and WDU) are evaluated for fixed levels of WDU (resp. IU). While FP (and PP) have the lowest IU for fixed and middle level of WDU, the Lasso  has the lowest WDU when IU is fixed to a medium value. Therefore, this experiment does not show a clear advantage of the FP and Fair Lasso over the standard Lasso. This could be due to the small amount of unfairness in the data or result from some mislabeling of the sensitive variable (sex) (as pointed out by \citep{gromping2019south}). Nonetheless, it demonstrates that group and individual unfairness can be reduced, with a moderate increase of BCE, by variable selection methods.}

%\ds{This data has some error on the sex variable}

% \begin{figure}[h!]
%     \centering
%     \includegraphics[width=0.5\linewidth, trim = 0.5cm 0cm 1cm 1.4cm, clip]{figures/germancredit_data.pdf}%
%     \caption{
%     \jack{Appendix}
%     Outcome (credit risk) distribution in the German credit data.  Group 1 (resp. group 0) gathers neighborhoods with a majority of white (resp. non-white) population.}
%     \label{fig:credit}
% \end{figure}

%Here we report the outcome distribution in the credit data set (Figure \ref{fig:credit}) as well as the methods' performance for fixed levels of DP unfairness (Figure \ref{fig:credit-barplot}) and individual unfairness (Figure \ref{fig:credit-barplot-iu})

% \section{Formatting Instructions}

% The appendices follow the same formatting instructions as in the main paper.
% The only difference is that the supplementary material must be in a \emph{single-column} format.

% Note that reviewers are under no obligation to examine your supplementary material.

% \section{Missing Proofs}

% The supplementary materials may contain detailed proofs of the results that are missing in the main paper.

% \subsection{Proof of Lemma 3}

% \textit{In this section, we present the detailed proof of Lemma 3 and then [ ... ]}

% \section{Additional Experiments}

% If you have additional experimental results, you may include them in the supplementary materials.

% \subsection{Effect of the Regularization Parameter}

% \textit{Our algorithm depends on the regularization parameter $\lambda$. Figure 1 below illustrates the effect of this parameter on the performance of our algorithm. As we can see, [ ... ]}

\end{document}